\documentclass[english,11pt]{article}

\usepackage{amsthm}
\usepackage[latin9]{inputenc}
\usepackage{babel}
\usepackage[hmargin=0.9in]{geometry}

\usepackage{subfig}
\usepackage{graphicx}
\usepackage[colorlinks=true,citecolor=blue,urlcolor=blue]{hyperref}
\usepackage{natbib}
\let\cite=\citep
\usepackage[dvipsnames]{xcolor}

\usepackage{float}
\usepackage{url}
\usepackage{bm}

\usepackage{amsmath}
\usepackage{amssymb,amsfonts}
\usepackage{upgreek}
\usepackage[textsize=footnotesize]{todonotes}
\usepackage{stmaryrd}

\usepackage{mwe,tikz}\usepackage[percent]{overpic}

\newtheorem{remark}{Remark}
\numberwithin{remark}{subsection}

\usepackage[export]{adjustbox}



\graphicspath{{./figures/}}

\newcommand{\bEta}{\bar{\eta}}
\newcommand{\bNu}{\bar{\nu}}
\newcommand{\bMu}{\bar{\mu}}

\newcommand{\bfn}{\mathbf{n}}

\newcommand{\order}{{\mathcal O}}
\newcommand{\yhat}{\hat{y}}

\def\avg[#1]{\left\langle #1 \right\rangle}
\def\intavg[#1]{\llbracket #1 \rrbracket}

\newcommand{\bB}{b_0}
\newcommand{\mwl}{\eta^*}
\newcommand{\bMwl}{\bar{\eta}^*}
\newcommand{\bigO}{\mathcal{O}}
\newcommand{\Rp}{R^+}
\newcommand{\Rn}{R^-}

\DeclareMathOperator{\sech}{sech}

\begin{document}
\title{Solitary water waves created by variations in bathymetry}
\author{
  Manuel Quezada de Luna
  \thanks{
    Computer, Electrical, and Mathematical Sciences \& Engineering Division,
    King Abdullah University of Science and Technology, 4700 KAUST, Thuwal
    23955, Saudi Arabia. (\{manuel.quezada, david.ketcheson\}@kaust.edu.sa).
  }
  \and
  David I. Ketcheson
  \footnotemark[1]
}
\maketitle

\begin{abstract}
  We study the flow of water waves over bathymetry that varies periodically along
  one direction.
  We derive a linearized, homogenized model and show that the periodic bathymetry induces
  an effective dispersion, distinct from the dispersion inherently present in
  water waves.
  We relate this dispersion to the well-known effective dispersion introduced
  by changes in the bathymetry in non-rectangular channels.
  Numerical simulations using the (non-dispersive) shallow water
  equations reveal that a balance between this effective dispersion and nonlinearity 
  can create solitary waves.
  We derive a KdV-type equation that approximates the behavior of these waves
  in the weakly-nonlinear regime.  We show that, depending on geometry,
  dispersion due to bathymetry can be much stronger than traditional water
  wave dispersion and can prevent wave breaking in strongly nonlinear regimes.
  Computational experiments using depth-averaged water wave models 
  confirm the analysis and suggest that
  experimental observation of these solitary waves is possible.
\end{abstract}

\section{Introduction}
\subsection{Wave propagation in media with a periodic structure}
The importance of laminates and composite materials in engineering led to the
study of elastic waves in periodically-varying media.  Long wavelength linear elastic waves
experience an effective dispersion that arises due to
the periodic variation in the material coefficients \cite{sunContTheoryLamMedium}.
Similar conclusions regarding general linear wave propagation were reached by
using Bloch expansions in \cite{santosa1991} and by using homogenization theory in \cite{Fish2001}.
In one dimension, this effective dispersion is the result of reflection, and its
strength is correlated with the degree of variation of the impedance in the medium.

For nonlinear elastic waves, a similar effect has been studied in \cite{leveque2003}.
The induced dispersion due to reflections leads to the formation of solitary waves
that behave similarly to those arising in nonlinear dispersive
wave equations like the Korteweg-de Vries (KdV) equation \cite{korteweg1895xli}.
This behavior also depends on the degree of variation in the impedance;
if it is not strong enough then there is little effective dispersion
and shocks tend to develop, as they would in a homogeneous medium 
\cite{2012_ketchesonleveque_periodic,ketcheson2020effRH}.
In the multidimensional setting, effective dispersion arises not only from
variation in the impedance, but also from variation in the
linearized sound speed \cite{quezada2014two}.
This latter source of effective dispersion can also lead to solitary wave
formation for nonlinear waves \cite{ketcheson2015diffractons}.
These waves have been called {\it diffractons} since they appear as a consequence of 
diffraction due to changes in the sound speed.

\begin{figure}[!h]
  \begin{centering}
    \subfloat[Bathymetry that changes periodically in one direction. \label{fig:periodic_channel}]{
      \includegraphics[scale=0.15,valign=c]{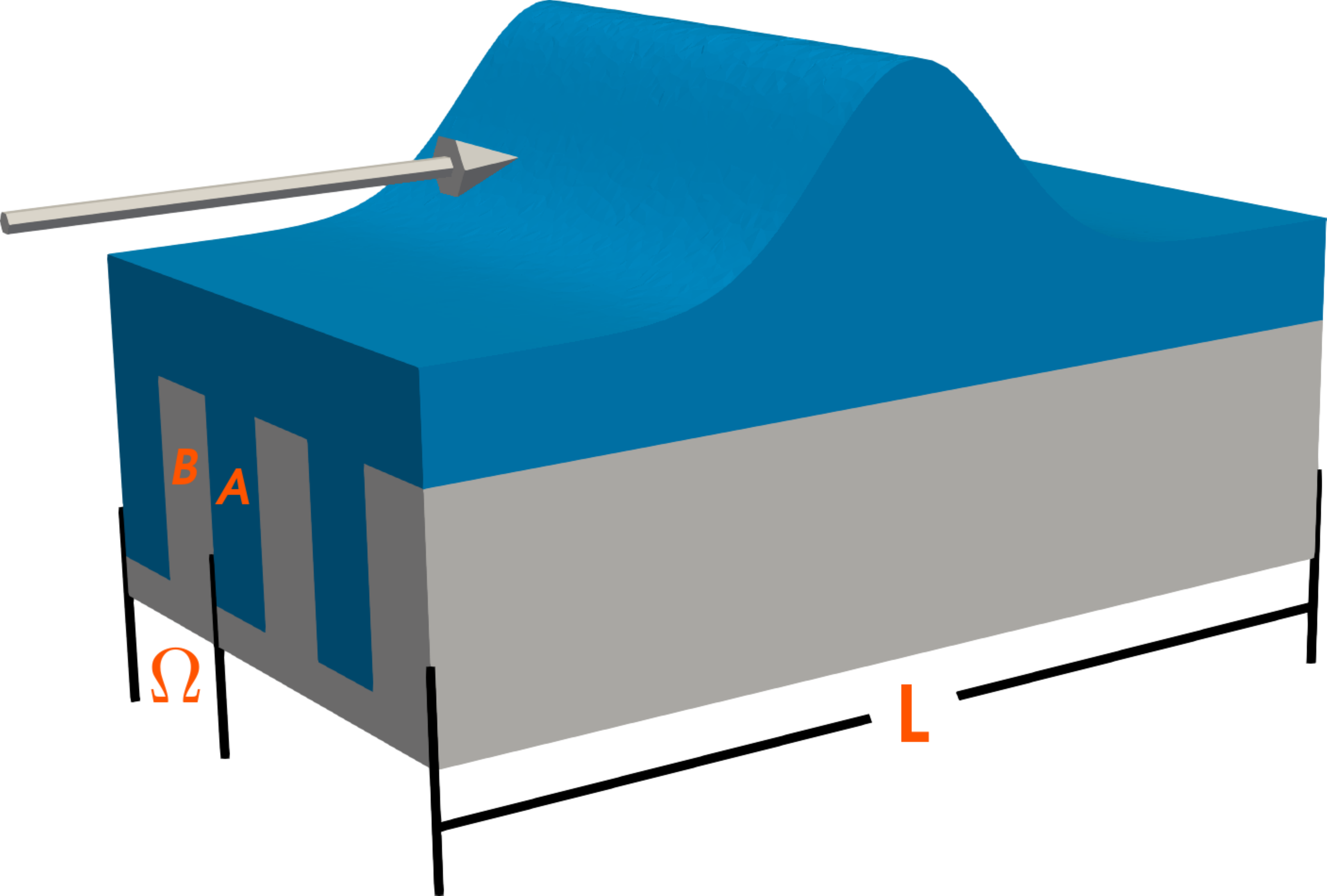}}
    \qquad\qquad
    \subfloat[Narrow channel with non-flat bathymetry. \label{fig:extracted_channel}]{
      \includegraphics[scale=0.15,valign=c]{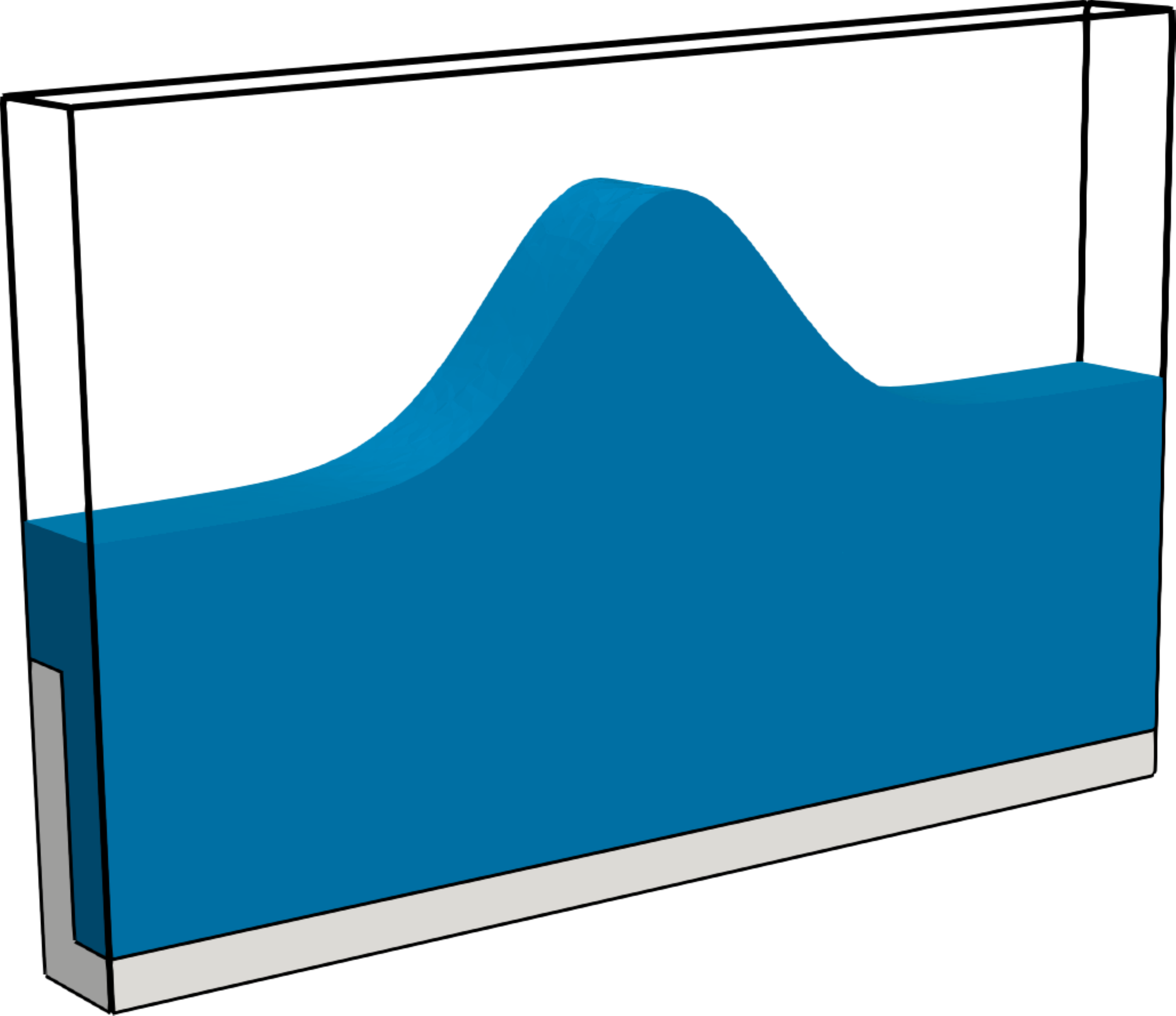}
    }
    \par
  \end{centering}
  \caption{An example of the kind of bathymetry studied in this work.
    \label{fig:channels_intro}}
\end{figure}

The first objective of the present work is to investigate similar effects for water waves.
Periodic variation of the medium is introduced through bathymetry
that varies periodically in one direction.  
In \S\ref{sec:dispersion_by_diffraction} we derive effective homogenized equations
for small-amplitude waves in this setting.  These equations show
that such waves experience an effective dispersion, similar in nature to the
effects discussed above for elastic waves.  We refer to this as \emph{bathymetric dispersion}.
These effective equations describe waves varying in two horizontal dimensions; if
restricted to plane waves propagating transverse to the variation in bathymetry
(as depicted in Figure~\ref{fig:periodic_channel}) they are similar to those
derived in \cite{chassagne2019dispersive}.
The amount of dispersion increases with the variation in the bathymetry, which is
also correlated with variation in the linearized impedance and sound speed for water waves.

In \S\ref{sec:SWEs_diffractons}, we perform numerical simulations
of the shallow water equations with periodic bathymetry and obtain solitary waves.
Our analysis and qualitative results
apply to general bathymetric profiles that are periodic in one direction.  For concrete illustration,
in most of the numerical examples we consider piecewise-constant bathymetry,
as depicted in Figure \ref{fig:periodic_channel}.

We refer to these solitary waves as {\it bathymetric solitary waves}, since they appear
only in the presence of varying bathymetry.
We compare some properties of these solitary waves with those of
the diffractons observed in \cite{ketcheson2015diffractons} and with KdV solitons.
In \S\ref{sec:kdv_for_diffractons}, based on the linear homogenized system, we propose a KdV-type equation,
which we solve numerically and compare versus direct simulations of the shallow water equations.
In \S\ref{sec:about_dispersive_effects} we compare our KdV model with other dispersive models. 
Finally, we make some concluding remarks in \S\ref{sec:conclusions}.

\subsection{Solitary water-waves in narrow non-rectangular channels}
\begin{figure}[!h]
  \begin{centering}
    \subfloat[Front view]{
      \includegraphics[scale=0.1]{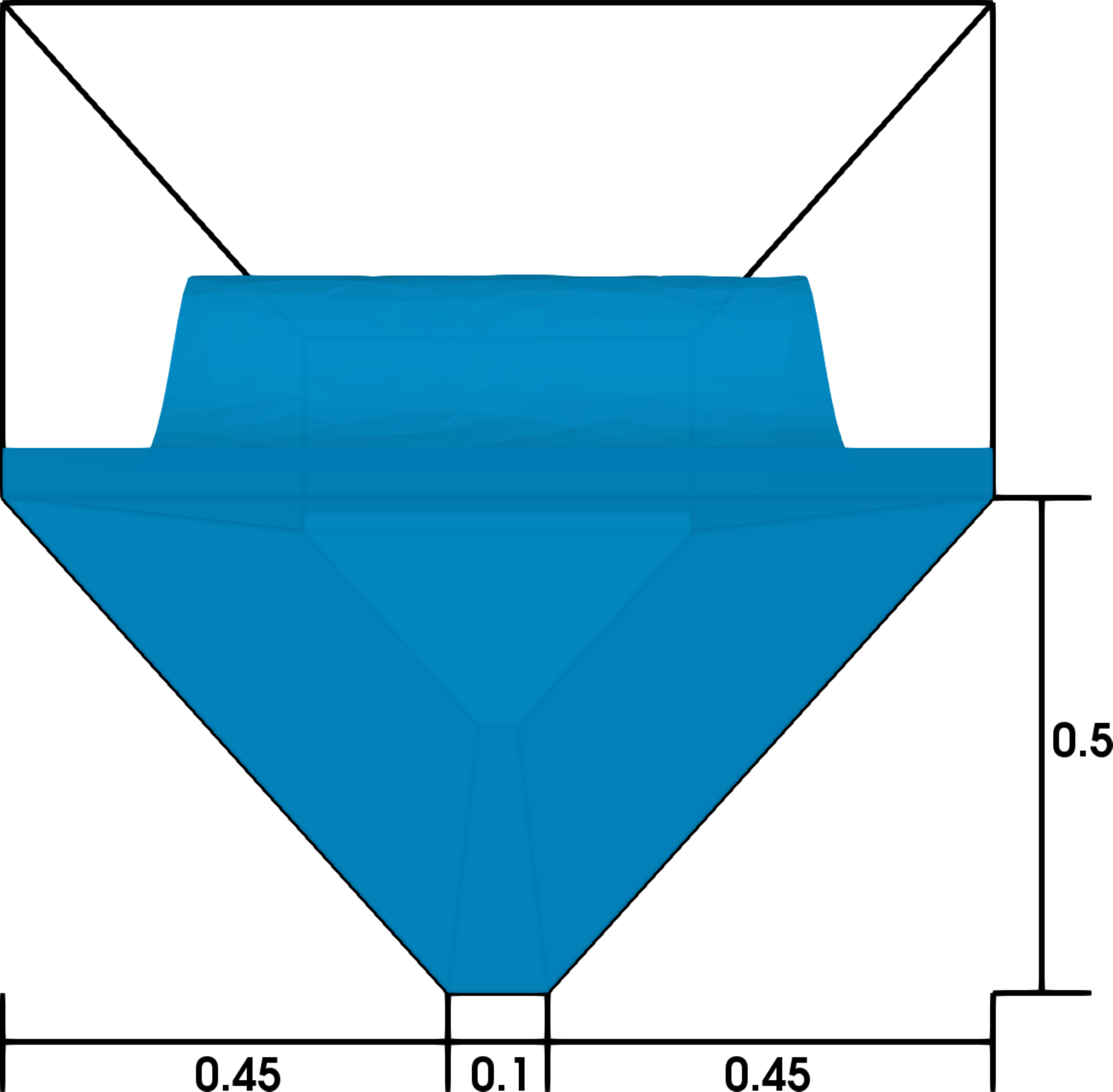}
    }
    \qquad
    \qquad
    \subfloat[Isometric view]{
      \includegraphics[scale=0.18]{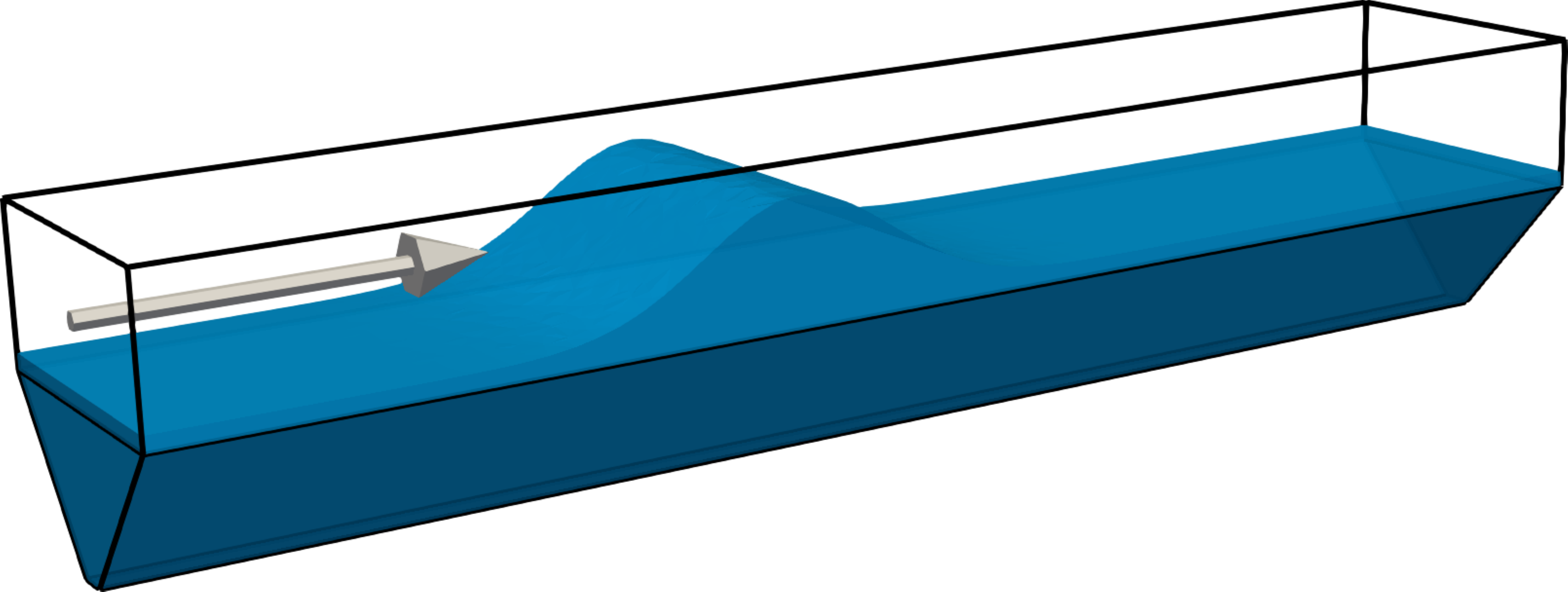}
    }
    \par
  \end{centering}
  \caption{Channel with inclined walls. The channel is infinitely long.
    \label{fig:channel}}
\end{figure}

Dispersion of water waves due to bathymetry has been analyzed and observed in a very different
setting: flow in a channel like that depicted in Figure \ref{fig:extracted_channel} 
or \ref{fig:channel}.
If the bottom of the channel is not flat,
small amplitude waves can be shown to obey an effective equation that
includes an additional dispersion, distinct from the inherent dispersion of
water waves and depending purely on the channel geometry
\cite{peregrine1968long,peters1966rotational,shen1969asymptotic,teng1997effects,chassagne2019dispersive}.
Just as in the presence of periodic bathymetry, here the effective dispersion
also increases with the amount of variation in the bathymetry.  
In fact, due to symmetry of the solution over periodic bathymetry, 
the two settings are equivalent. For example, consider the periodic bathymetry in 
Figure \ref{fig:periodic_channel} and the non-rectangular channel in Figure \ref{fig:extracted_channel}.
The solution in the periodic channel restricted to the domain extending
from the middle of segment A to the middle of segment B is identical
to the solution in the non-rectangular channel (with slip boundary conditions at the sides of the channel).
Flow in channels with other shapes, such as that shown in Figure \ref{fig:channel}
can also be viewed equivalently as flow over an infinite periodic bathymetry, as
long as the channel does not have sloping sides that extend above the water surface.
Thus, bathymetric solitary waves arise also in narrow non-rectangular channel flow.
This equivalence also establishes a connection between our homogenized effective equations
for the infinite domain and the effective equations for channels derived in \cite{chassagne2019dispersive}.

Before moving on, let us mention a few additional works on how bathymetry
influences wave propagation.  Rosales \& Papanicolau analyzed the case of weakly
nonlinear shallow water waves with small bathymetry changes,
\cite{rosales1983gravity}, while Nachbin \& Papanicolau studied the case of
small (linear) waves with large variations in bathymetry, \cite{nachbin1992water}.
Both works focused on waves in one horizontal space dimension.
An extension to two dimensions was presented in \cite{craig2005hamiltonian}, with similar results.
Based on the Green-Naghdi model \cite{green1976derivation}, Chassagne et. al.
\cite{chassagne2019dispersive} studied the dispersive effects of bores in
non-rectangular channels, with an emphasis on the influence of sloping banks.

Thus far we have dealt with the shallow water model, which neglects important physical effects
such as dispersion, dissipation and surface tension.
In \S\ref{sec:about_dispersive_effects}
we concentrate on waves in non-rectangular channels,
with the goal of determining whether it is feasible to experimentally observe solitary waves
that are created primarily by bathymetric dispersion.  To do so we must go beyond
the shallow water model.  The shallow water equations neglect the inherent dispersive
nature of water waves.  If solitary waves over periodic bathymetry
are observed experimentally, can one distinguish them from other
solitary waves arising from other dispersive effects (such as KdV solitons)?

The model most directly relevant to the present work is perhaps that of Peregrine
\cite{peregrine1968long} (see also \cite{peters1966rotational,shen1969asymptotic,teng1997effects}),
which leads to an effective KdV equation in which the dispersion coefficient is modified
by the channel geometry.
In \S\S \ref{sec:peregrine} and \ref{sec:about_dispersive_effects} 
we compare the dispersion in these three models (KdV, Peregrine's, and ours)
and show that in certain regimes bathymetric dispersion can be the dominant effect.
This answers the question above in the affirmative.
In \S\ref{sec:peregrine}, we also show that our model and Peregrine's model, despite being based on
different assumptions, agree well in a certain physical regime.

The code and instructions to create every plot and all the results in this work
are available at \url{https://github.com/manuel-quezada/water_wave_diffractons_RR}.

\subsection{Objectives and our contribution}
We have two objectives in this work.
The first objective, which we tackle in \S\S\ref{sec:dispersion_by_diffraction} and \ref{sec:SWEs_diffractons},
is to extend the results in \cite{ketcheson2015diffractons} to obtain bathymetric solitary waves and study some
of their properties. In \S\ref{sec:kdv_for_diffractons} we go beyond the work in that reference and
obtain a KdV-type equation valid for weakly nonlinear regimes. 
The second objective, which is addressed in \S\ref{sec:about_dispersive_effects},
is to assess the feasibility of observing these waves in physical experiments.
Our contribution is to:
\begin{itemize}
\item derive an effective model for bathymetry-induced dispersion of waves in two
  horizontal dimensions;
\item connect two distinct areas of study: hyperbolic equations with periodic coefficients
  and water wave models in channels with non-flat bathymetry;
\item show that this (bathymetric) dispersion alone can lead to the formation of solitary waves;
\item derive a KdV-type equation that models these solitary waves;
\item show that bathymetric dispersion can be the dominant source of dispersion for
  certain classes of waves.
\end{itemize}

\section{Effective dispersion due to periodic bathymetry}\label{sec:dispersion_by_diffraction}
Water waves are inherently dispersive, and this is represented through dispersive terms
or non-hydrostatic pressure terms in models such as the KdV equation \cite{korteweg1895xli},
the Boussinesq equations \cite{Boussinesq1872} and the Green-Naghdi model \cite{green1976derivation}.
In this section, we demonstrate that small-amplitude shallow water waves
propagating over periodic bathymetry undergo an effective
dispersion.  In order to clearly distinguish this source of dispersion from
the dispersion present over flat bathymetry, we focus on the --dispersionless--
shallow water equations over variable bathymetry:
\begin{subequations} \label{shallow_water_equations}
  \begin{align}
    h_t + (hu)_x + (hv)_y & = 0, \\
    (hu)_t + (hu^2)_x+gh(h+b)_x + (huv)_y & = 0, \\
    (hv)_t + (huv)_x + (hv^2)_y+gh(h+b)_y & = 0.
  \end{align}
\end{subequations}
Here $h$ is the height of the column of water,
$u$ and $v$ are the $x$- and $y$-velocities respectively,
$g$ is the magnitude of the gravitational force and $b(x,y)$ is the periodic bathymetry.
The reference point $z=0$ is chosen to coincide with the lowest point of the bathymetry.
Unless otherwise noted, we use $g=9.8\left(\frac{m}{s^2}\right)$; hereafter, we do not
explicitly reference units of measure but use SI units throughout.
We consider a domain that extends infinitely in both $x$ and $y$,
with bathymetry periodic in $y$.
We let $\Omega$ denote the period
and use it as the unit of measurement so that $\Omega=1$.

The analysis and results of this
section are similar to those presented in \cite{quezada2014two},
which treated the acoustic wave equation in a periodic medium.
These results are also connected with the work by \cite{chassagne2019dispersive},
where the authors derived an effective model for the shallow water equations that captures
the dispersive effects when one-dimensional waves travel over non-flat channels.
The main differences between the second reference and the results we present in this section is that
the bathymetry that we consider is assumed to be changing periodically in one
direction and that our effective equations are valid for propagation in two-dimensions. 

\begin{figure}[!h]
  \begin{center}
    \includegraphics[scale=0.5]{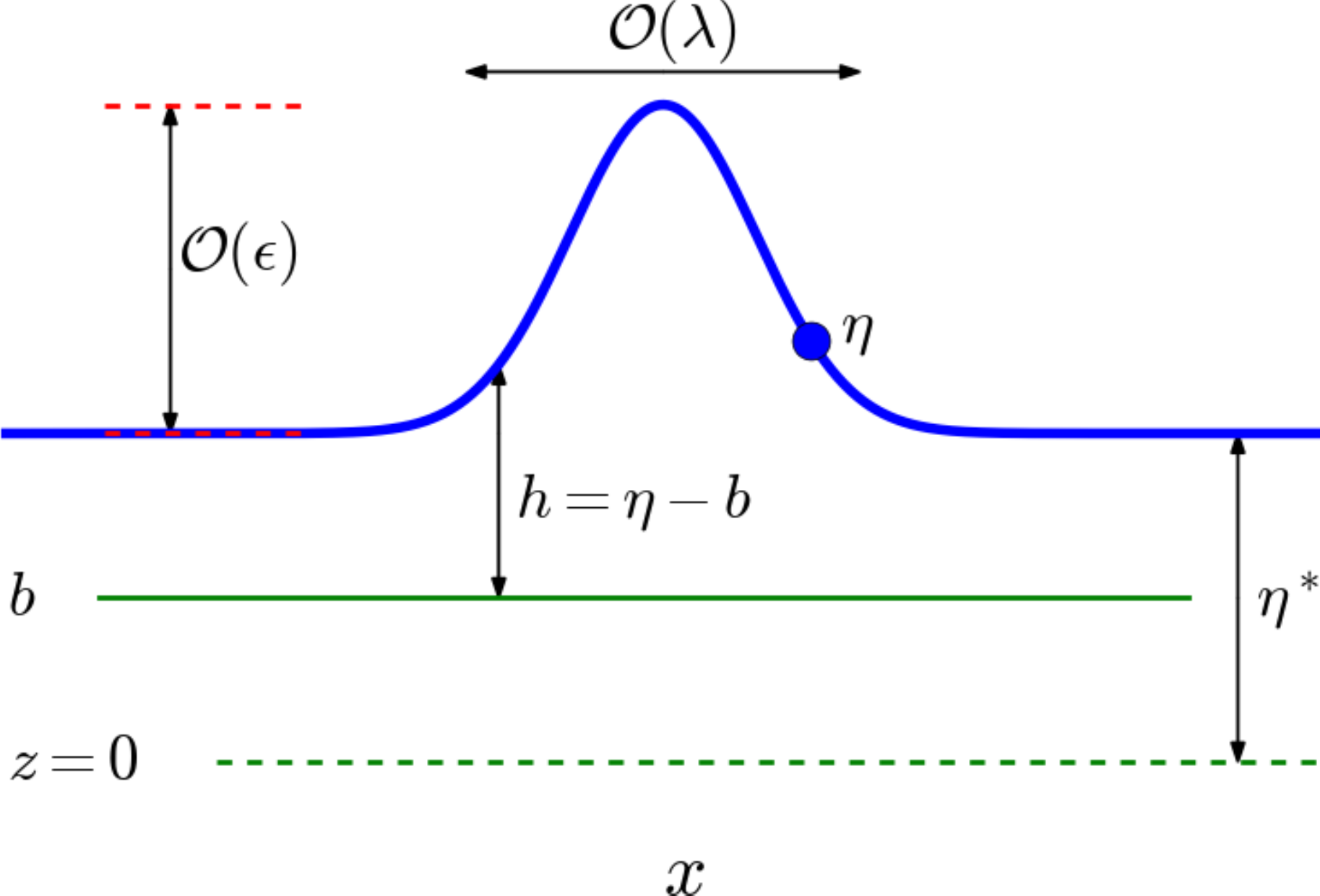}
  \end{center}
  \caption{Notation for shallow water equations with variable bathymetry.
    The reference point $z=0$ is located at the bottom of the bathymetry.
    The surface elevation is denoted by $\eta(x,y)$ and
    the undisturbed surface elevation is denoted by $\mwl$. 
    The water depth is denoted by $h(x,y)=\eta(x,y)-b(y)$. 
    The bathymetry $b(y)$ varies along the axis that points
    into the page.
    Figure \ref{fig:periodic_channel} presents a 3D view for one
    specific bathymetric profile.
    In this diagram, the bathymetry $b$ is piecewise constant; 
    however, in general we assume it to be $y$-periodic.
    \label{fig:notation}}
\end{figure}

\subsection{Linearization and homogenization}\label{sec:homogenization}
We aim to obtain a constant-coefficient homogenized system that approximates
\eqref{shallow_water_equations} for small-amplitude, long-wavelength perturbations.
To do this we follow \cite{quezada2014two} and references therein and perform a
homogenization, which is valid for small-amplitude waves.
We consider waves with characteristic wavelength $\lambda$ propagating
in the presence of periodic bathymetry with period $\Omega$, where $\Omega \ll \lambda$.
By letting $\delta:=\Omega/\lambda$, we introduce a fast scale $\hat{y}:=\delta^{-1}y$
in the $y$-direction. We assume that the solution $h$, $u$ and $v$ depend also on this
fast scale; i.e., $h=h(x,y,t,\hat{y})$ and similarly for $u$ and $v$.
Finally, we assume that the bathymetry depends only on the fast scale; i.e., $b=b(\hat{y})$.
These are key steps in the homogenization process; see e.g., \cite{Fish2001}.
Note that, by these assumptions, $(\cdot)_y\mapsto (\cdot)_y +\delta^{-1}(\cdot)_{\hat{y}}$. 
Now we obtain a dimensionless version of \eqref{shallow_water_equations}
(after the homogenization process we go back to the variables with dimensions).
This can be done by introducing the following dimensionless variables:
\begin{align}\label{dimensionless_variables}
  x^\prime=\frac{x}{\lambda}, \quad
  y^\prime=\frac{y}{\lambda}, \quad
  \hat{y}^\prime=\frac{\hat{y}}{\lambda}, \quad
  t^\prime=\frac{c}{\lambda}t, \quad
  \eta^\prime=\frac{\eta}{\mwl}, \quad
  h^\prime=\frac{h}{\mwl}, \quad
  u^\prime=\frac{u}{c}, \quad
  v^\prime=\frac{u}{c}, \quad
  b^\prime=\frac{b}{\mwl},
\end{align}
where $c:=\sqrt{g\mwl}$. We remark that we scale $\hat{y}$ by $\lambda$
since the fast variation in $\hat{y}$ is introduced via its definition ($\hat{y}=\delta^{-1}y$). 
After considering the system in non-conservative form and, for simplicity, dropping the tildes we get
\begin{align*}
  \eta_t+[(\eta-b)u]_x+[(\eta-b)v]_y + \delta^{-1}[(\eta-b)v]_{\hat{y}}&= 0, \\
  u_t+uu_x+v(u_y+\delta^{-1}u_{\hat{y}})+\eta_x &=0, \\
  v_t+uv_x+v(v_y+\delta^{-1}v_{\hat{y}})+\eta_y + \delta^{-1}\eta_{\hat{y}}&=0, 
\end{align*}
where $\eta=h+b$ denotes the surface elevation.
We consider small-amplitude waves and perform an asymptotic expansion around
$\eta=\mwl$, $u=0$ and $v=0$.
The linear system is 
\begin{subequations}
  \begin{align*}
    \eta_t + [(\mwl-b)u]_x + [(\mwl-b)v]_y + \delta^{-1}[(\mwl-b)v]_{\hat{y}} &= 0, \\
    u_t + \eta_x &= 0, \\
    v_t + \eta_y + \delta^{-1}\eta_{\hat{y}}& = 0. 
  \end{align*}
\end{subequations}
Now, let $\mu:=(\mwl-b)u$ and $\nu:=(\mwl-b)v$ denote the (linearized) $x$- and $y$-momentum, respectively.
Doing so, we get 
\begin{subequations} \label{linear_swe}
  \begin{align}
    \eta_t + \mu_x + \nu_y  + \delta^{-1} \nu_{\hat{y}} &= 0, \\
    \mu_t + (\mwl-b(\hat{y})) \eta_x & = 0, \\
    \nu_t + (\mwl-b(\hat{y})) (\eta_y +\delta^{-1} \eta_{\hat{y}}) &= 0.
  \end{align}
\end{subequations}
We have explicitly noted the spatial dependence of the bathymetry $b$
in order to emphasize that \eqref{linear_swe} is a first-order linear
hyperbolic system with spatially-varying coefficients.

In the following equations and in Sections \ref{sec:homogenized_order_1} and \ref{sec:homogenized_corrections}, to avoid
confusion with subindices, we use use
$(\cdot)_{i,x}$ to denote differentiation of $(\cdot)_i$ with respect to $x$, and similarly for the other derivatives.
Using the formal expansion $\eta(x,y,\hat{y},t)=\sum_{i=0}^\infty\delta^i\eta_i(x,y,\hat{y},t)$
and similarly for $\mu$ and $\nu$, we get
\begin{subequations} \label{expanded_system}
  \begin{align}
    \sum_{i=0}^\infty \delta^i\eta_{i,t} + \sum_{i=0}^\infty \delta^i\mu_{i,x}
    + \sum_{i=0}^\infty \delta^i\nu_{i,y} + \delta^{-1}\sum_{i=0}^\infty\delta^i\nu_{i,\hat{y}}& = 0, \\
    \sum_{i=0}^\infty \delta^i\mu_{i,t} + (\mwl-b) \sum_{i=0}^\infty\delta^i\eta_{i,x} & = 0, \\
    \sum_{i=0}^\infty \delta^i\nu_{i,t}
    + (\mwl-b) \left(\sum_{i=0}^\infty\delta^i\eta_{i,y} + \delta^{-1} \sum_{i=0}^\infty\delta^i\eta_{i,\hat{y}}\right) & = 0.
  \end{align}
\end{subequations}
The next step is to equate terms of the same order. At each order we apply the average operator
$\langle\cdot\rangle:=\frac{1}{|\Omega|}\int_\Omega(\cdot)dy=\frac{1}{\lambda}\int_0^\lambda(\cdot)d\hat{y}$
to obtain the homogenized leading-order system and corrections to it.
In addition, we obtain expressions for the non-homogenized variables. 
In the next sections we present details for the derivation of the homogenized leading-order system
and for the first correction. Then we present the results considering one more correction. 

\subsubsection{Homogenized $\order(1)$ system}\label{sec:homogenized_order_1}
From \eqref{expanded_system}, we equate terms of $\order(\delta^{-1})$ and conclude that
$\nu_0=:\bar{\nu}_0(x,y,t)$ and $\eta_0=:\bar{\eta}_0(x,y,t)$ (where the bar is used to denote
variables that are independent on the fast scale $\hat{y}$).
From \eqref{expanded_system}, we take terms of $\order(1)$ to obtain
\begin{subequations}\label{system_order_1}
  \begin{align}
    \bEta_{0,t}+\mu_{0,x}+\bNu_{0,y}+\nu_{1,\yhat} &= 0, \\
    \mu_{0,t}+(\mwl-b)\bEta_{0,x} &= 0, \label{system_order_1_eqn2} \\
    \bNu_{0,t}+(\mwl-b)\left(\bEta_{0,y}+\eta_{1,\yhat}\right) &= 0. \label{system_order_1_eqn3}
  \end{align}
\end{subequations}
Divide \eqref{system_order_1_eqn3} by $\mwl-b$, apply the average operator to \eqref{system_order_1} and
assume that $\nu_1$ and $\eta_1$ are $\yhat$-periodic to get 
\begin{subequations}\label{hom_system_order_1}
  \begin{align}
    \bEta_{0,t}+\bMu_{0,x}+\bNu_{0,y} &= 0, \\
    \bMu_{0,t}+(\mwl-b)_m\bEta_{0,x} &= 0, \label{hom_system_order_1_eqn2} \\
    \bNu_{0,t}+(\mwl-b)_h\bEta_{0,y} &= 0,
  \end{align}
\end{subequations}
where 
$(\cdot)_m:=\langle\cdot\rangle=\frac{1}{|\Omega|}\int_\Omega(\cdot)dy$ 
and 
$(\cdot)_h:=\langle(\cdot)^{-1}\rangle^{-1}=\left[\frac{1}{|\Omega|}\int_\Omega (\cdot)^{-1} dy\right]^{-1}$
denote the mean and harmonic averages respectively.
System \eqref{hom_system_order_1} is the homogenized leading-order system.
It has the same form as the variable bathymetry system \eqref{linear_swe} but with effective constant bathymetry coefficients.

From \eqref{system_order_1_eqn2} and \eqref{hom_system_order_1_eqn2}
(and by choosing an appropriate initial condition for $\bMu_0$) we obtain
\begin{align}\label{mu0}
  \mu_0=\frac{\mwl-b}{(\mwl-b)_m}\bMu_0.
\end{align}

Now we obtain expressions for the non-averaged $\order(1)$ terms in \eqref{system_order_1}.
To do this we make the following ansatz:
\begin{subequations}\label{ansatz_order_1}
  \begin{align}
    \nu_1&=\bNu_1+A(\yhat)\bMu_{0,x}, \\
    \eta_1&=\bEta_1+B(\yhat)\bEta_{0,y},
  \end{align}
\end{subequations}
which is chosen to reduce \eqref{system_order_1} to a system of ODEs.
By substituting the ansatz \eqref{ansatz_order_1}, the homogenized leading-order system
\eqref{hom_system_order_1} and the relation for $\mu_0$ \eqref{mu0} into the $\order(1)$ system
\eqref{system_order_1} and by requiring that the fast variable coefficients vanish we get
\begin{subequations}\label{fast_functions_order_1}
  \begin{align}
    A_{,\yhat}+(\mwl-b)(\mwl-b)_m^{-1}-1 &= 0, \\
    B_{,\yhat}-(\mwl-b)^{-1}(\mwl-b)_h+1 &= 0.
  \end{align}
\end{subequations}
We look for solutions of \eqref{fast_functions_order_1} with the normalization condition
$\langle A\rangle=\langle B\rangle=0$. Note that
$\langle A_{,\yhat}\rangle=\langle B_{,\yhat}\rangle=0$, which implies that
$A$ and $B$ are $\yhat$-periodic. 

\subsubsection{Homogenized corrections}\label{sec:homogenized_corrections}
From \eqref{expanded_system} we collect $\order(\delta)$ terms, plug the ansatz
\eqref{ansatz_order_1} and apply the average operator $\langle\cdot\rangle$ to get
\begin{subequations}\label{hom_system_order_delta}
  \begin{align}
    \bEta_{1,t}+\bMu_{1,x}+\bNu_{1,y} &= 0, \\
    \bMu_{1,t}+(\mwl-b)_m\bEta_{1,x} &= -\langle B(\mwl-b)\rangle\bEta_{0,xy}, \\
    \bNu_{1,t}+(\mwl-b)_h\bEta_{1,y} &= -(\mwl-b)_h\langle A(\mwl-b)^{-1}\rangle\bMu_{0,xt}.
  \end{align}
\end{subequations}
For the bathymetry that we consider in this work $\langle B(\mwl-b)\rangle=\langle A(\mwl-b)^{-1}\rangle=0$.
Following similar (but considerably more algebraic intense) steps we obtain the homogenized second correction
\begin{subequations}\label{hom_system_order_delta2}
  \begin{align}
    \bEta_{2,t}+\bMu_{2,x}+\bNu_{2,y} &= 0, \\
    \bMu_{2,t}+(\mwl-b)_m\bEta_{2,x} &= -\langle(\mwl-b)F\rangle \bEta_{0,xyy} - \langle(\mwl-b)E\rangle \bEta_{0,xxx},\\
    \bNu_{2,t}+(\mwl-b)_h\bEta_{2,y} &= (\mwl-b)_h^2\langle(\mwl-b)^{-1}D\rangle \bEta_{0,yyy} + (\mwl-b)_m(\mwl-b)_h\langle(\mwl-b)^{-1}C\rangle \bEta_{0,xxy},
  \end{align}
\end{subequations}
where $C,~D,~E$ and $F$ are fast variable functions that are given by the following ODEs:
\begin{subequations}\label{fast_functions_order_2}
  \begin{align}
    C_{,\yhat}-\left[1-(\mwl-b)_m^{-1}(\mwl-b)\right]B+A &= 0, \\
    D_{,\yhat}-B &= 0,\\
    E_{,\yhat}-(\mwl-b)_m(\mwl-b)^{-1}A &= 0, \\
    F_{,\yhat}+B &= 0,
  \end{align}
\end{subequations}
with the normalization conditions
$\langle C\rangle = \langle D\rangle = \langle E\rangle = \langle F\rangle = 0$.
It can be easily shown that, for the periodic bathymetry that we consider,
the coefficients in the right hand side of \eqref{hom_system_order_delta2} do not vanish.

Finally, given the homogenized leading-order system \eqref{hom_system_order_1}
and the homogenized corrections \eqref{hom_system_order_delta} and \eqref{hom_system_order_delta2},
we combine them into a single system by defining $\bEta:=\langle\bar{\eta}_0+\delta\bar{\eta}_1+\dots\rangle$
and similarly for $\bar{\mu}$ and $\bar{\nu}$. We obtain
\begin{subequations}\label{hom_system}
  \begin{align}
    \bEta_{,t}+\bMu_{,x}+\bNu_{,y} &= 0, \\
    \bMu_{,t}+(\mwl-b)_m\bEta_{,x} &= \delta^2(\mwl-b)_m\left[\beta_1\bEta_{,xyy}  + \beta_2 \bEta_{,xxx}\right],\\
    \bNu_{,t}+(\mwl-b)_h\bEta_{,y} &= \delta^2(\mwl-b)_h\left[\gamma_1\bEta_{,yyy} + \gamma_2\bEta_{,xxy}\right],
  \end{align}
\end{subequations}
where
\begin{subequations}\label{hom_coeffs}
  \begin{align}
    \beta_1   &= -(\mwl-b)_m^{-1}\langle(\mwl-b)F\rangle,
    & \beta_2 &= -(\mwl-b)_m^{-1} \langle(\mwl-b)E\rangle, \\
    \gamma_1 &= (\mwl-b)_h\langle(\mwl-b)^{-1}D\rangle,
    & \gamma_2 &= (\mwl-b)_m\langle(\mwl-b)^{-1}C\rangle.
  \end{align}
\end{subequations}

The homogenized system \eqref{hom_system} is an effective approximation of the two dimensional
linearized shallow water equations over $y$-periodic bathymetry whose period is much smaller
than the characteristic wave length.  

If we assume also that the initial data does not depend on $y$,
let $\bMwl:=\langle\mwl-b\rangle$
denote the average undisturbed surface elevation,
drop the bars in \eqref{hom_system} and go back to the dimension variables,
we obtain
\begin{align}\label{homogenized_linear_system}
  \eta_{,tt}-g\bMwl\eta_{,xx}=-g\bMwl\delta^2\beta_2\eta_{,xxxx},
\end{align}
which models propagation only in the $x$-direction.
The speed of small-amplitude, long wavelength perturbations is,
as one might expect, $c_{\text{eff}}:=\sqrt{g\bMwl}$.
Equations \eqref{hom_system}-\eqref{hom_coeffs} and \eqref{homogenized_linear_system} are
valid for small-amplitude waves over arbitrary bathymetry that is periodic in $y$.
Below, we will specialize to the piecewise-constant case.

\subsection{Piecewise-constant bathymetry}
Now we consider a specific case of study with piecewise-constant bathymetry
\begin{align} \label{bathymetry}
  b(y) & = 
  \begin{cases}
    0, & \mbox{ if } ~ n + \frac{1}{2} \le y/\Omega < n+1, \\
    \bB, & \mbox{ if } ~ n \le y/\Omega < n + \frac{1}{2},
  \end{cases}
\end{align}
where $n$ is any integer. This bathymetry profile is depicted in Figure \ref{fig:periodic_channel}.
The coefficients \eqref{hom_coeffs} are then
\begin{subequations}\label{hom_coeffs2}
  \begin{align}
    \beta_1&=\frac{\bB^2}{48(2\eta^*-\bB)^2}\lambda^2,
    &
    \beta_2&=\frac{-\bB^2}{192\mwl(\mwl-\bB)}\lambda^2, \\
    \gamma_1&=-\beta_1,
    &
    \gamma_2&=-\beta_2.
  \end{align}
\end{subequations}
The term appearing on the right hand side of \eqref{homogenized_linear_system} is dispersive;
the coefficient of dispersion is in this case given by
\begin{align}
  -g\bMwl \delta^2\beta_2 &= g\bMwl\left[\frac{\bB^2\Omega^2}{192\mwl(\mwl-\bB)}\right].
\end{align}
It is evident that this dispersion is purely an effect of the bathymetric
variation; notice that it increases as 
$\bB$ grows, and vanishes as $\mwl\rightarrow \infty$ (keeping the bathymetry fixed).
We remark that the dispersion in equation \eqref{homogenized_linear_system} 
is a consequence of changes in the bathymetry and not due to non-hydrostatic pressure effects. 
These dispersive effects are different from those present in dispersive water wave models, 
like the KdV equation, in which dispersion is present even when the bathymetry profile is flat.

\begin{figure}[!h]
  \begin{center}
    \includegraphics[scale=0.2]{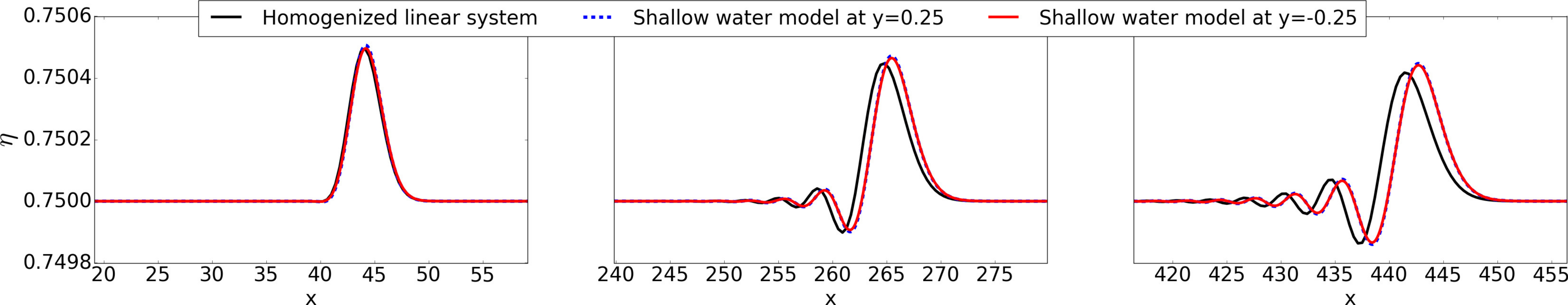}
  \end{center}
  \caption{Solution of the shallow water equations
    \eqref{shallow_water_equations}
    with periodic bathymetry versus solution of the linearized homogenized approximation
    \eqref{homogenized_linear_system}.
    We show the surface elevation $\eta$
    at (from left to right) $t=20, ~120$ and $t=200$.
    We plot two different slices of the solution at different $y$ values;
    however, because $\delta \ll 1$ there is almost no phase difference
    and the two plots are nearly exactly aligned.
    \label{fig:linearization}}
\end{figure}

In Figure \ref{fig:linearization} we compare the solution of
the (nonlinear) shallow water system with variable bathymetry \eqref{shallow_water_equations}
to that of the homogenized linear system \eqref{homogenized_linear_system}.
We take initial data 
\begin{align} \label{eta0}
  \eta(x,y,t=0) & =\mwl+\epsilon \exp{\left(-\frac{x^2}{2\alpha}\right)}, &  u(x,y,0) =v(x,y,0)=0
\end{align}
with $\epsilon=0.001, \alpha=2$ and $\mwl=0.75$.  The bathymetry is given by
\eqref{bathymetry} with
\begin{align} \label{params}
  \bB=0.5.
\end{align}
The dispersion predicted by the linearized, homogenized model is also
clearly evident in the nonlinear, variable-coefficient solution.
Both models are solved to very high precision; the differences
between the solutions are primarily due to the nonlinear effects
that are neglected in \eqref{homogenized_linear_system}.
The shallow water equations \eqref{shallow_water_equations} are solved using 
the finite volume code PyClaw \cite{pyclaw-sisc}, 
with the Riemann solver developed in \cite{GEORGE20083089}.
The mesh resolution is $\Delta x=\Delta y=1/128$.
The linear homogenized equations \eqref{homogenized_linear_system} are solved using a
Fourier spectral collocation method in space and a fourth order Runge-Kutta
method in time; see \cite{trefethen2000spectral}.

\section{Bathymetric solitary waves}\label{sec:SWEs_diffractons}
\begin{figure}[!h]
  \begin{centering}
    \subfloat[Solitary waves resulting from initial Gaussian pulse with $\alpha=2$\label{fig:three_diffractons}]{
      \includegraphics[scale=0.22]{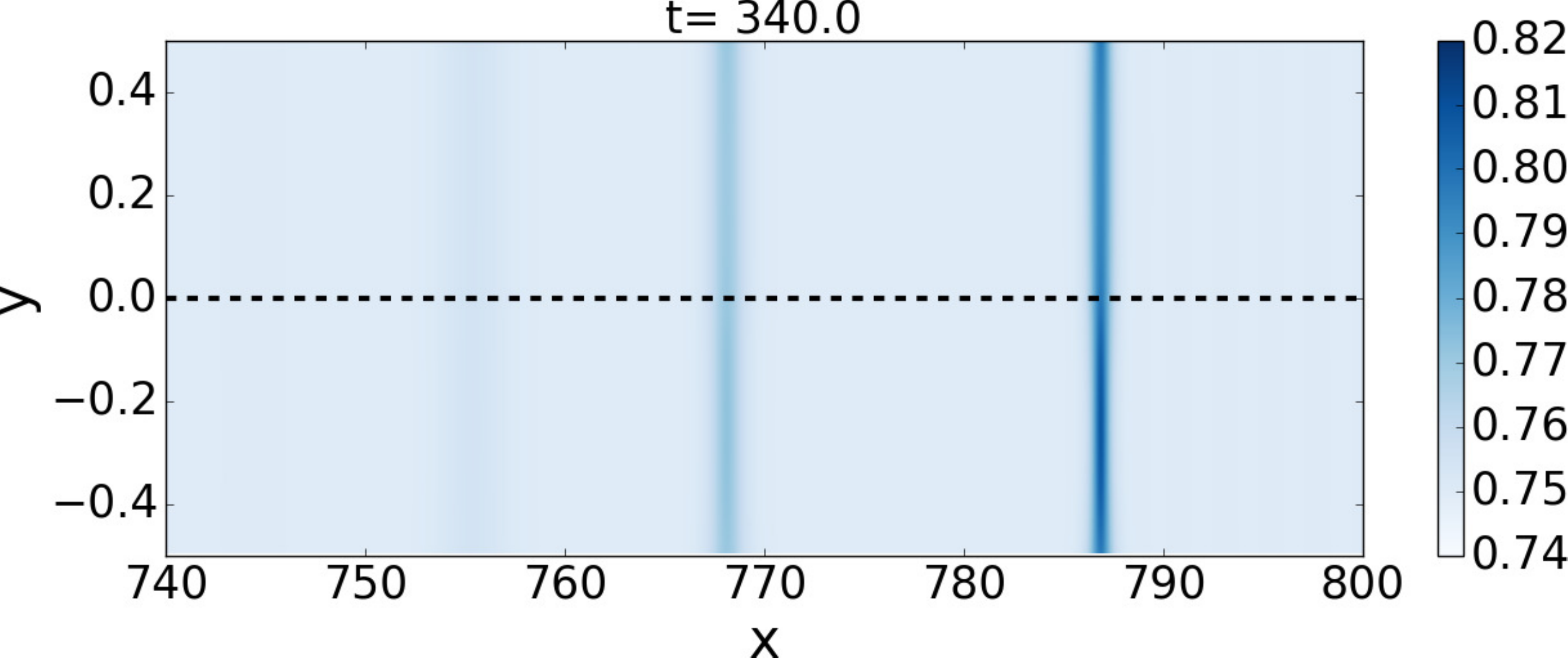}\qquad\quad
      \includegraphics[scale=0.22]{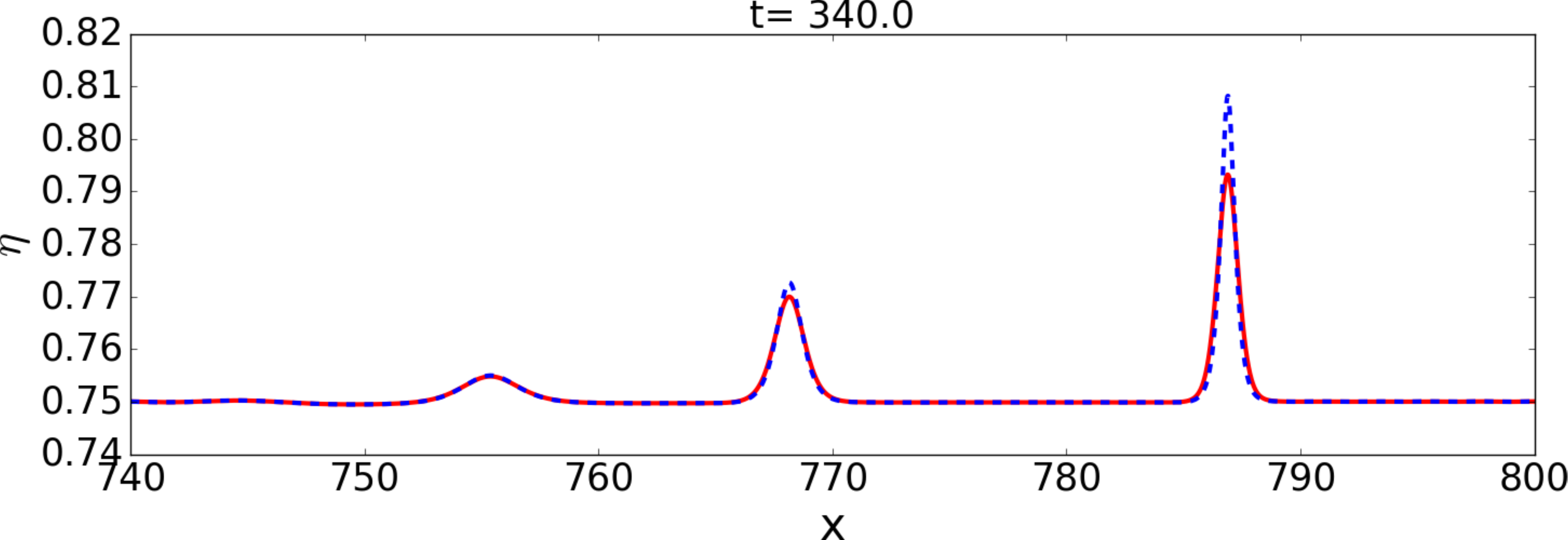}}
    
    \subfloat[Solitary waves resulting from initial Gaussian pulse with $\alpha=10$\label{fig:more_diffractons}]{
      \includegraphics[scale=0.22]{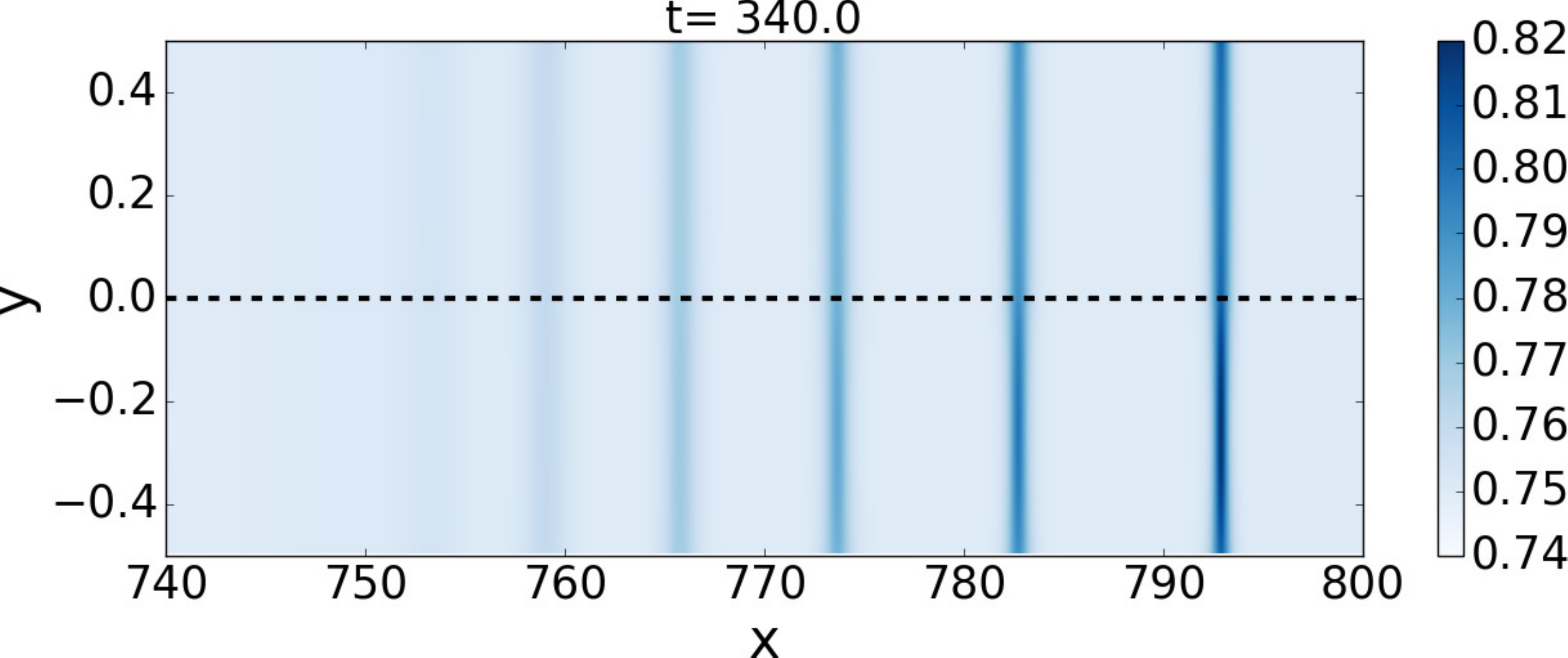}\qquad\quad
      \includegraphics[scale=0.22]{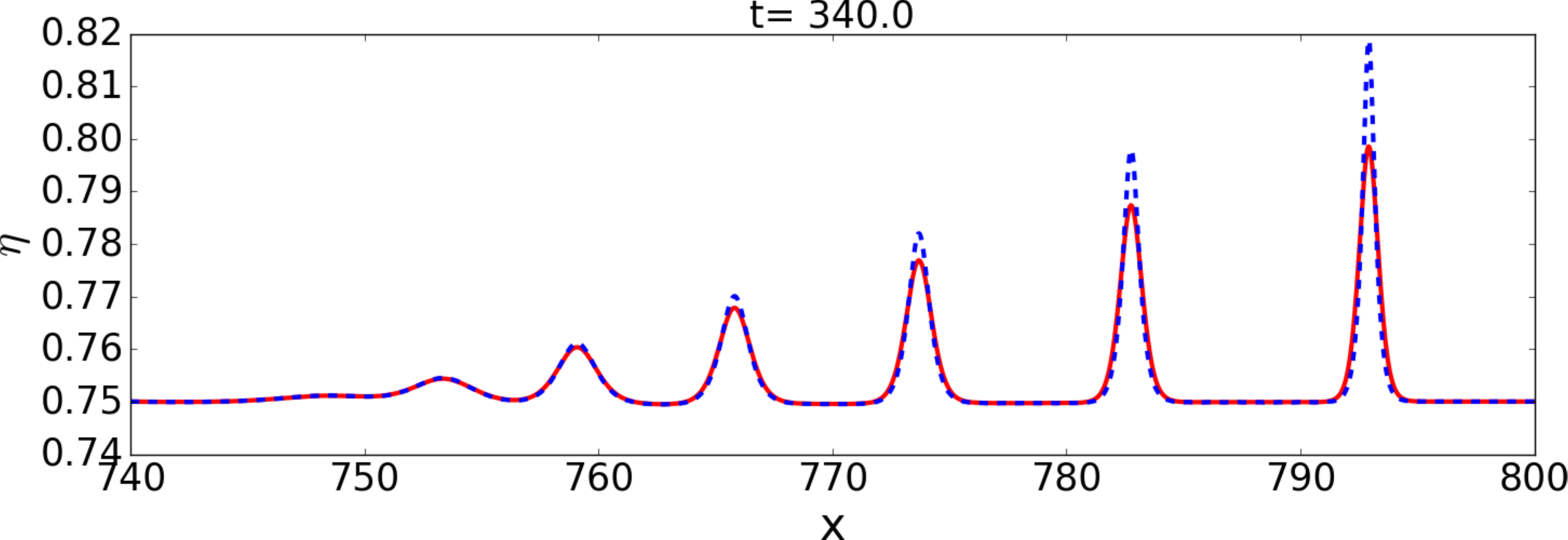}}
    \par
  \end{centering}
  \caption{Bathymetric solitary waves at $t=340$.
    The initial condition is given by \eqref{eta0}
    with $\epsilon=0.05$, $\mwl=0.75$ and (a) $\alpha=2$ and (b) $\alpha=10$. 
    In the left panels we show the surface elevation plots 
    (where the dashed line represents the location of the jump in the bathymetry)
    and in the right panels we show slices along $y=0.25$ (in dashed blue) and $y=-0.25$ (in solid red).
    \label{fig:diffractons}}
\end{figure}

Let us now study solutions of the nonlinear, variable-coefficient shallow
water model \eqref{shallow_water_equations} in a more strongly nonlinear regime.
We repeat the experiment above, taking \eqref{eta0} and \eqref{params} but
with a much larger perturbation given by $\epsilon=0.05$ and $\alpha=2$ or
$\alpha=10$.
We again solve the equations using the finite volume solver PyClaw.
The results are shown in Figure \ref{fig:diffractons}.
The mass of the initial pulse determines the number of solitary waves created. 
In the rest of this section we use the solitary waves shown in Figure \ref{fig:diffractons}
and (following \cite{ketcheson2015diffractons}) study some properties for these solitary waves. 
In particular, we investigate the long-time stability and shape evolution,
the scaling and speed-amplitude relations and the interaction of bathymetric solitary waves.
The results from these experiments suggest that although bathymetric solitary waves are
similar to KdV solitons they possess fundamental differences.
We explore these similarities and differences in \S\ref{sec:kdv_for_diffractons}.
Moreover, we derive a KdV-type equation that approximates the solution of
\eqref{shallow_water_equations} for $x$-propagation of plane waves over
periodic bathymetry like the one depicted in Figure \ref{fig:periodic_channel}.

\subsection{Long-time stability and shape evolution}
We first investigate the long-time behavior and the shape evolution 
of these solitary waves.  To do this we isolate the first solitary wave in Figure 
\ref{fig:three_diffractons}, which corresponds to $t=340$,
use it as initial condition for a new simulation and propagate it by itself until a 
final time of $t=1000$.
Let $X(t)$ denote the location of the solitary wave's peak at time $t$;
we compute
\begin{align}
  \Delta \eta =\max_{t\in[1,\dots,1000]}\left\{ \frac{|| \eta\left(x-X(0),y,t=0\right)-\eta\left(x-X(t),y,t\right) ||_{2(x,y)}}{|| \eta\left(x-X(0),y,t=0\right) ||_{2(x,y)}} \right\},
\end{align}
which is the largest relative difference between the shape of the solution at 
$t\in[1,\dots,1000]$ and the initial condition.
We perform this experiment on a grid with $\Delta x = \Delta y = 1/64$ and obtain 
$\Delta \eta\approx 7.66\times 10^{-4}$. 
Afterwards, we refine the grid so that $\Delta x = \Delta y = 1/128$ and obtain 
$\Delta \eta\approx 1.79\times 10^{-4}$.
The results indicate that these solutions are indeed solitary waves that propagate
with a fixed shape, up to numerical errors.

\subsection{Scaling and speed-amplitude relations} \label{sec:speed-amplitude}
Many solitary waves, including the diffractons found in
\cite{ketcheson2015diffractons}, have a shape that is exactly or nearly that of
a $\text{sech}^2$ function.  Here we investigate the shape of bathymetric solitary waves.
Although these are two-dimensional waves, they vary more strongly
with respect to $x$ than $y$.  Our expectation
(based on \cite{ketcheson2015diffractons})
is that the cross section of a bathymetric solitary wave for a fixed value of $y$ should be
close to a $\text{sech}^2$ function.

\begin{figure}[!h]
  \begin{centering}
    \subfloat[Amplitude as a function of $y$.\label{fig:y_profiles}]{\includegraphics[scale=0.35]{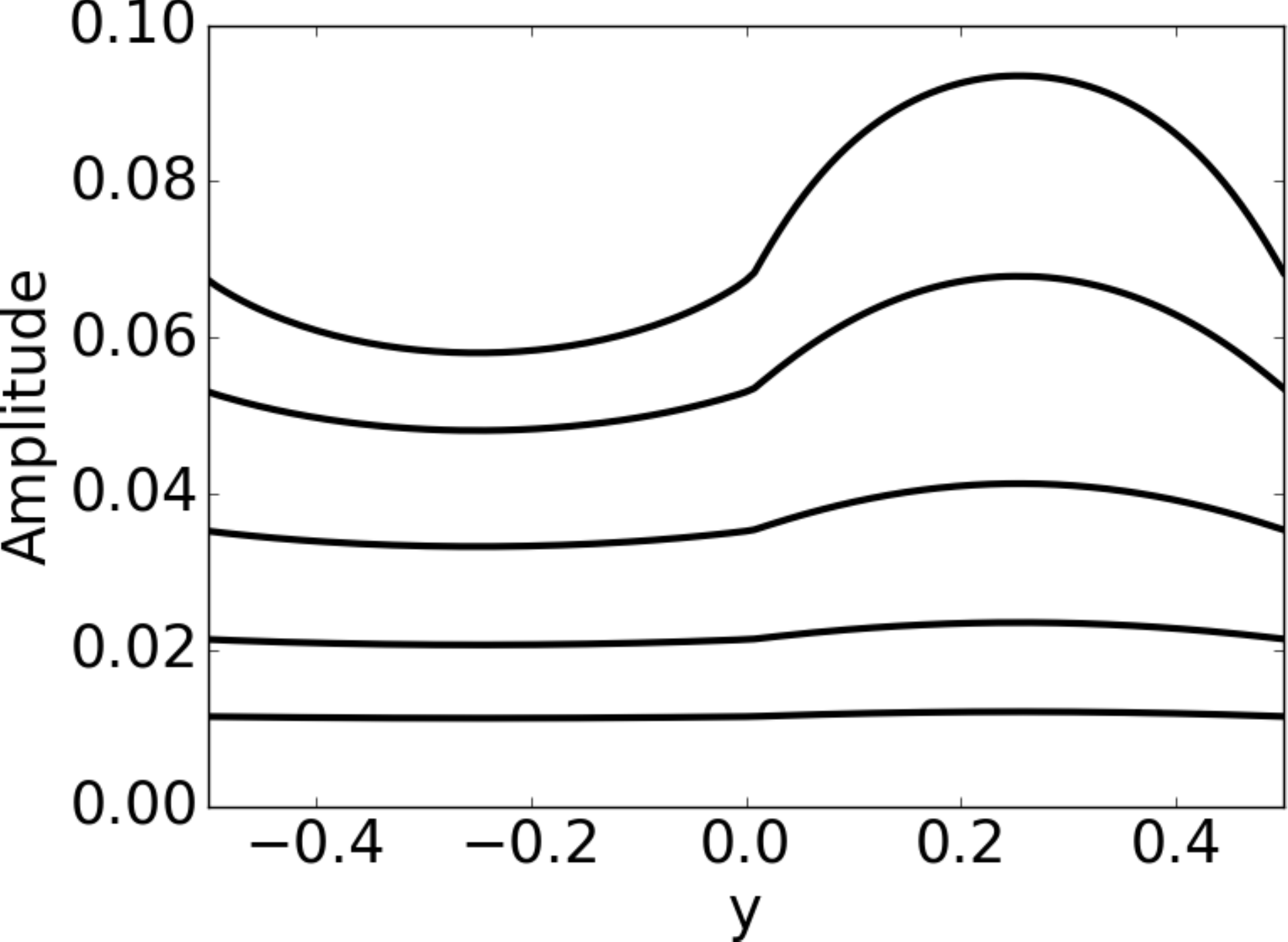}}
    \qquad\qquad
    \subfloat[Variation $\Delta A$ versus mean amplitude $\bar A$.\label{fig:var_vs_amp}]{\includegraphics[scale=0.35]{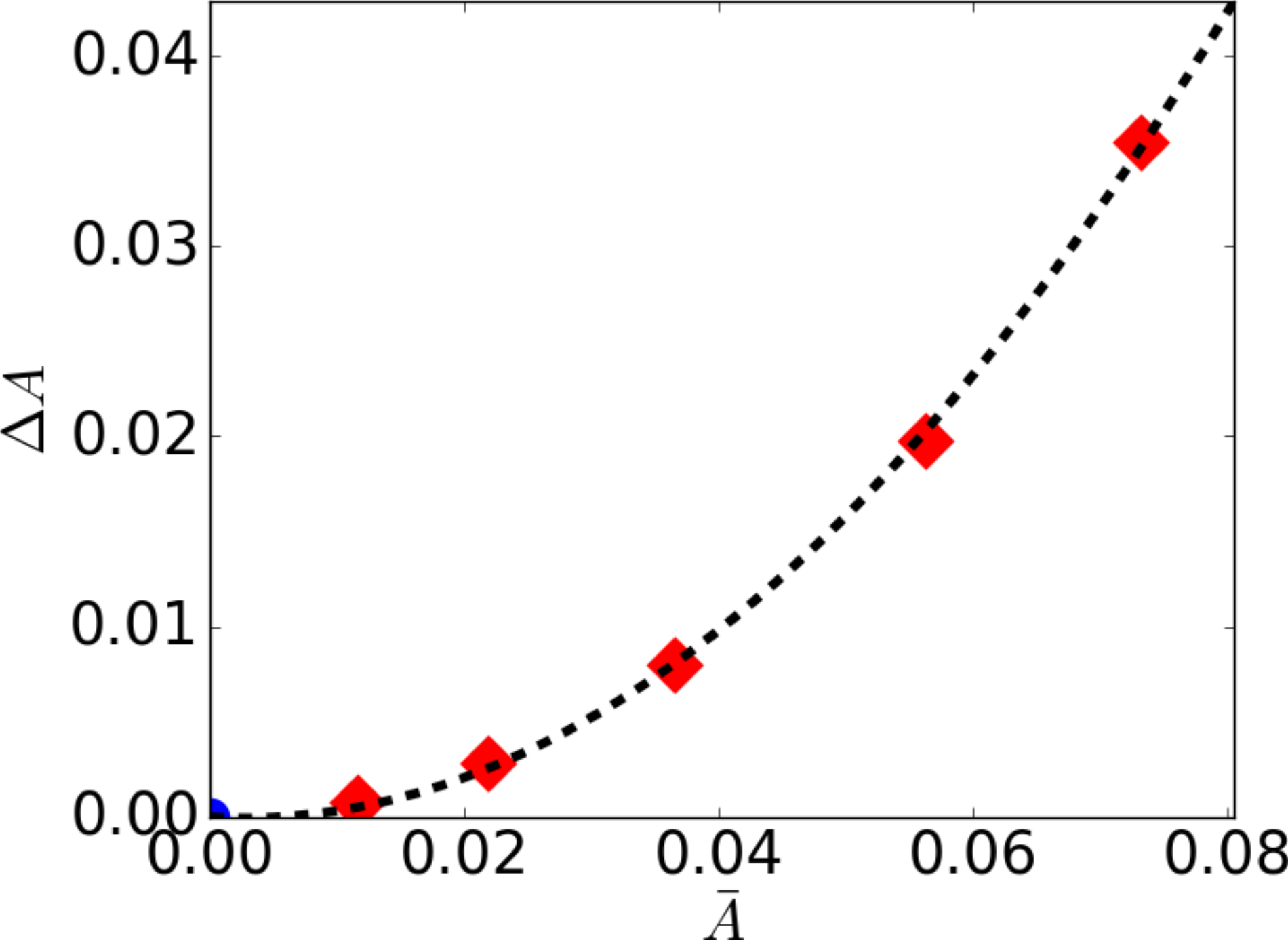}}
    \par
  \end{centering}
  \caption{
    For each solitary wave in Figure \ref{fig:more_diffractons}, we plot 
    (a) the amplitude as a function of $y$, and 
    (b) the variation $\Delta A:=\max_y A(y)-\min_yA(y)$ as a function of 
    mean amplitude $\bar A:=\frac{1}{\Omega}\int_{-\Omega/2}^{\Omega/2}A(y)dy$.
    The black dashed line in (b) is a quadratic least-squares curve fitted to the data 
    and constrained to pass through the known value $(0,0)$.
    In both cases, the amplitude is measured relative to the undisturbed water level $\mwl$.}
\end{figure}

To illustrate the two-dimensional structure of these waves,
in Figure \ref{fig:y_profiles} we plot, for the first five solitary waves from
Figure \ref{fig:more_diffractons}, the peak amplitude as a function of $y$; i.e.
$A(y):=\max_x \{\eta(x,y)-\mwl\}$.
Observe that the variation in $y$ is stronger for larger amplitude solitary waves;
for the smallest ones the wave is nearly uniform in $y$.
This suggests that the $y$-variation is a nonlinear effect.
To strengthen this argument, we plot in Figure \ref{fig:var_vs_amp} the variation in amplitude 
$\Delta A:=\max_y A(y)-\min_yA(y)$
versus the mean amplitude
$\bar A:=\frac{1}{\Omega}\int_{-\Omega/2}^{\Omega/2}A(y)dy.$
In addition, we plot a quadratic least-squares curve fitted to the data
and constrained to pass through the known value $(0,0)$.
The profiles plotted here are qualitatively similar to what is predicted
by the model derived in \cite{peregrine1969solitary}, though they differ in magnitude.

\begin{figure}[!h]
  \begin{centering}
    \includegraphics[scale=0.215]{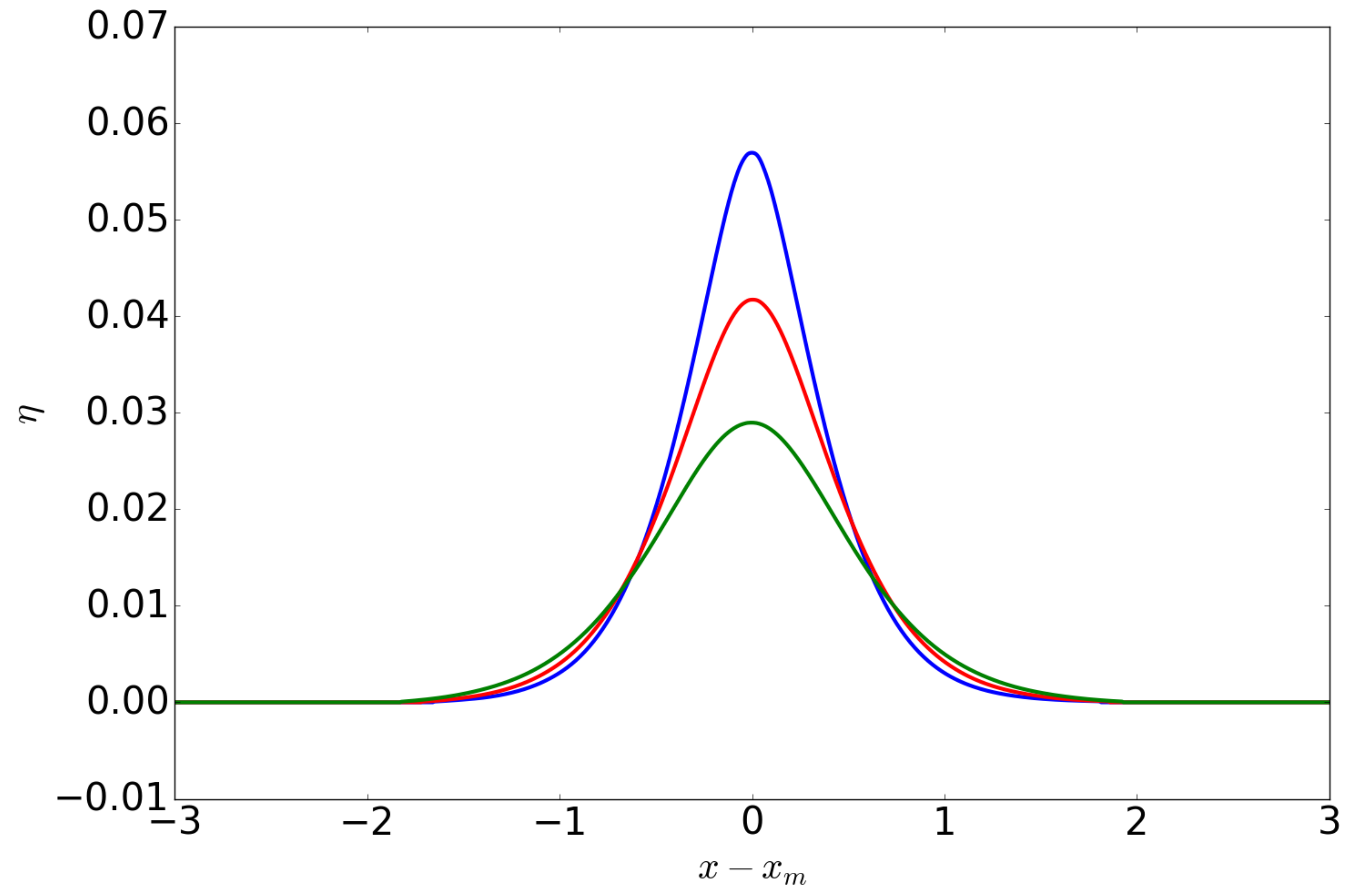}\qquad
    \includegraphics[scale=0.215]{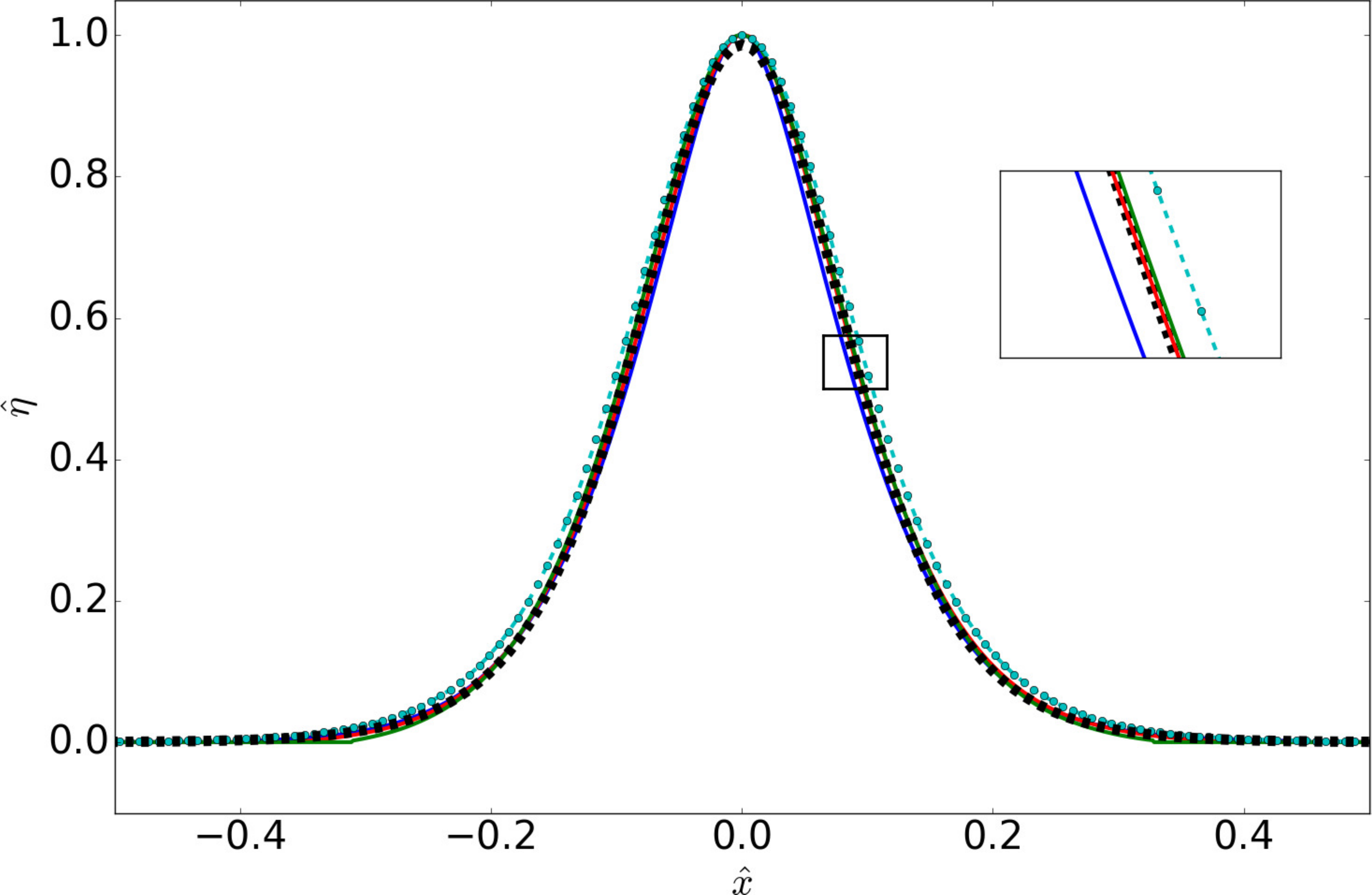}
    \par
  \end{centering}
  \caption{Scaling relation for bathymetric solitary waves.
    In the left panel we show the first three $y$-averaged solitary waves (given by \eqref{mean_amplitude_diffracton}) from Figure \ref{fig:more_diffractons}
    centered at the origin. In the right panel we show the same solitary waves after the scaling given by
    \eqref{scaling}. The black dashed line in the right panel is a $\text{sech}^2$ function fitted to the 
    data and the dotted cyan line is a soliton solution of a KdV-type equation that we derive in 
    \S\ref{sec:kdv_for_diffractons}. 
    \label{fig:scaling}}
\end{figure}

We consider again the first three solitary waves in Figure \ref{fig:more_diffractons}, computing
the $y$-averaged surface height
\begin{align}\label{mean_amplitude_diffracton}
  \eta_m(x,t=340)\approx\frac{1}{\Omega}\int_{-\Omega/2}^{\Omega/2} [\eta(x,y,t=340) - \mwl] dy,
\end{align}
and rescaling each wave by its amplitude
\begin{align}\label{scaling}
  \hat{\eta}_m(\hat{x}) := \frac{\eta_m\left(\hat{x}\right)}{A_m},
\end{align}
where $A_m=\max_x \eta_m(x)$, $\hat{x}=\sqrt{A_m}(x-x_m)$, with
$x_m=\text{argmax}_x \eta_m(x)$.
In the left panel of Figure \ref{fig:scaling} we show the three non-scaled, 
$y$-averaged solitary waves (given by \eqref{mean_amplitude_diffracton})
and in the right panel we show the same averaged solitary waves after the scaling defined by \eqref{scaling}.
After this scaling, all of the bathymetric solitary waves look similar, 
which is a common property of many solitary waves. 
The black dashed line in the right panel is a $\text{sech}^2$ function with amplitude and width fitted 
(in a least squares sense) to all of the solitary waves after the scaling.
In \S\ref{sec:kdv_for_diffractons} we derive a KdV-type equation that approximates 
the right-going part of the solution of \eqref{shallow_water_equations} with variable bathymetry 
\eqref{bathymetry}. The dotted cyan line in the right panel of Figure \ref{fig:scaling} 
is the solution of the aforementioned KdV-type equation with $A_m=1$; i.e., 
it is a soliton that approximates the bathymetric solitary waves for the configuration that we consider in Figure 
\ref{fig:more_diffractons}. 

\begin{figure}[!h]
 \begin{centering}
   \includegraphics[scale=0.35]{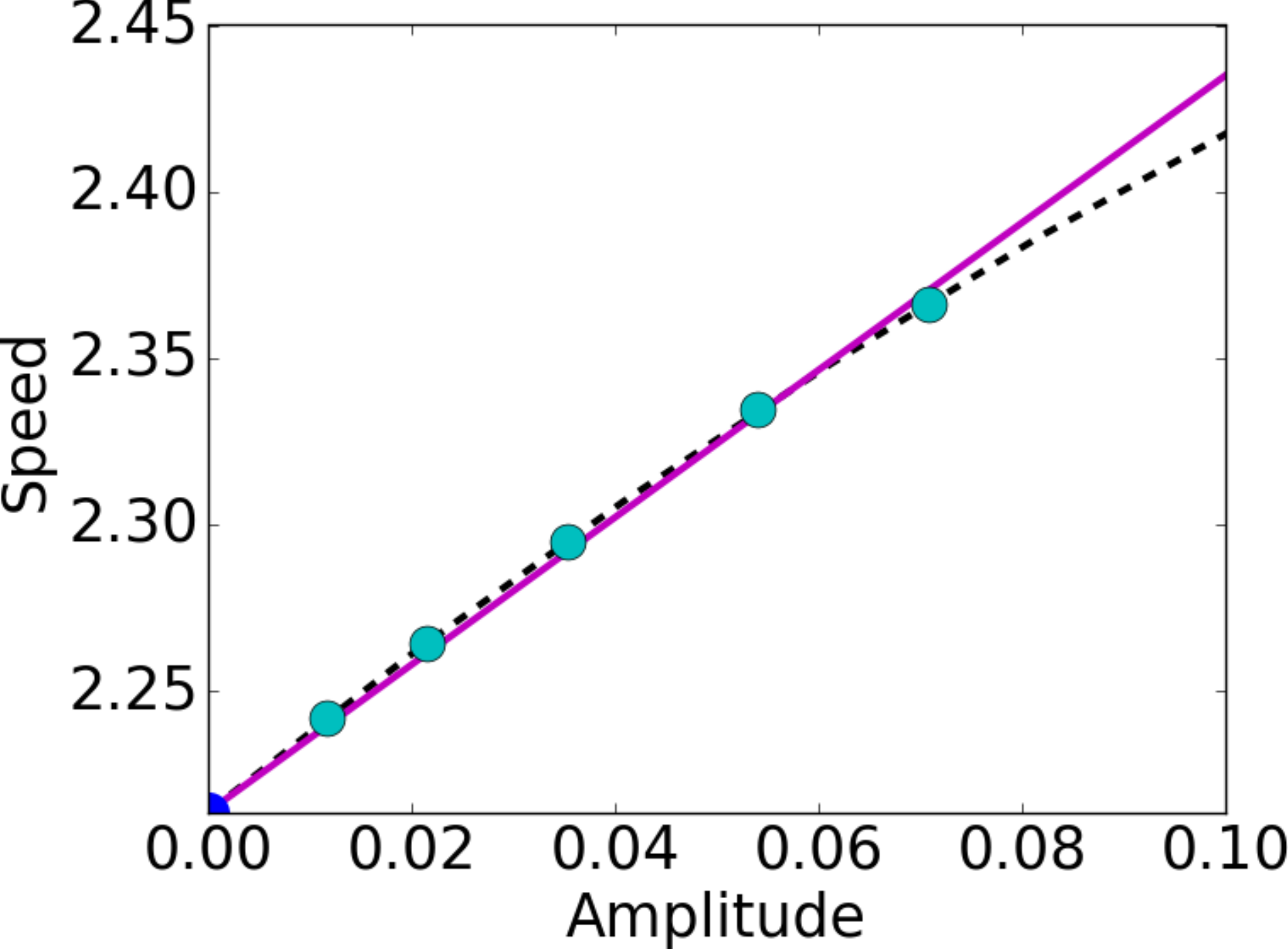}
   \par
 \end{centering}
 \caption{
   Speed-amplitude relation for bathymetric solitary waves.
   We measure the amplitude based on \eqref{mean_amplitude_diffracton}.
   The solid purple line is based on a KdV-type equation that we derive in \S\ref{sec:kdv_for_diffractons}
   and the dashed black line is the quadratic least-squares fitted curve to the cyan circles and constrained
   to pass through the known value $c_{\text{eff}}=\sqrt{g\langle\mwl-b\rangle}$ for zero-amplitude waves.
   The amplitude is measured relative to the undisturbed water level $\mwl$.
   \label{fig:speed-amplitude}}
\end{figure}

KdV solitons have a linear speed-amplitude relation \cite{korteweg1895xli,zabusky1965interaction}.
This is also true for other solitary waves.
For example, stegotons, which are solitary waves created due to effective
dispersion introduced by reflections in periodic media,
also have a linear speed-amplitude relation \cite{leveque2003}.
As we discussed before, for a given bathymetric solitary wave, the amplitude is $y$-dependent.
Therefore, to define the speed-amplitude relation, we must first define the amplitude of a bathymetric solitary wave.
If we use the $y$-averaged wave peak amplitude, we obtain a nearly linear relation,
as shown by the blue circles in Figure~\ref{fig:speed-amplitude}.
This relation also agrees well with the predicted speed-amplitude relationship
for small amplitude soliton solutions of a KdV-type model that we derive in
\S\ref{sec:kdv_for_diffractons}; this relation is shown with a solid purple line.
If we use instead the minimum or maximum amplitude (which occur at $y=0.25$ and $y=-0.25$, respectively),
we obtain nonlinear relationships, as shown also in Figure~\ref{fig:speed-amplitude}.

\subsection{Interaction of bathymetric solitary waves}\label{sec:interaction}
Another well-known property of many solitary waves is their tendency to interact
with one another only through a phase shift.
In this section we study the interaction of two solitary waves that are propagating
in either the same direction or opposite directions.
In both situations, the solitary waves are taken from 
the results shown in Figure \ref{fig:three_diffractons}. 
In all plots we show slices of the surface elevation 
along $y=0.25$ and $y=-0.25$.

\begin{figure}[!h]
 \begin{centering}
   \includegraphics[scale=0.21]{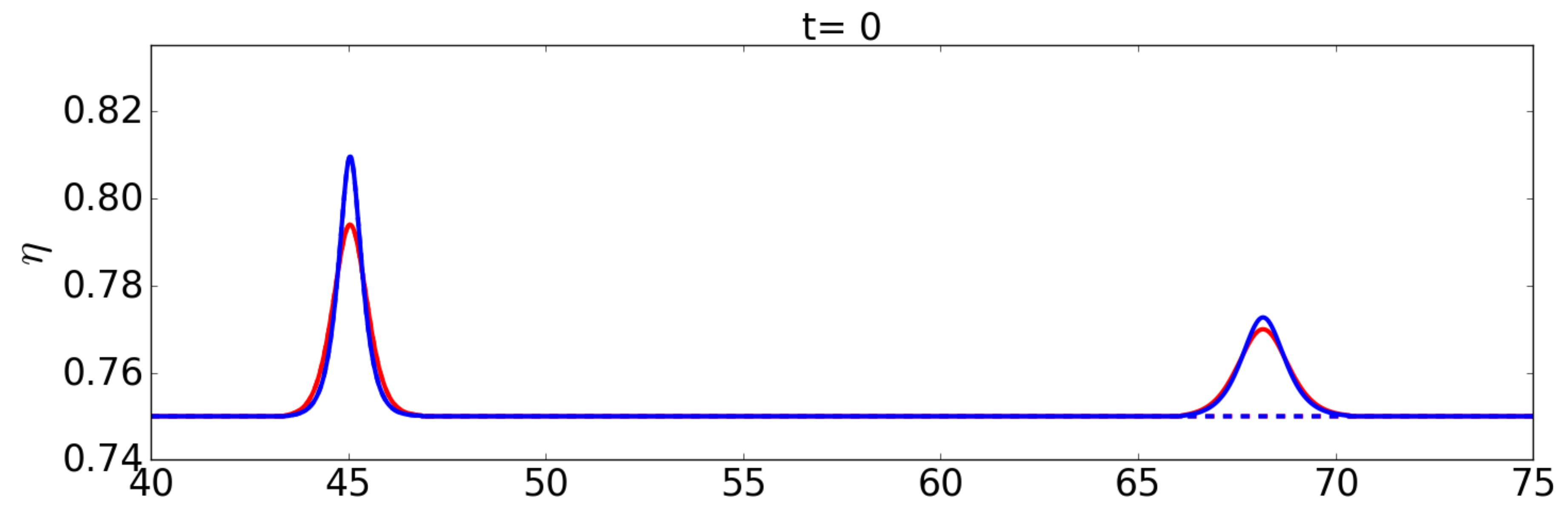}\qquad
   \includegraphics[scale=0.21]{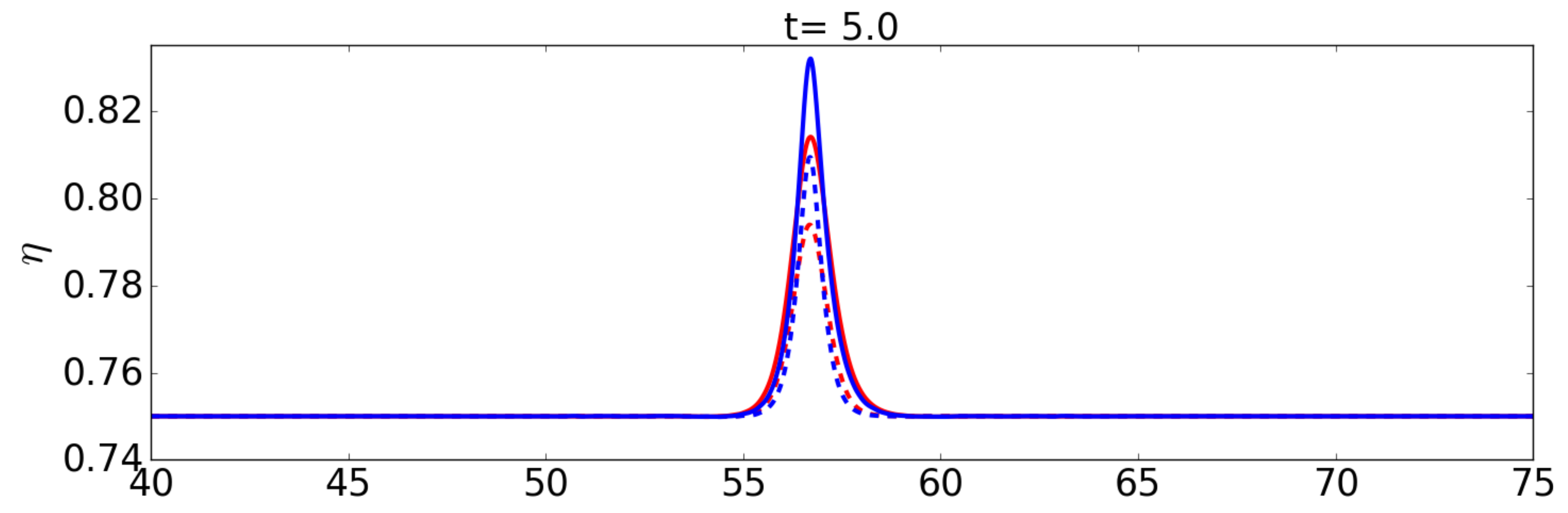}
   
   \includegraphics[scale=0.21]{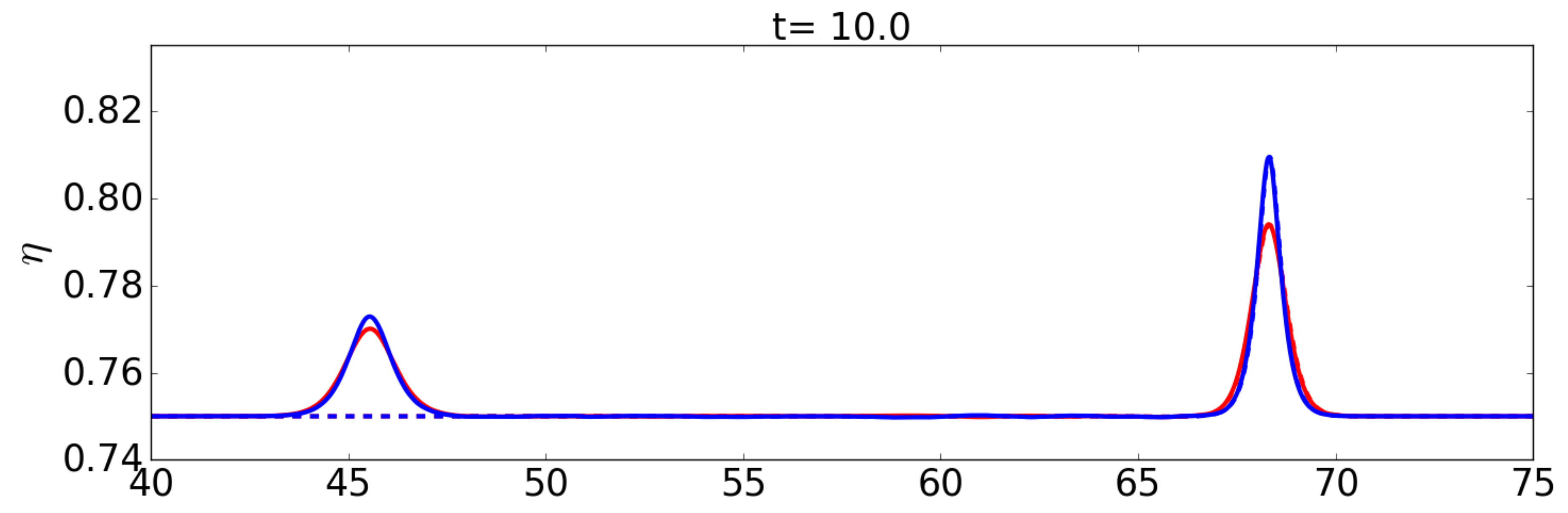}\qquad
   \includegraphics[scale=0.21]{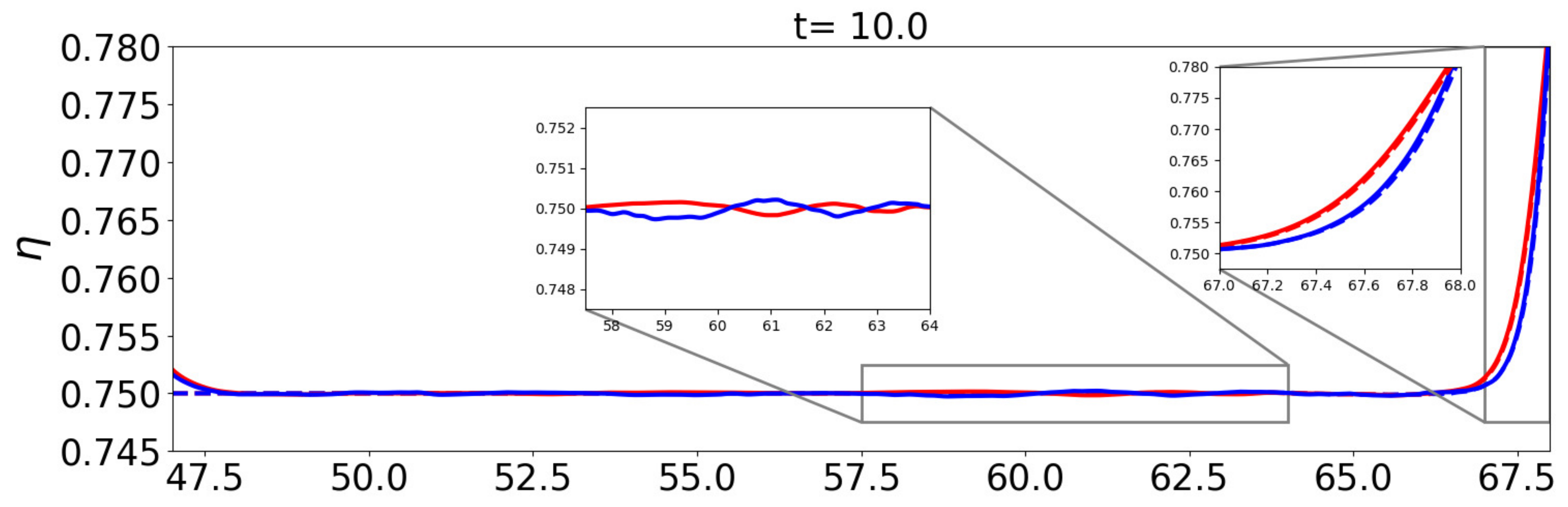}
   \par
 \end{centering}
 \caption{Counter-propagating collision.
   We show (in different color) slices at the middle of each bathymetry section. 
   As a reference, we plot the propagation of the taller solitary wave by itself.  
   In the bottom-right panel we zoom to the tails of the solitary waves after 
   the interaction to notice the oscillations and the slight change in phase. 
   \label{fig:counter_propagation_collision}}
\end{figure}

To produce a counter-propagating collision we negate the velocity of the shorter wave.
The initial condition is shown in the top-left panel of Figure \ref{fig:counter_propagation_collision}. 
Here the taller wave propagates to the right while the smaller one moves to the left.
We show the solution at different times during and after the interaction. 
As a reference, we propagate the taller wave by itself and superimpose the solution using dashed lines.
After the interaction very small oscillations are visible (see the bottom-right panel),
suggesting that the interaction is not elastic.
This has been reported before for diffractons \cite{ketcheson2015diffractons};
however, in this case, the oscillations in the tail are much weaker.
Note that there is an almost unnoticeable change in the phase for the taller
solitary wave with respect to the propagation without interaction; this is due
to the relatively short time of interaction. 
Although the phase shift is also small for diffractons \cite{ketcheson2015diffractons}, 
the phase shift in our numerical experiments with bathymetric solitary waves is much smaller.

\begin{figure}[!h]
 \begin{centering}
   \includegraphics[scale=0.21]{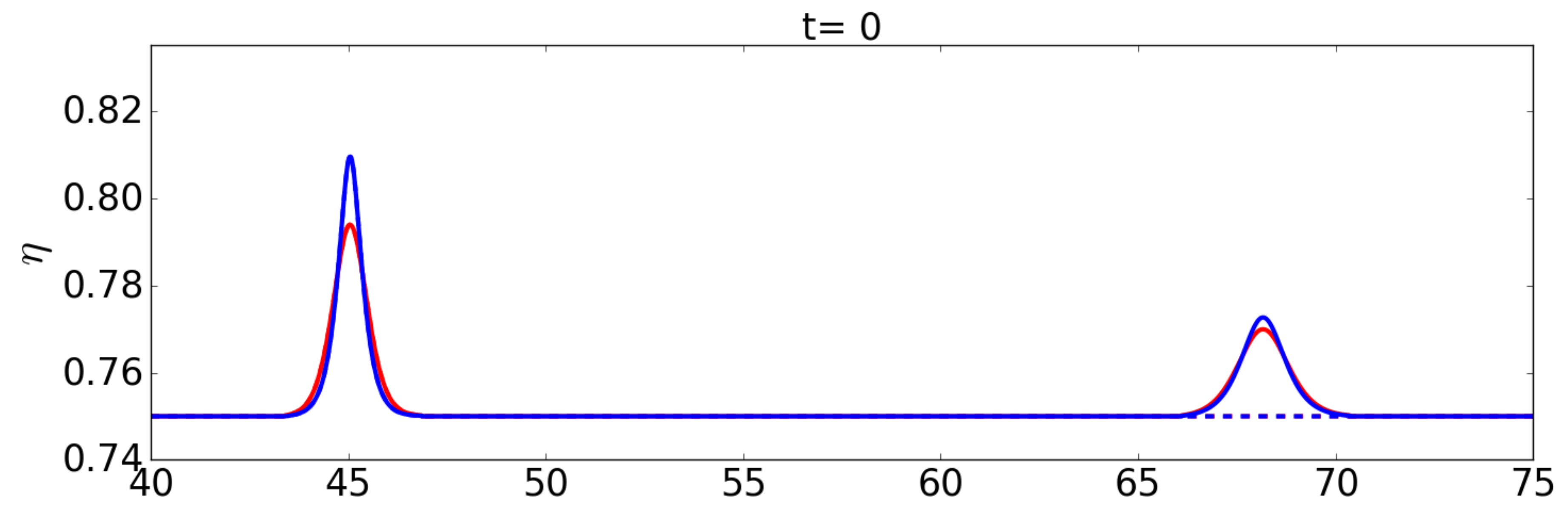}\qquad
   \includegraphics[scale=0.21]{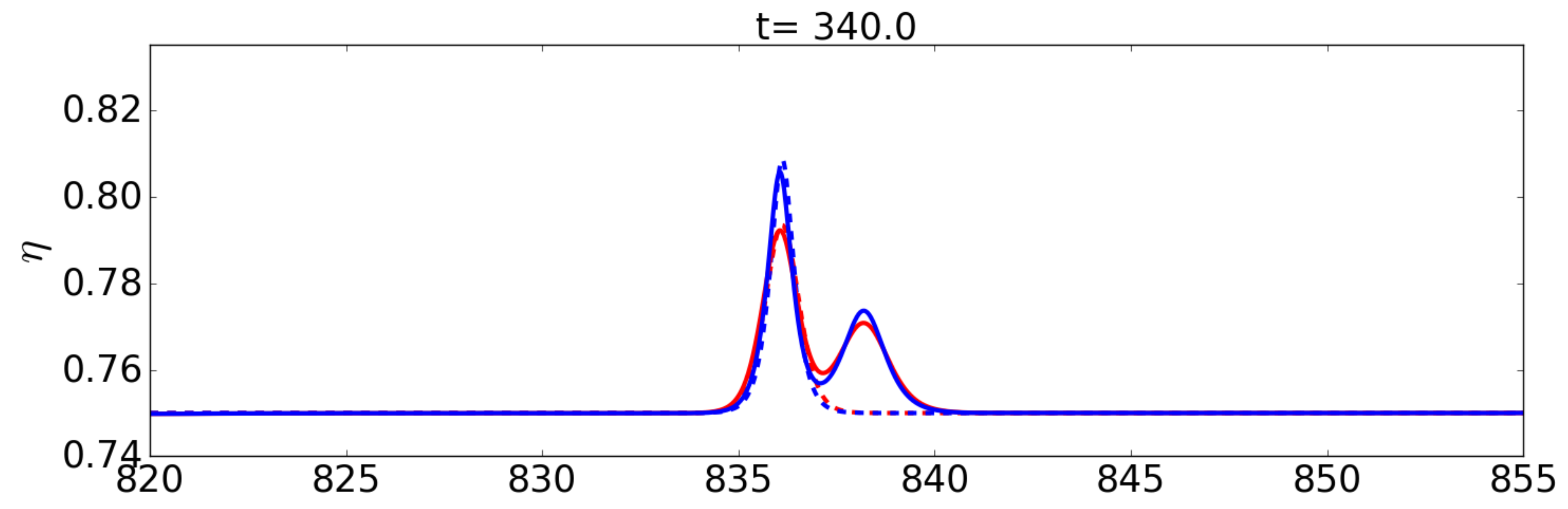}
   
   \includegraphics[scale=0.21]{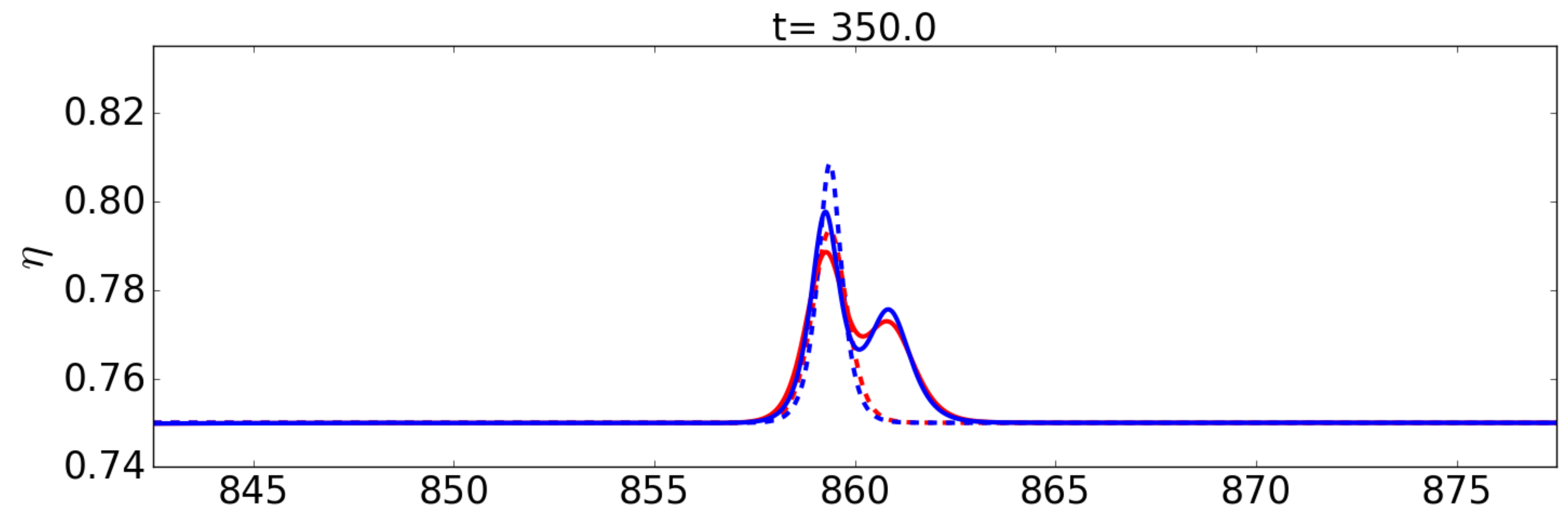}\qquad
   \includegraphics[scale=0.21]{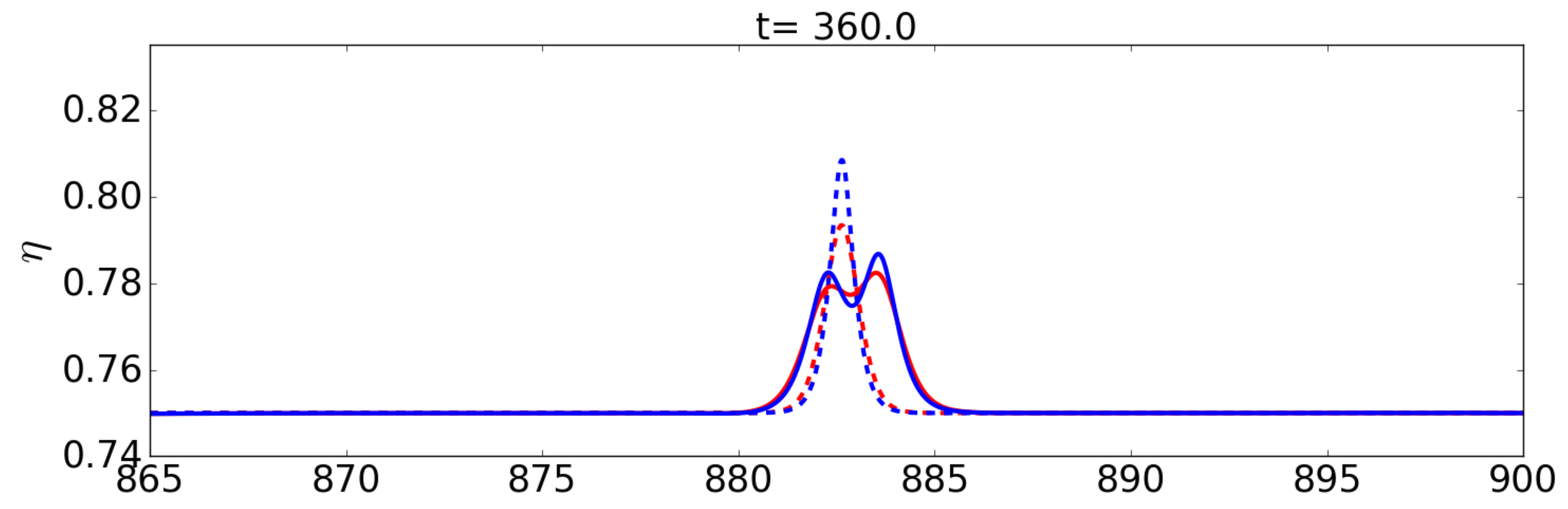}
   
   \includegraphics[scale=0.21]{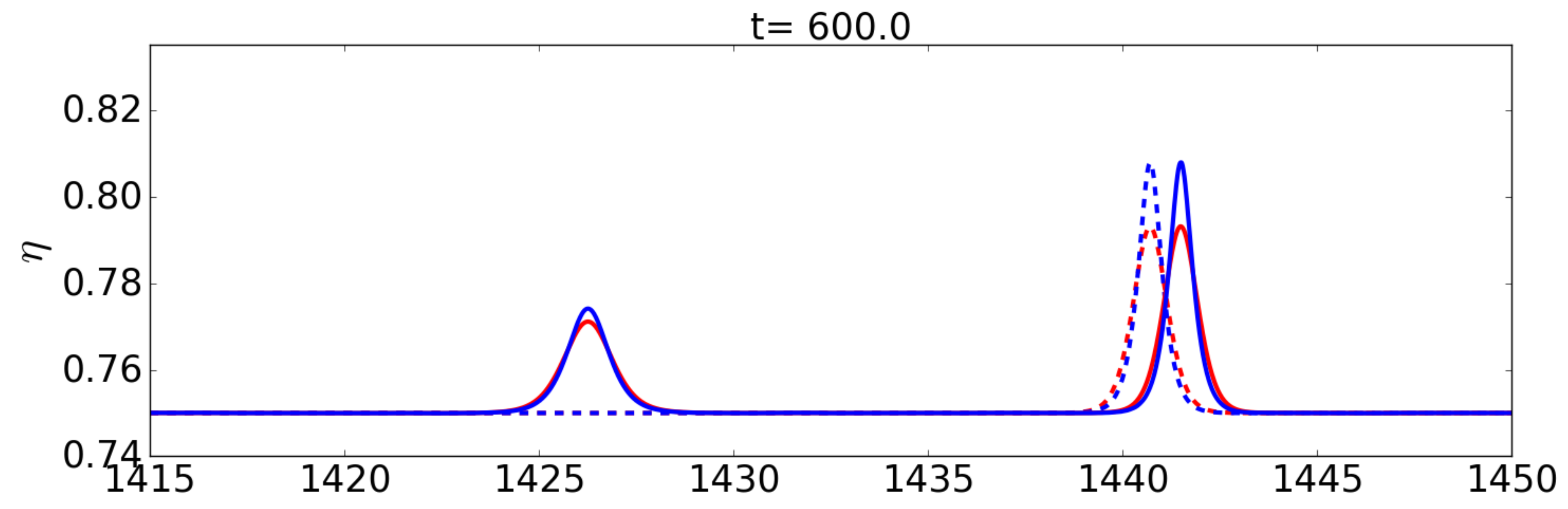}\qquad
   \includegraphics[scale=0.21]{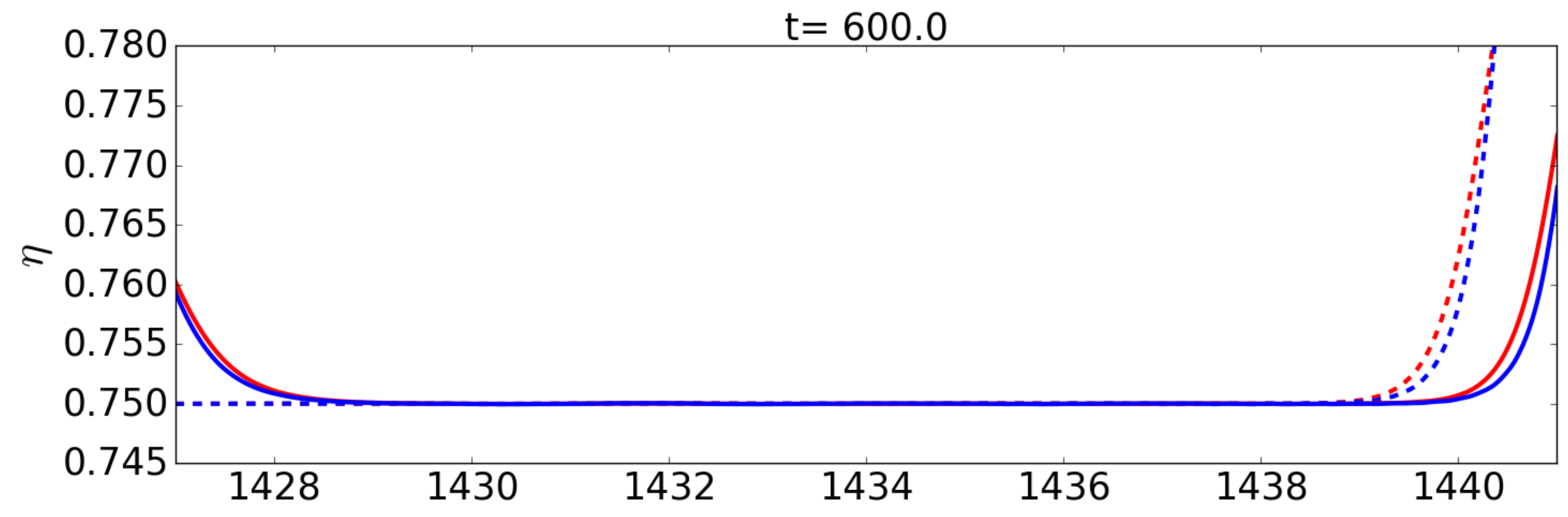}
   \par
 \end{centering}
 \caption{Co-propagating collision at different times. 
   We show (in different color) slices at the middle of each bathymetry section. 
   As a reference, we plot the propagation of the taller solitary wave by itself.  
   In the bottom-right panel we zoom to the tails of the solitary waves after 
   the interaction.
   \label{fig:co_propagation_collision}}
\end{figure}

Now we consider a collision where both waves move in the same direction. 
The initial condition is shown in the top-left panel of Figure \ref{fig:co_propagation_collision}. 
Here both waves move to the right.
Since the taller wave moves faster (see \S\ref{sec:speed-amplitude}) it eventually reaches and passes the smaller one.
Again we show the solution at different times during and after the interaction. 
As a reference, we propagate the taller wave by itself and superimpose the solution using dashed lines.
In this case there are no visible oscillations after the collision,
suggesting an elastic collision. In contrast to the counter-propagating collision, 
the interaction time is larger, which leads to a noticeable shift in phase.
This is a common feature for other solitary waves and for diffractons.

\section{KdV model for weakly nonlinear bathymetric solitary waves}\label{sec:kdv_for_diffractons}
In the previous section we explored some properties of bathymetric solitary waves.
The shape of any given of these waves is not $y$-independent; i.e., these are truly
two-dimensional waves that travel in one direction. However, their shape is closely
connected with a one-dimensional $\text{sech}^2$ function. In addition, bathymetric solitary waves
travel without significant change in the shape and interact similar to KdV-solitons.
All these properties strongly suggest that one might model small amplitude bathymetric 
solitary waves via a KdV equation. We do not expect such a model to be completely accurate 
for large-amplitude bathymetric solitary waves since as the amplitude increases the waves 
behave less like solitons; see \S\ref{sec:speed-amplitude}.
In this section, we obtain a KdV-type equation that accounts for bathymetric dispersion and is 
valid for weakly nonlinear waves. 
Before doing that, however, in the next section we compare the shallow water solutions
with Peregrine's equation that accounts for additional sources of water wave dispersion.

\subsection{Bathymetric solitary waves via an inherently dispersive water wave model} \label{sec:peregrine}
Another water wave model that accounts for dispersive effects due to changes in the bathymetry
has been derived and analyzed in \cite{peregrine1968long,teng1997effects}.  That model for flow
in non-rectangular channels takes the form
\begin{align}\label{peregrine}
  \eta_t+\sqrt{g\bMwl}\eta_x+\frac{3}{2}\sqrt{\frac{g}{\bMwl}}(\eta-\bMwl)\eta_x+\frac{\kappa^2}{6}(\bMwl)^2\sqrt{g\bMwl}\eta_{xxx}=0.
\end{align}
Equation \eqref{peregrine} is a KdV-type equation with a modified dispersion coefficient.
Here $\kappa^2$ is called the channel shape factor; it depends on the cross section of the channel and is given by 
\begin{align}\label{kappa2}
  \kappa^2=\frac{3}{(\bMwl)^2}\left[\frac{1}{|D|}\int_D\Psi(y,z)dydz-\frac{1}{|L|}\int_L\Psi(y,\mwl)dy\right]
\end{align}
where $D\subset \mathbb{R}^2$ is the cross section of the channel, $L\subset D$ is the top boundary of the cross section
(i.e., the undisturbed free surface) and $\Psi$ is the solution of the following elliptic boundary value problem:
\begin{subequations}\label{elliptic_pde_peregrine}
  \begin{align}
    \Psi_{yy}+\Psi_{zz} &= 1, \\
    \Psi_z|_{z=\mwl} &= \bMwl, \\
    \Psi_{\bfn} &= 0.
  \end{align}
\end{subequations}
Here $\bfn$ is the unit vector normal to the boundary of the cross section of the channel. 
For a rectangular channel, $\kappa=1$ so \eqref{peregrine} becomes the standard KdV equation.

It is natural to ask how Peregrine's model \eqref{peregrine} compares with solutions of the
shallow water equations in a non-rectangular channel -- or equivalently, over the kind of
periodic bathymetry we have considered.  To answer this, we first observe that we cannot
expect agreement between the models if the shallow water model leads to shock formation;
this can occur if the initial data is very large or if the bathymetry variation is small
(i.e. if $\kappa \approx 1$) \cite{ketcheson2020effRH}.

\begin{figure}[!h]
  \begin{centering}
    \subfloat[$\bB=0.01$,
      $\mwl=0.015$ and $\epsilon=0.001$. \label{fig:peregrine_regime1}]{
      \includegraphics[scale=0.1525]{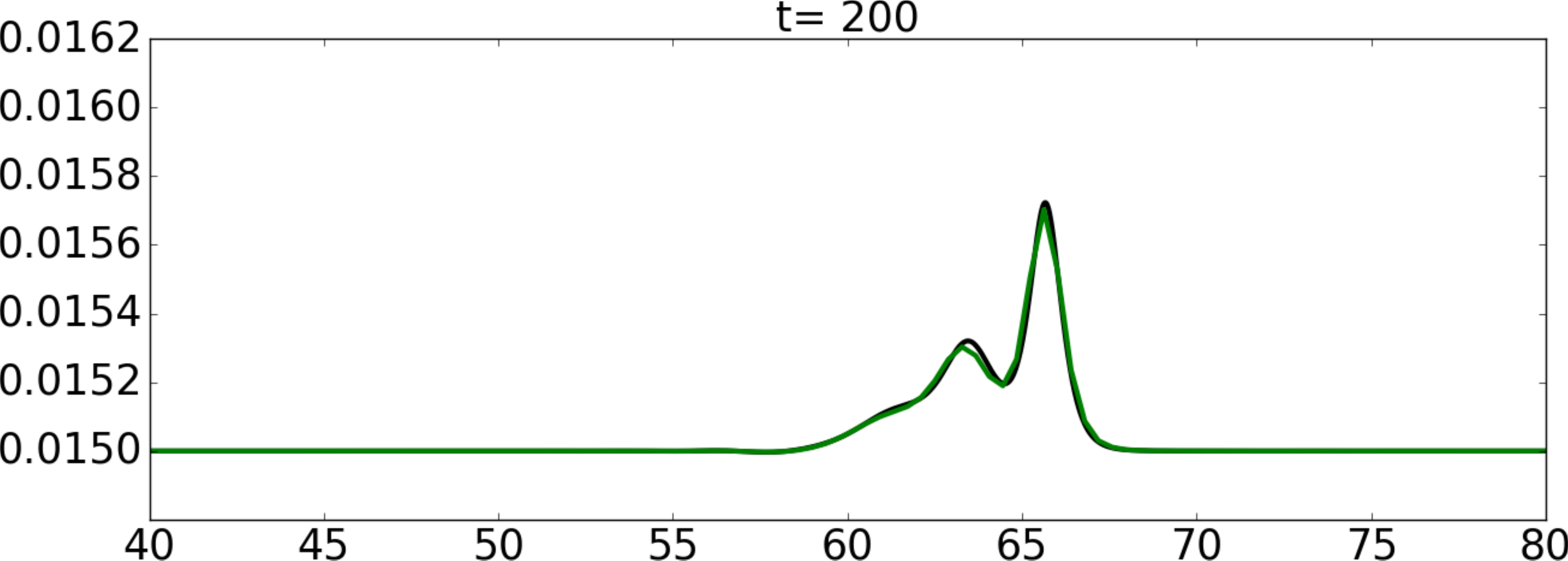} \quad
      \includegraphics[scale=0.1525]{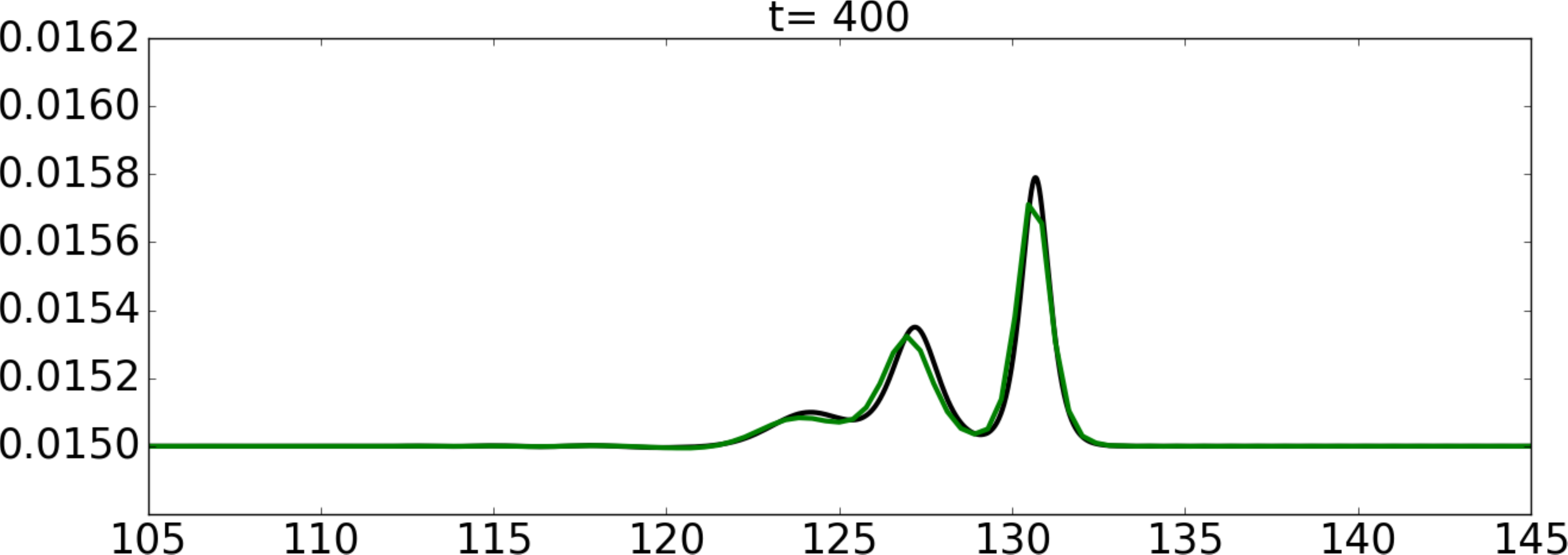} \quad
      \includegraphics[scale=0.1525]{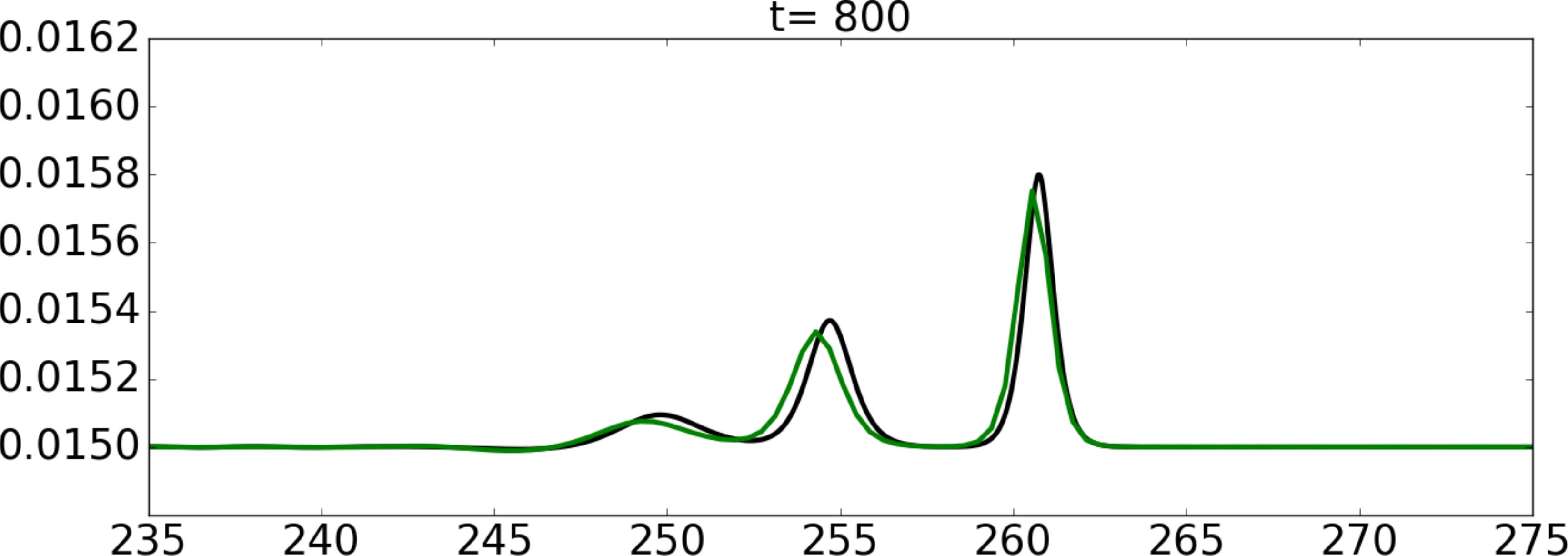} 
    }

    \subfloat[$\bB=0.5$,
      $\mwl=0.75$ and $\epsilon=0.05$. \label{fig:peregrine_regime2}]{
      \includegraphics[scale=0.16]{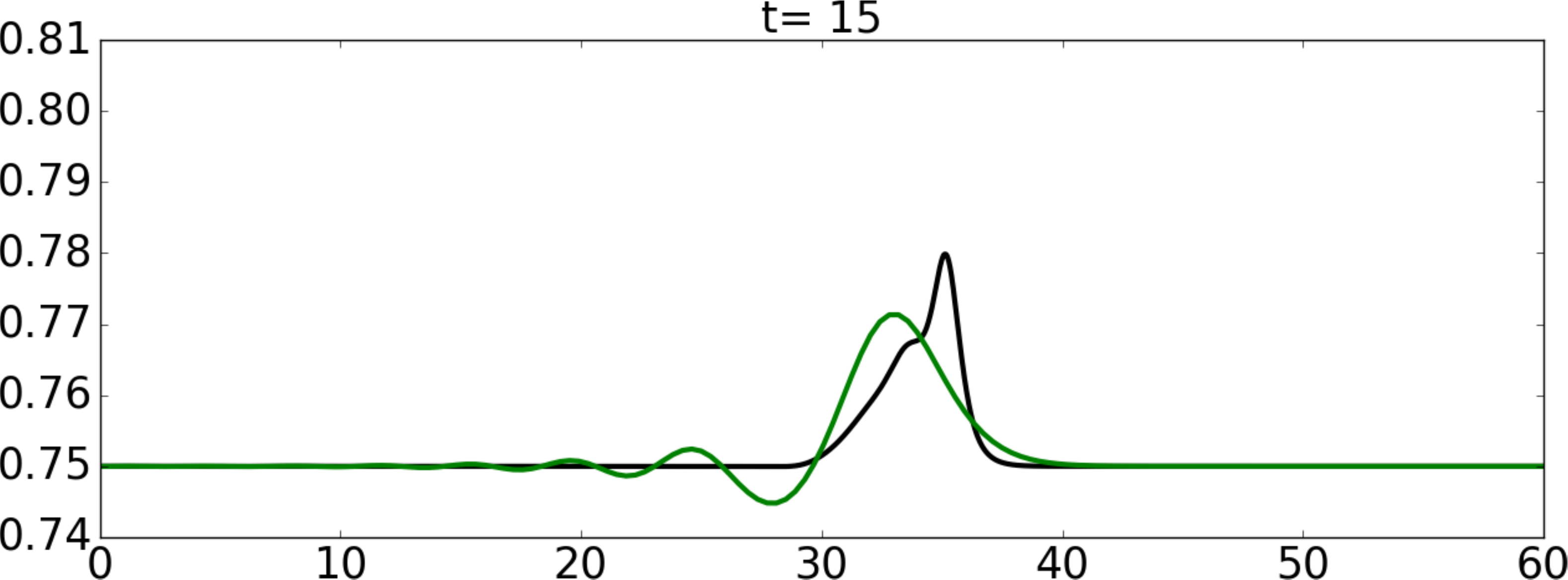} \quad
      \includegraphics[scale=0.16]{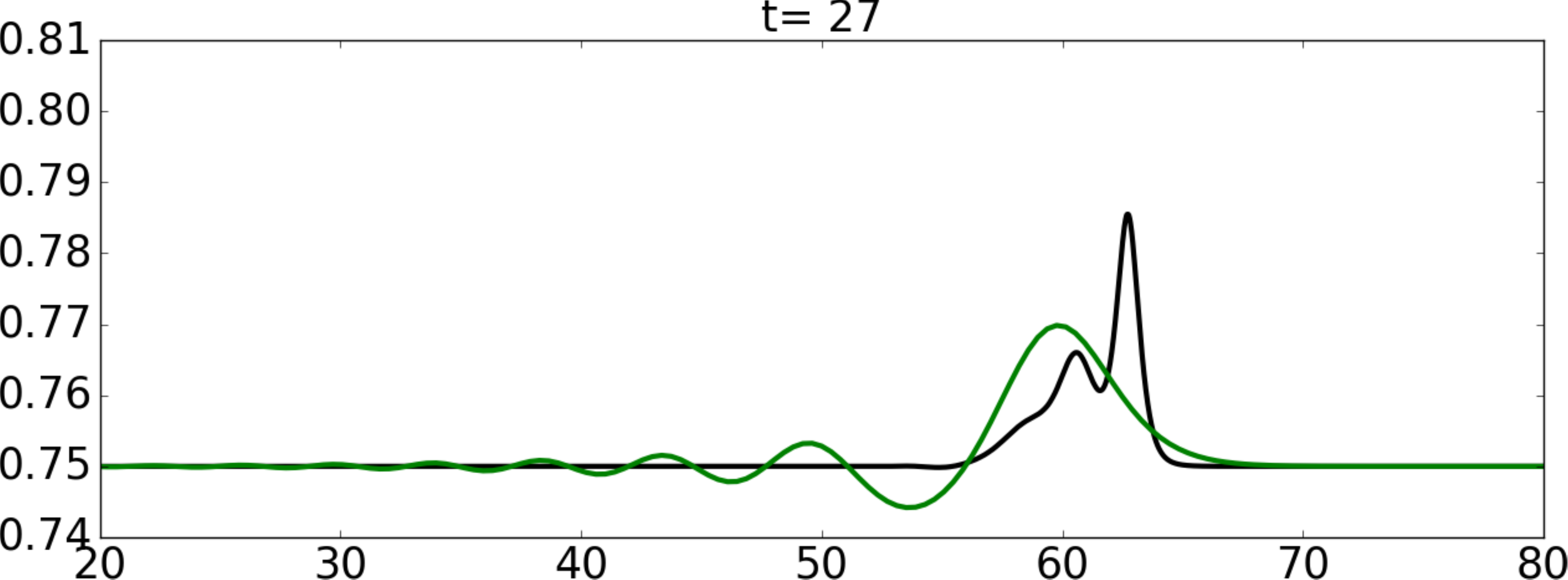} \quad
      \includegraphics[scale=0.16]{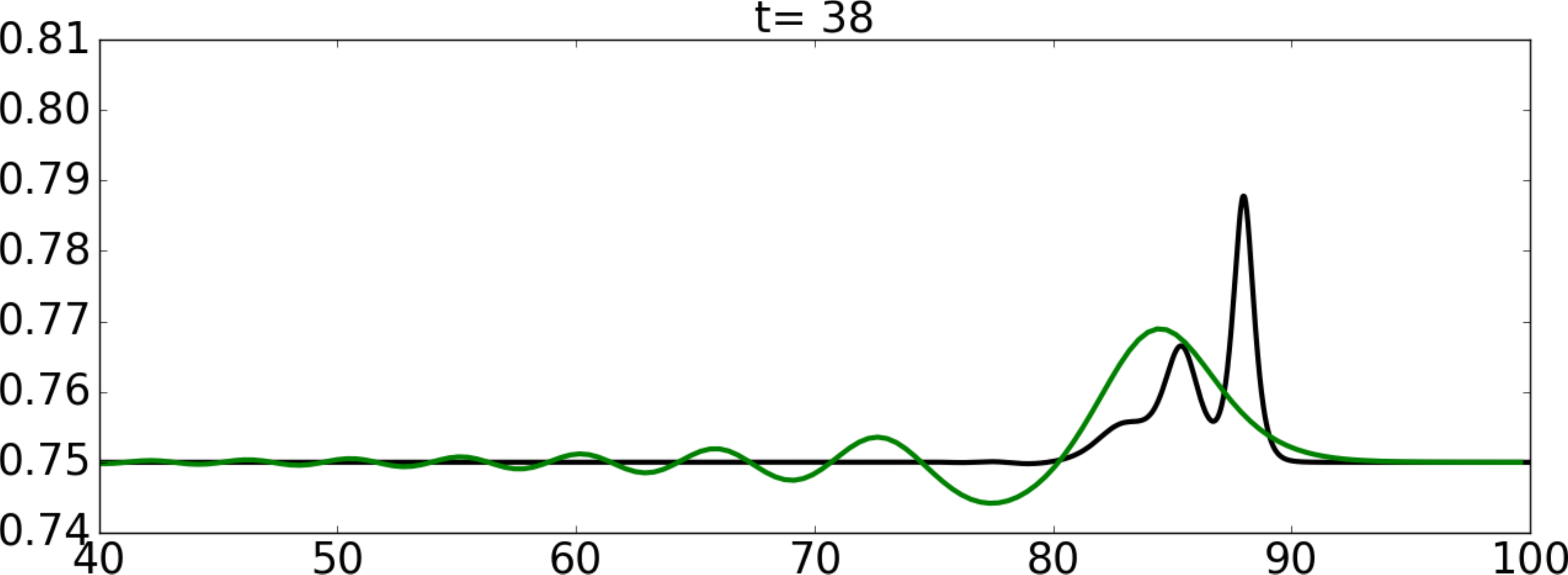}
    }
    \par
  \end{centering}
  \caption{
    In green, solution of Peregrine's model \eqref{peregrine}. 
    In black, $y$-averaged solution of the shallow water equations \eqref{shallow_water_equations} 
    with periodic bathymetry. 
    The bathymetry is given by \eqref{bathymetry} with $\bB$ as indicated in the subfigures. 
    The initial condition is given by \eqref{eta0} with $\alpha=2$ and $\eta^*$ and $\epsilon$ as indicated 
    in the subfigures.\label{fig:peregrine}}
\end{figure}

On the other hand, if the bathymetry variation is relatively large and the initial data
is not too large, the two models can be in relatively close agreement for fairly long times.
An example of this is shown in Figure \ref{fig:peregrine_regime1}.  Here we take
bathymetry given by \eqref{bathymetry} with $\bB=0.01$.  The initial data is
given by \eqref{eta0} with $\eta^*=0.015$, $\epsilon=0.001$, and $\alpha=2$.
Note that for this case $\kappa^2\approx 214.14$.  
We solve \eqref{peregrine} using a Fourier pseudospectral collocation method.  
We see very close agreement
between the solutions, even after the waves have traveled a distance equal to
hundreds of times the channel width.

In Figure \ref{fig:peregrine_regime2} we show another example, in which
bathymetric dispersion is much less dominant.  The bathymetry and the amplitude
of the initial data are 50 times larger, with $\bB=0.5$, $\eta^*=0.75$, and
$\epsilon=0.05$.  Note that for this case $\kappa^2\approx 1.58$.  We see
that the resulting solutions are completely different, even from a qualitative
perspective.
This is expected, since the model leading to \eqref{peregrine}
includes additional dispersion beyond that caused by
bathymetry variation.  The shallow water model neglects that additional
dispersion and may therefore be less accurate in any regime
(like that of the latter scenario) where it is important.
In Section \ref{sec:about_dispersive_effects}, we compare the strength of
these two types of dispersion independently and compare each against that
predicted by Peregrine's model; see Figure \ref{fig:dispCoeff}.

\begin{remark}[About the amplitude of the initial data]
  The shallow water equations \eqref{shallow_water_equations} with the initial condition \eqref{eta0}
  model the propagation of a left- and a right-going wave. On the other hand, the KdV-type equation 
  \eqref{peregrine} models the propagation of a right-going wave. For this reason, the amplitude $\epsilon$
  of the initial condition \eqref{eta0} differs by a factor of two between \eqref{shallow_water_equations}
  and \eqref{peregrine}. For simplicity, we report the 
  amplitude used for the shallow water equations, for the KdV-type equations we use half of that amplitude. 
\end{remark}

\subsection{KdV-type equation with purely bathymetric dispersion}
Now we derive a KdV-type equation whose dispersive effects are only a consequence of changes in bathymetry. 
This equation models small amplitude bathymetrtic solitary waves. 
In the work by \cite{leveque2003}, the authors considered a one-dimensional nonlinear system and
derived homogenized equations with a dispersive correction.
Later in \cite{ketcheson2015diffractons}, the authors extended the results to two-dimensions.
In these references, the nonlinearity was chosen to facilitate the homogenization process. 
Ideally, we would aim to proceed as in \S\ref{sec:dispersion_by_diffraction}
and obtain a nonlinear homogenized system with constant coefficients and a dispersive correction.
Unfortunately, the nonlinear terms in the shallow water equations complicate the process.
Based on these references and the results in \S\ref{sec:dispersion_by_diffraction},
it is possible to make an ansatz for what the homogenized nonlinear system should look like. 
We hypothesize that
\begin{subequations}\label{hom_nonlinear_system}
\begin{align}
  \eta_t+\left[(\eta-b_m)u\right]_x &= \delta^2\Phi,\\
  u_t+uu_x+g\eta_x &= g\delta^2\beta_2\eta_{xxx}
\end{align}
\end{subequations}
is a homogenized system that models the $x$-propagation of shallow water waves over
general $y$-periodic bathymetry.
Here $b_m=\langle b\rangle=\frac{1}{2}\bB$, $\beta_2$ is the coefficient of dispersion defined in \eqref{hom_coeffs}
and $\delta^2\Phi=\delta^2\Phi(\eta,\eta_x,\eta_{xx},\eta_{xxx},u,u_{x},u_{xx},u_{xxx})$ 
is an $\bigO(\delta^2)$ nonlinear term that depends not only on the solution but
also on its derivatives. 
We do not know a closed form for $\Phi$; however, we expect it to 
introduce (potentially nonlinear) dispersive effects.

From \eqref{hom_nonlinear_system} we derive a KdV-type equation. To do this we
follow e.g. \cite{garnier2007effective}. First we consider the Riemann invariants of the left hand side of \eqref{hom_nonlinear_system}:
\begin{subequations}\label{r-invariants}
  \begin{align}
    \Rn &=2\sqrt{g(\eta-b_m)}-u,
    \quad \text{ on } \quad dx/dt=u-\sqrt{g(\eta-b_m)}, \\
    \Rp &=2\sqrt{g(\eta-b_m)}+u,
    \quad \text{ on } \quad dx/dt=u+\sqrt{g(\eta-b_m)}.
  \end{align}
\end{subequations}
By plugging \eqref{r-invariants} into \eqref{hom_nonlinear_system} we obtain a system
of PDEs for the space-time evolution of the Riemann invariants $\Rn$ and $\Rp$.
We focus on the right going invariant $\Rp$ and choose $\Rn=2\sqrt{g\bMwl}$, 
which is set constant to match with the solution as $|x|\rightarrow\infty$.
Here $\bMwl=\mwl-b_m$.
Doing this, we obtain
\begin{align*}
  \Rp_t-\frac{1}{4}(\Rn-3\Rp)\Rp_x=\frac{\delta^2\beta_2}{8}[(\Rn+\Rp)\Rp_{xxx}+3\Rp_x\Rp_{xx}] + \delta^2 C,
\end{align*}
where $C$ is an unknown function of $\Rn$, $\Rp$ and the derivatives of $\Rp$; i.e., 
$C=C(\Rn,\Rp,\Rp_x,\Rp_{xx},\Rp_{xxx})$. 
By using the Riemann invariants \eqref{r-invariants}, we go back to the physical variables and obtain
\begin{align}\label{before_KdV}
  \eta_t-\sqrt{g\bMwl}\eta_x+3\sqrt{g(\eta-b_m)}\eta_x=\delta^2\frac{\beta_2}{2}\sqrt{g(\eta-b_m)}\eta_{xxx} + \delta^2\hat{\Phi},
\end{align}
where $\hat{\Phi}$ is an unknown nonlinear function of $\eta$ and its derivatives;
i.e., $\hat{\Phi}=\hat{\Phi}(\eta,\eta_x,\eta_{xx},\eta_{xxx})$.
Although, we do not know the exact form of $C$ and $\hat{\Phi}$, 
these terms appear as a consequence of the manipulations of $\Phi$ via the Riemann invariants.
Based on Section \ref{sec:homogenization} and the references therein, we expect and assume that 
$\hat{\Phi}$ is a dispersive term.
Finally, we expand the nonlinear term $\sqrt{\eta-b_m}$ around $\mwl$ and drop the terms that
are of size $\bigO(\delta^2\varepsilon)$, where $\varepsilon\sim \eta-\eta^*$. We obtain
\begin{subequations}\label{sw_KdV}
  \begin{align}\label{sw_KdV_eqn1}
    \eta_t+\sqrt{g\bMwl}\eta_x+\frac{3}{2}\sqrt{\frac{g}{\bMwl}}(\eta-\mwl)\eta_x+\sigma(\gamma)\sqrt{g\bMwl}\eta_{xxx} = 0,
  \end{align}
  with
  \begin{align}\label{sw_KdV_dispCoeff}
    \sigma(\gamma)=\delta^2\frac{|\beta_2|}{2}(1+\gamma),
  \end{align}
\end{subequations}
where $\gamma$ is a constant (to be determined) that accounts for the linear dispersive part of $\delta^2\hat{\Phi}$. 
Equation \eqref{sw_KdV_eqn1} is a KdV-type equation with a dispersion coefficient given by \eqref{sw_KdV_dispCoeff}.
It models the $x$-propagation of shallow water waves over two-dimensional 
$y$-periodic bathymetry. Different types of bathymetry are modeled via 
$\beta_2$, which is given by \eqref{hom_coeffs}.
Just like the linear dispersive equation derived in \S \ref{sec:dispersion_by_diffraction}, 
the KdV-type equation \eqref{sw_KdV} 
only accounts for dispersive effects due to changes in the bathymetry.

It is important to remark that \eqref{sw_KdV} is a weakly nonlinear approximation of (the right-going part of)
\eqref{hom_nonlinear_system}; therefore, we expect it to be a good approximation only for
small-amplitude waves.
Indeed, solitary wave solutions of \eqref{sw_KdV} travel with a speed proportional to their
amplitude (see \S\ref{sec:profile_and_speed_sw_diffractons}); however, from \S\ref{sec:speed-amplitude},
we know that the speed-amplitude relation for bathymetric solitary waves
is approximately linear only for small-amplitude waves.

\subsubsection{Profile and speed of weakly nonlinear bathymetric solitary waves}\label{sec:profile_and_speed_sw_diffractons}
\begin{figure}[!h]
  \begin{center}
    \subfloat[$A_m=6.25\times 10^{-5}$]{
      \includegraphics[scale=0.145]{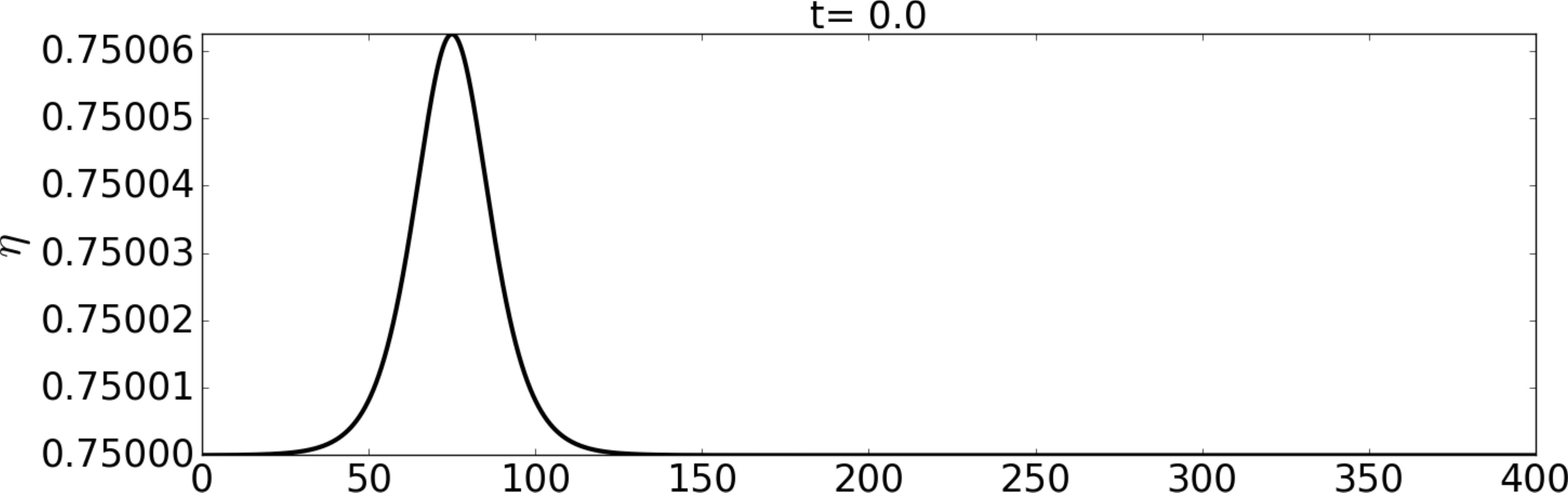}
      \includegraphics[scale=0.145]{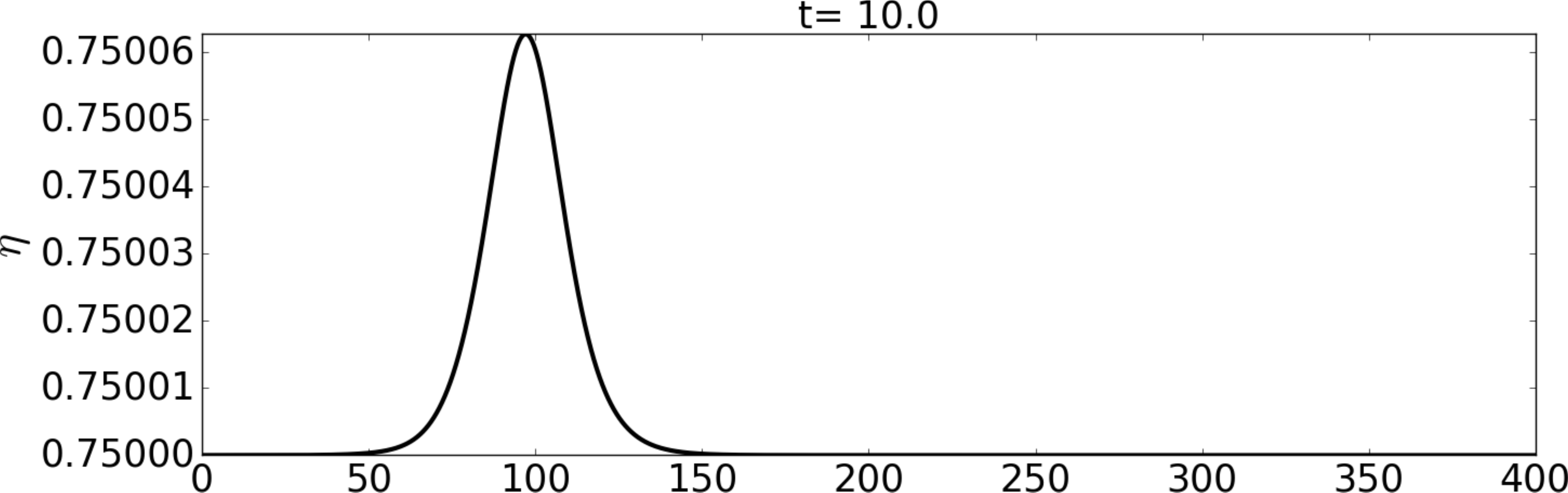}
      \includegraphics[scale=0.145]{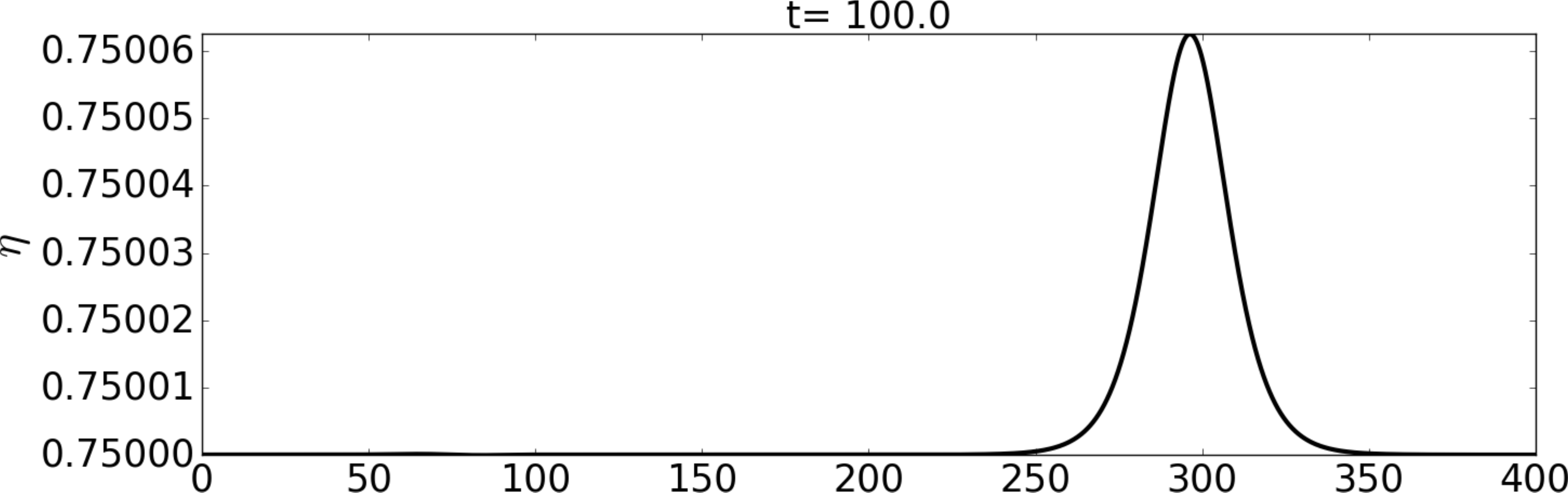}
    }
    
    \subfloat[$A_m=2\times 10^{-3}$]{
      \includegraphics[scale=0.145]{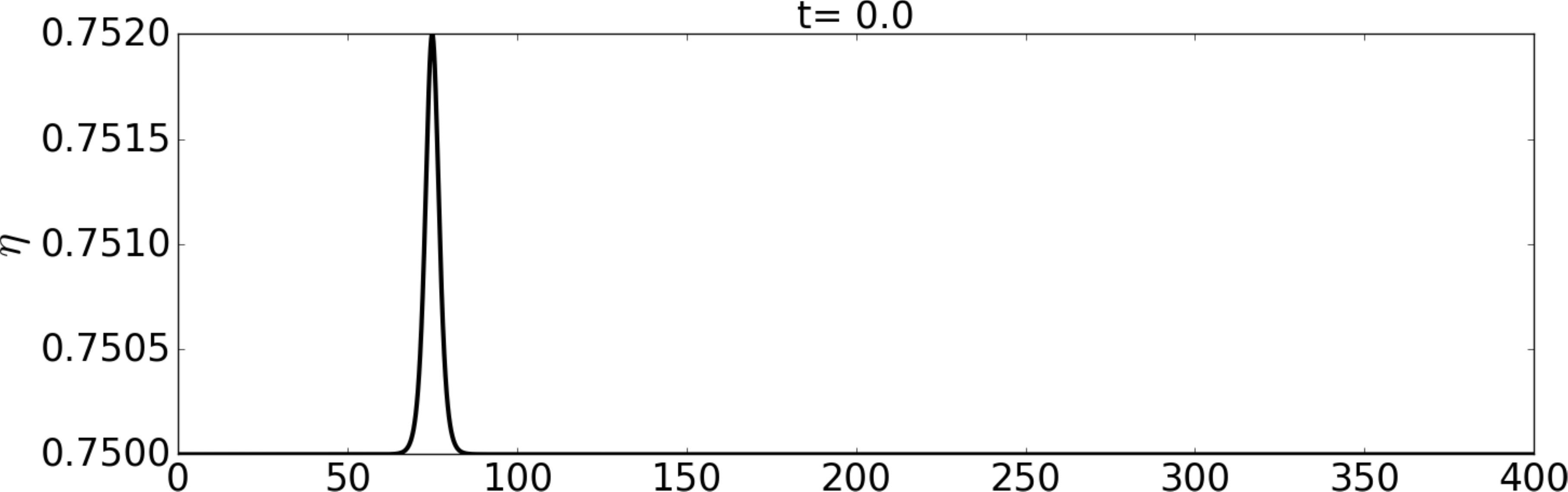}
      \includegraphics[scale=0.145]{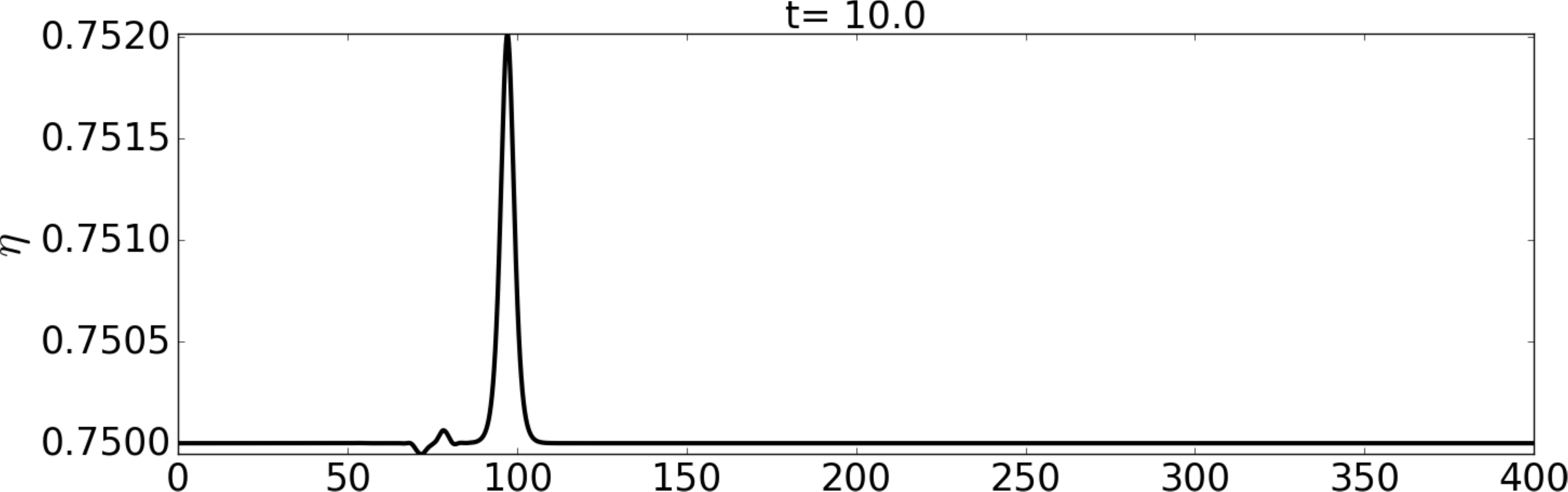}
      \includegraphics[scale=0.145]{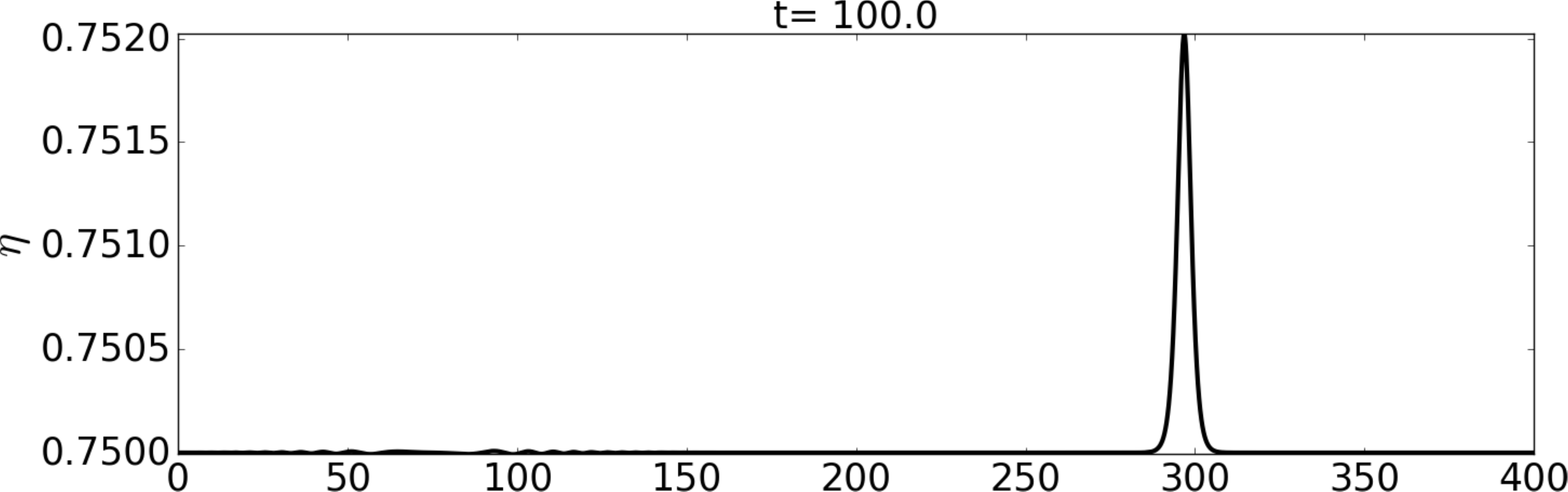}
    }    
  \end{center}
  \caption{
    Solution of the shallow water equations \eqref{shallow_water_equations}
    with periodic bathymetry given by \eqref{bathymetry} with $\bB=0.5$. 
    The initial condition is given by \eqref{sw_kdv_diffracton_shape} with $\gamma=0$
    and \eqref{sw_kdv_diffracton_uv_IC}. We show a slice along $y=-0.25$.
    \label{fig:sw_kdv_diffractons_evolution}}
\end{figure}

\begin{figure}[!h]
  \begin{center}
    \includegraphics[scale=0.145]{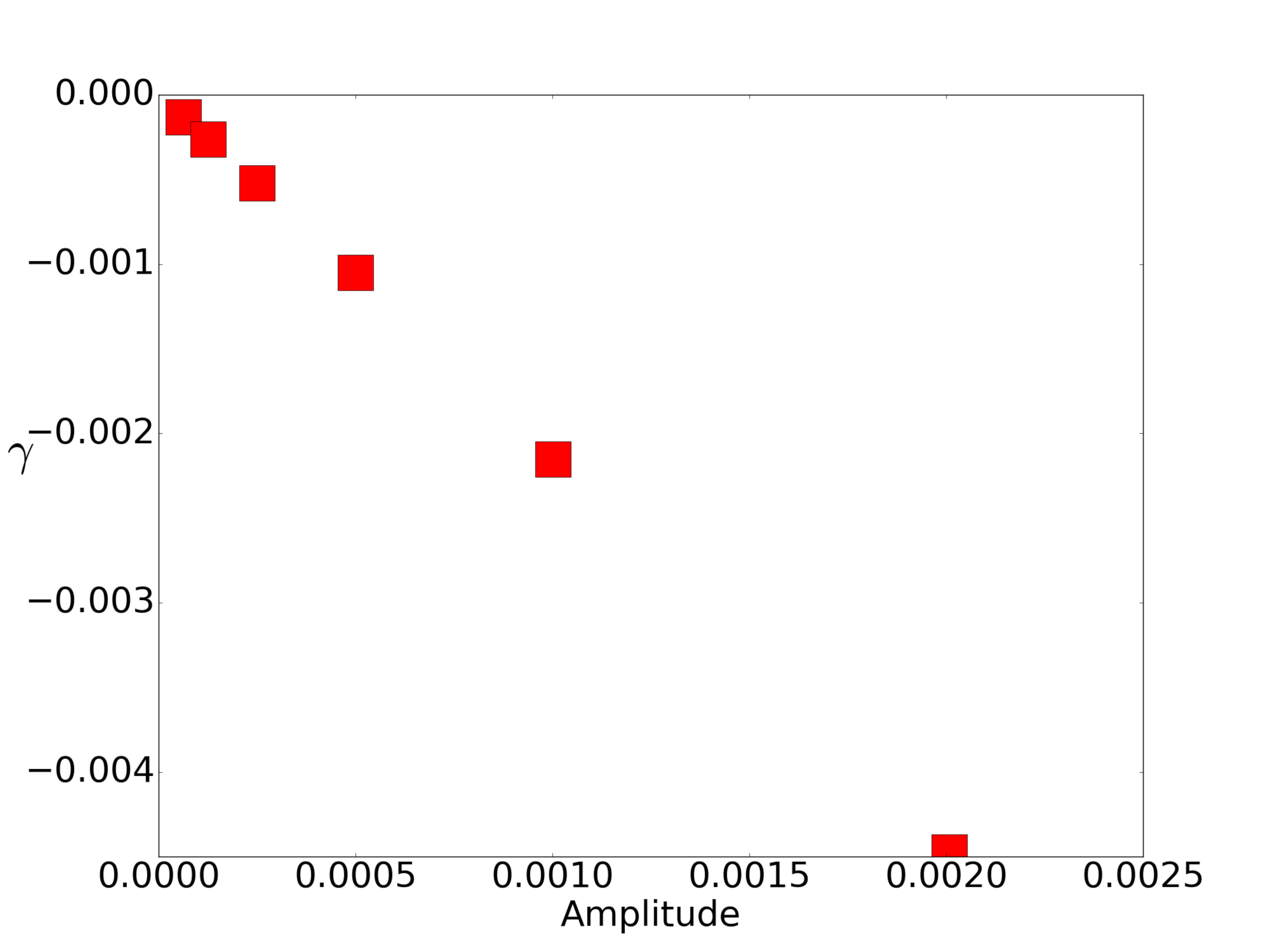}
  \end{center}
  \caption{
    Estimated $\gamma$ such that \eqref{sw_kdv_diffracton_shape} is the closest
    (in a least squares sense) to a given bathymetric solitary wave.
    The bathymetric solitary waves correspond to the solution of the shallow water equations \eqref{shallow_water_equations}
    (at $t=100$) with periodic bathymetry given by \eqref{bathymetry} with $\bB=0.5$. 
    The initial condition is given by \eqref{sw_kdv_diffracton_shape} with $\sigma=\sigma(0)$,
    \eqref{sw_kdv_diffracton_uv_IC} and amplitude $A_m$ given by \eqref{sw_kdv_diffracton_A_IC}. 
    \label{fig:sw_kdv_diffractons_gammas}}
\end{figure}

We now look for a traveling wave solution of \eqref{sw_KdV} by substituting into
that equation the ansatz $\eta=\bMwl f(x-Vt)$, where
$V$ is the speed of the traveling wave.
By doing so, we obtain an ODE that we can integrate twice to get
that the shape of weakly nonlinear bathymetric solitary waves is given by
\begin{subequations}\label{sw_kdv_diffracton}
\begin{align}\label{sw_kdv_diffracton_shape}
  \eta-\mwl = A_m\sech^2\left[\left(\frac{A_m}{8\sigma(\gamma)\bMwl}\right)^{1/2}(x-Vt)\right],
\end{align}
where $A_m$ is the amplitude of the solitary wave and
\begin{align}\label{sw_kdv_diffracton_spee}
  V=\sqrt{g\bMwl}\left(1+\frac{A_m}{2\bMwl}\right)
\end{align}
\end{subequations}
is its speed.
Let us now estimate the correction value $\gamma$ in \eqref{sw_KdV_dispCoeff}.
To do this we use \eqref{sw_kdv_diffracton_shape} with $\gamma=0$ to generate multiple initial conditions for the 
shallow water equations \eqref{shallow_water_equations} with bathymetry given by \eqref{bathymetry} with $\bB=0.5$.
The initial condition for the velocity is
\begin{align}\label{sw_kdv_diffracton_uv_IC}
  u(x,y,0)=2\sqrt{g(\eta(x,y,0)-b_m)}-2\sqrt{g\bMwl}, \qquad v(x,y,0)=0,
\end{align}
where $b_m=\frac{1}{2}\bB$.
Let us use six initial conditions with
\begin{align}\label{sw_kdv_diffracton_A_IC}
  A_m=6.25\times 10^{-5}, ~~1.25\times 10^{-4}, ~~2.5\times 10^{-4}, ~~5\times 10^{-4}, ~~1\times 10^{-3} \text{ and } ~2\times 10^{-3}.
\end{align}
Then we propagate the waves until a final time of $t=100$.
For each simulation, the initial condition quickly evolves into a solitary wave
after some mass is left behind. Initial conditions that are close to being
solitary waves undergo only small changes.
In Figure \ref{fig:sw_kdv_diffractons_evolution}, we show the evolution of two of these waves
(using $A_m=6.25\times 10^{-5}$ and $A_m=2\times 10^{-3}$).
Finally, for each simulation, we isolate the solitary wave at $t=100$ and compute
$\gamma$ such that \eqref{sw_kdv_diffracton_shape} is the closest (in a least squares sense) to the
corresponding isolated solitary wave.
In Figure \ref{fig:sw_kdv_diffractons_gammas} we plot the value of $\gamma$ as a function of amplitude.
It is clear that $\gamma\approx 0$ for arbitrarily small-amplitude waves.

Finally, we compare the solution of \eqref{sw_KdV} with $\gamma=0$ versus the 
numerical solution of the shallow water equations \eqref{shallow_water_equations}. 
We consider the same two scenarios shown in Figure \ref{fig:peregrine} from \S\ref{sec:peregrine}. 
We solve \eqref{sw_KdV} using a Fourier pseudospectral collocation method.

\begin{figure}[!h]
  \centering
  \subfloat[
    Simulation of the same scenario as in Figure \ref{fig:peregrine_regime1},
    this time comparing \eqref{sw_KdV} with the shallow water equations \eqref{shallow_water_equations}.
    We use \eqref{bathymetry} with $\bB=0.01$ and \eqref{eta0} with $\mwl=0.015$ and $\epsilon=0.001$.
    \label{fig:swes_vs_sw_kdv_diffractons_regime1}]{
    \begin{tabular}{ccc}
      \includegraphics[scale=0.152]{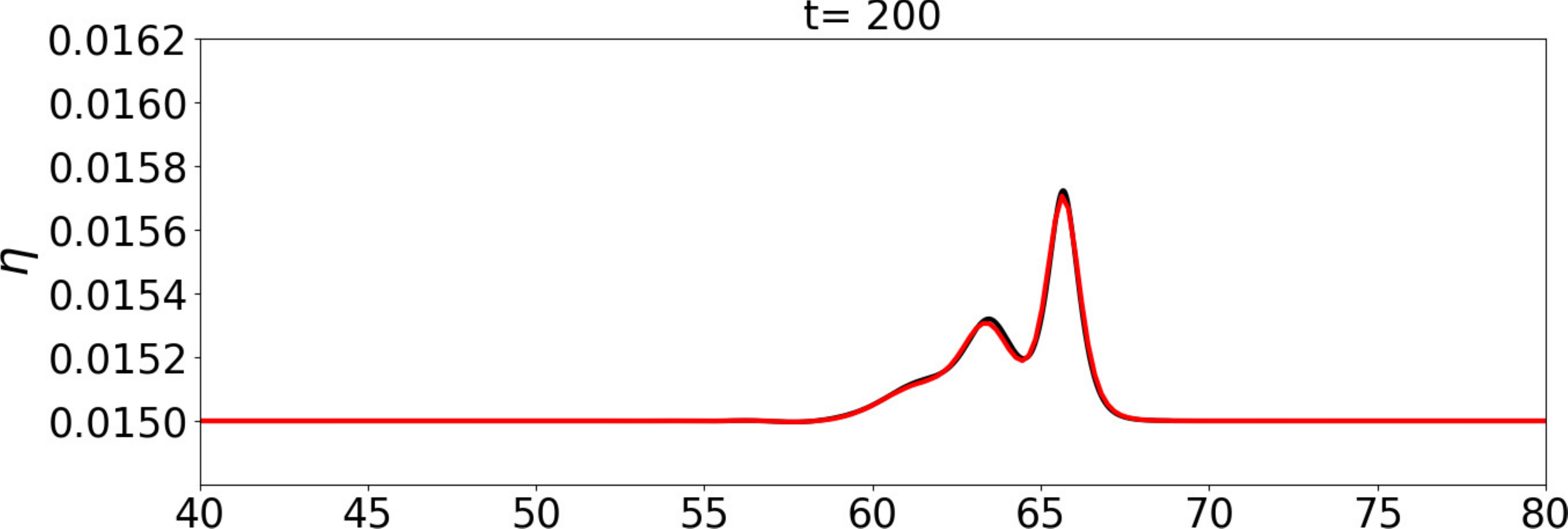} &
      \includegraphics[scale=0.152]{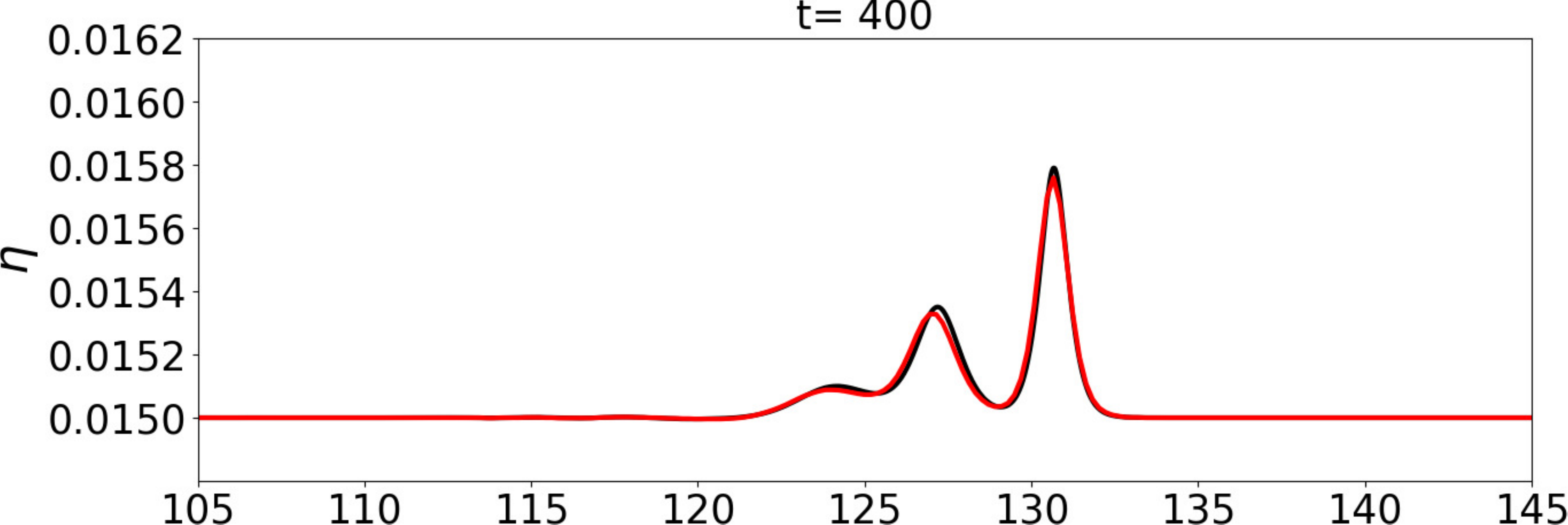} & 
      \includegraphics[scale=0.152]{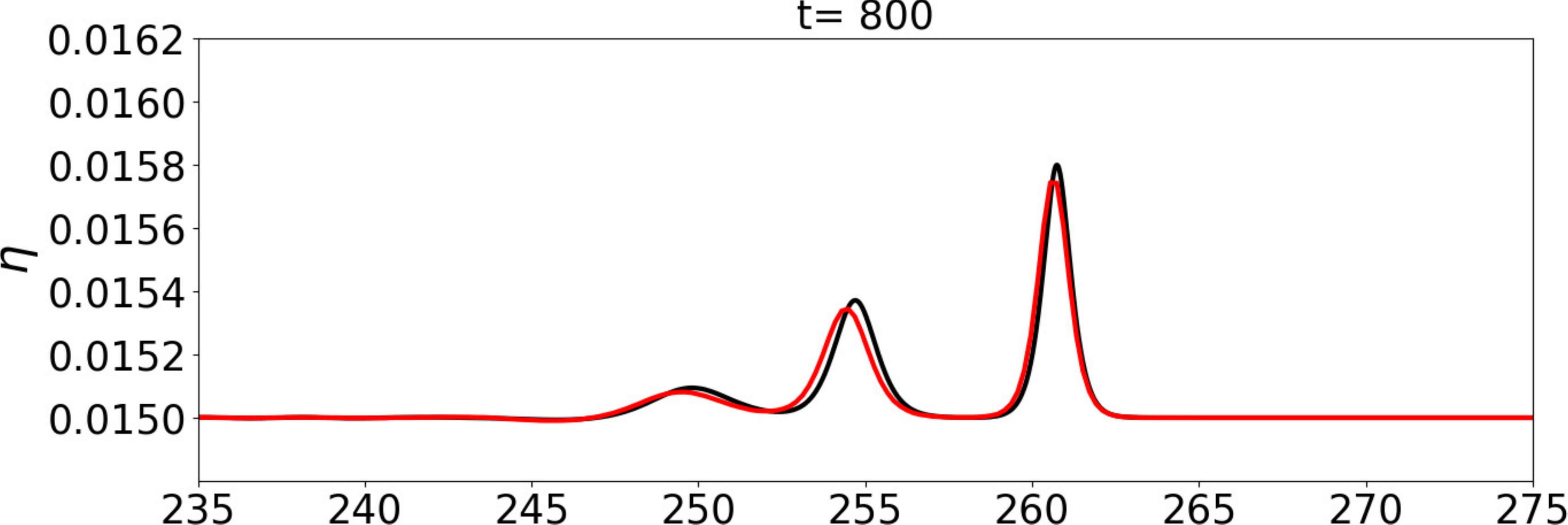} 
    \end{tabular}
  }
  
  \subfloat[
    Simulation of the same scenario as in Figure \ref{fig:peregrine_regime2},
    this time comparing \eqref{sw_KdV} with the shallow water equations \eqref{shallow_water_equations}.
    We use \eqref{bathymetry} with $\bB=0.5$ and \eqref{eta0} with $\mwl=0.75$ and $\epsilon=0.05$.
    \label{fig:swes_vs_sw_kdv_diffractons_regime2}]{
    \begin{tabular}{ccc}
      \includegraphics[scale=0.14]{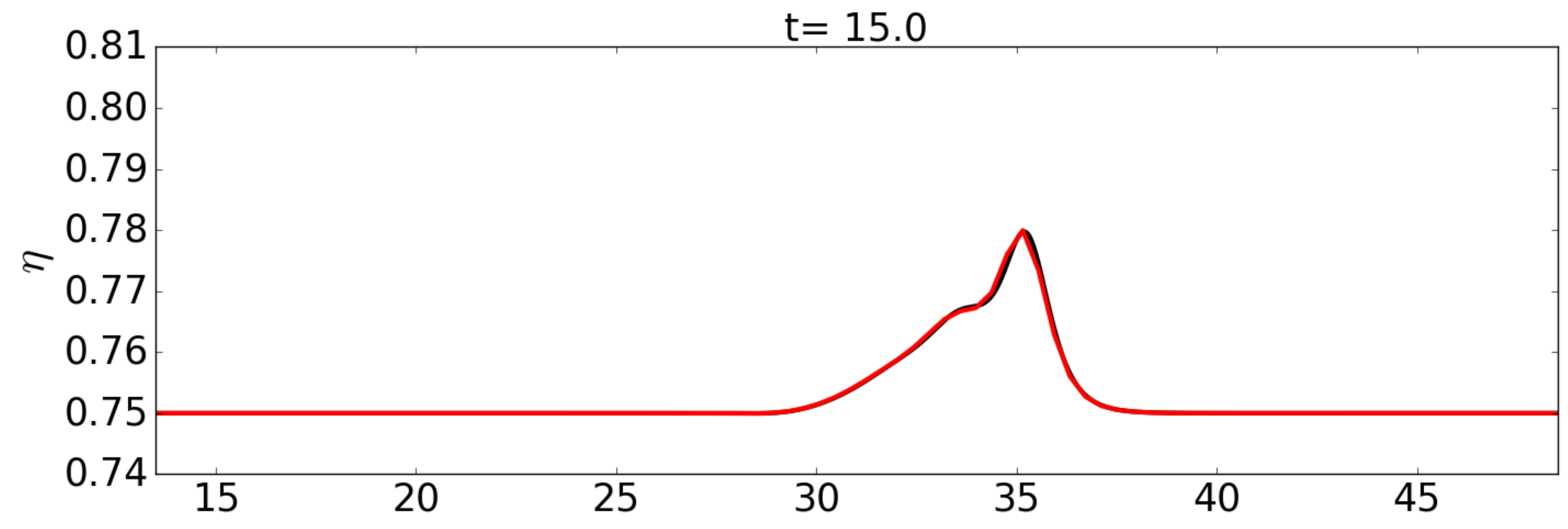} &
      \includegraphics[scale=0.14]{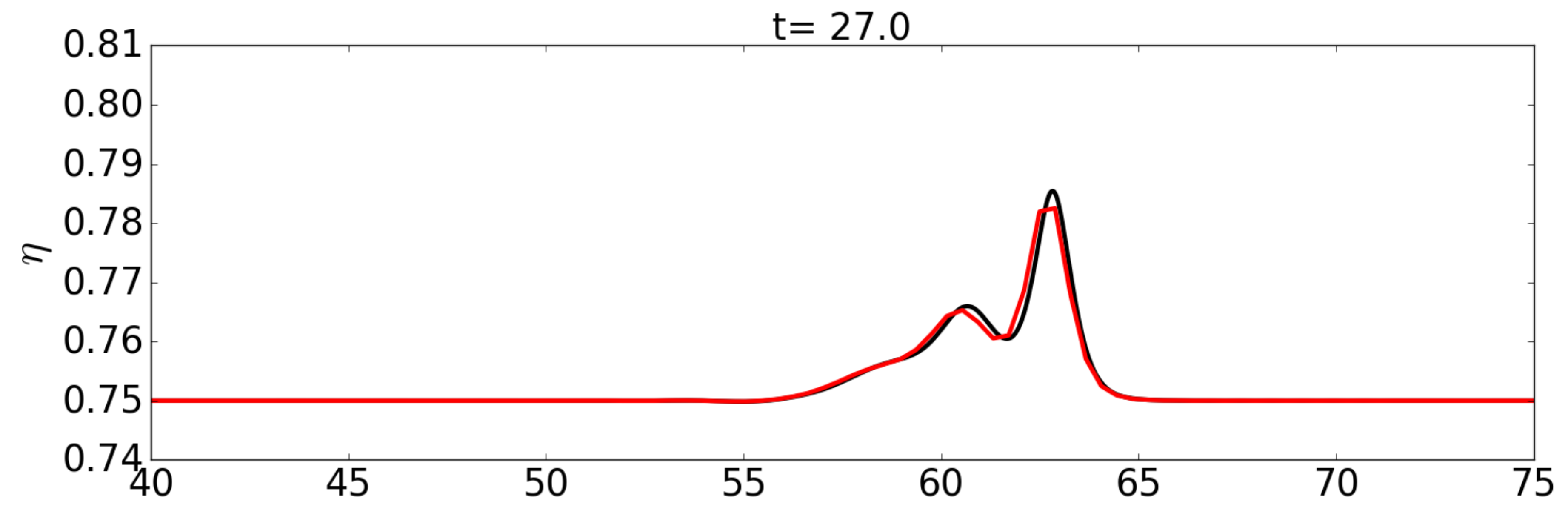} & 
      \includegraphics[scale=0.14]{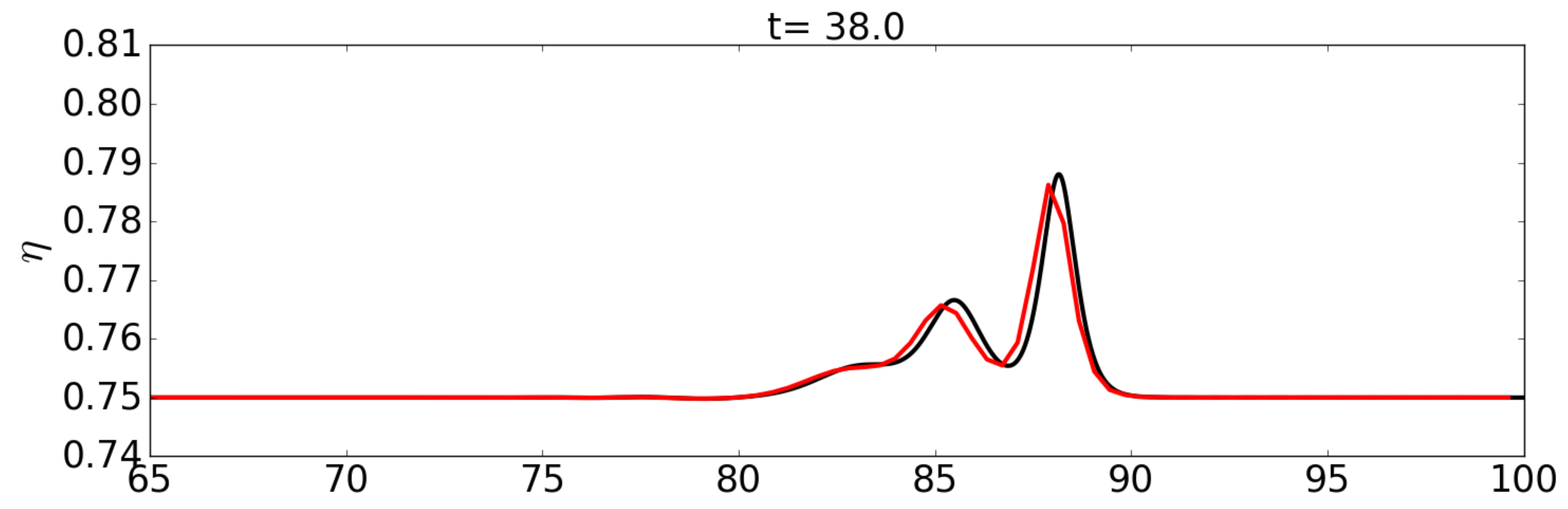} 
      \\
      \includegraphics[scale=0.14]{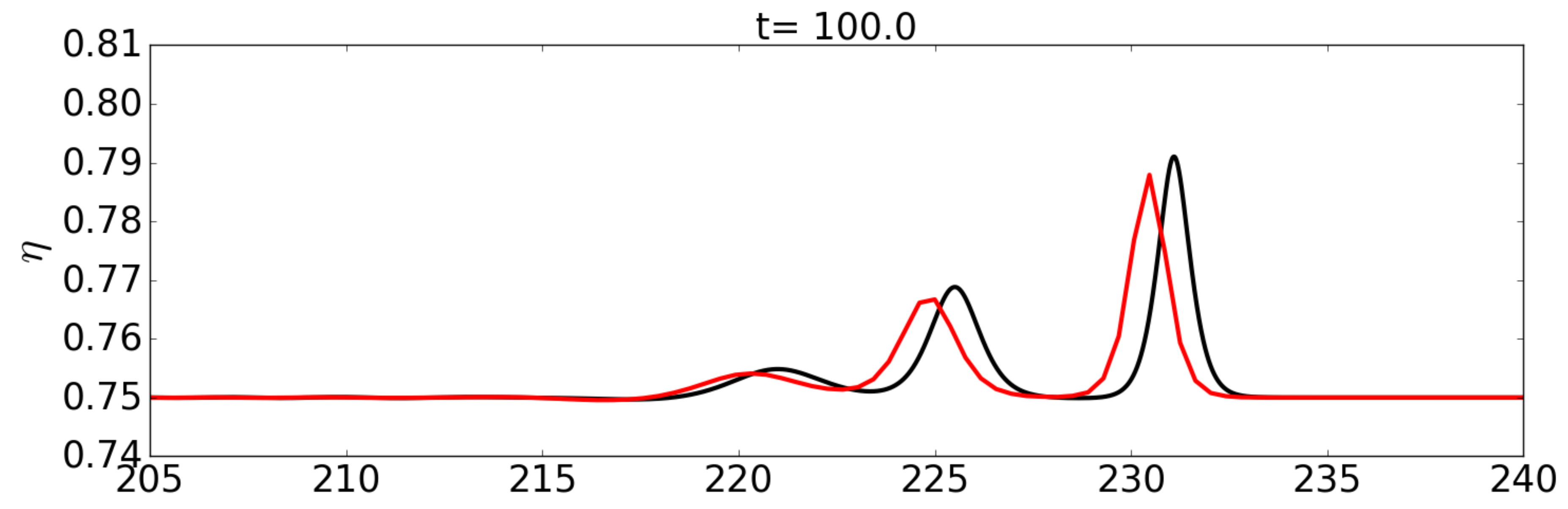} &
      \includegraphics[scale=0.14]{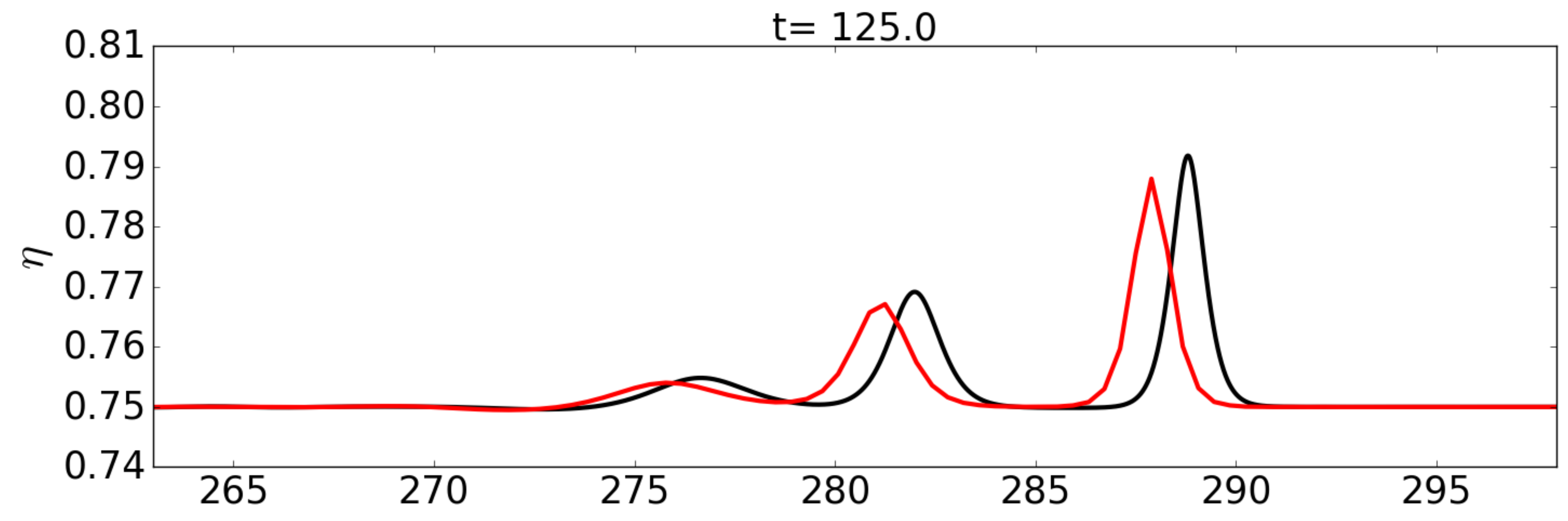} &
      \includegraphics[scale=0.14]{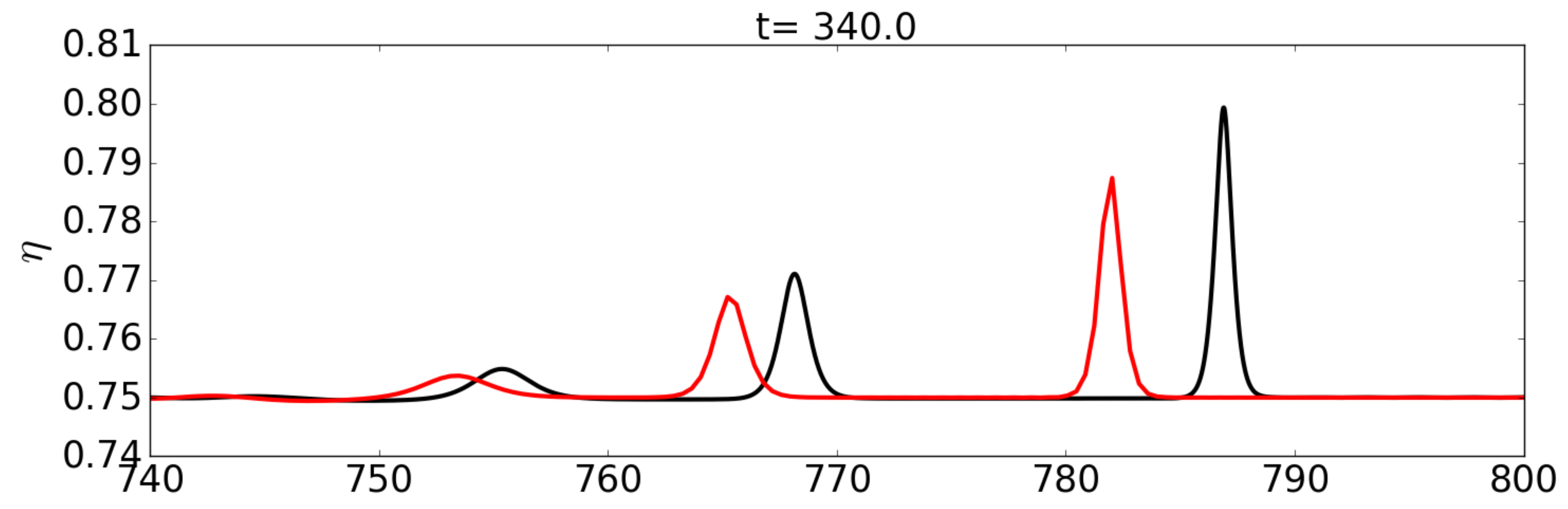}
    \end{tabular}
  }
  
  \subfloat[The same as Figure \ref{fig:swes_vs_sw_kdv_diffractons_regime2}, but
    with an initial wave that is twice as tall ($\epsilon=0.1$). \label{fig:swes_vs_sw_kdv_diffractons_v2}]{
    \begin{tabular}{ccc}
      \includegraphics[scale=0.14]{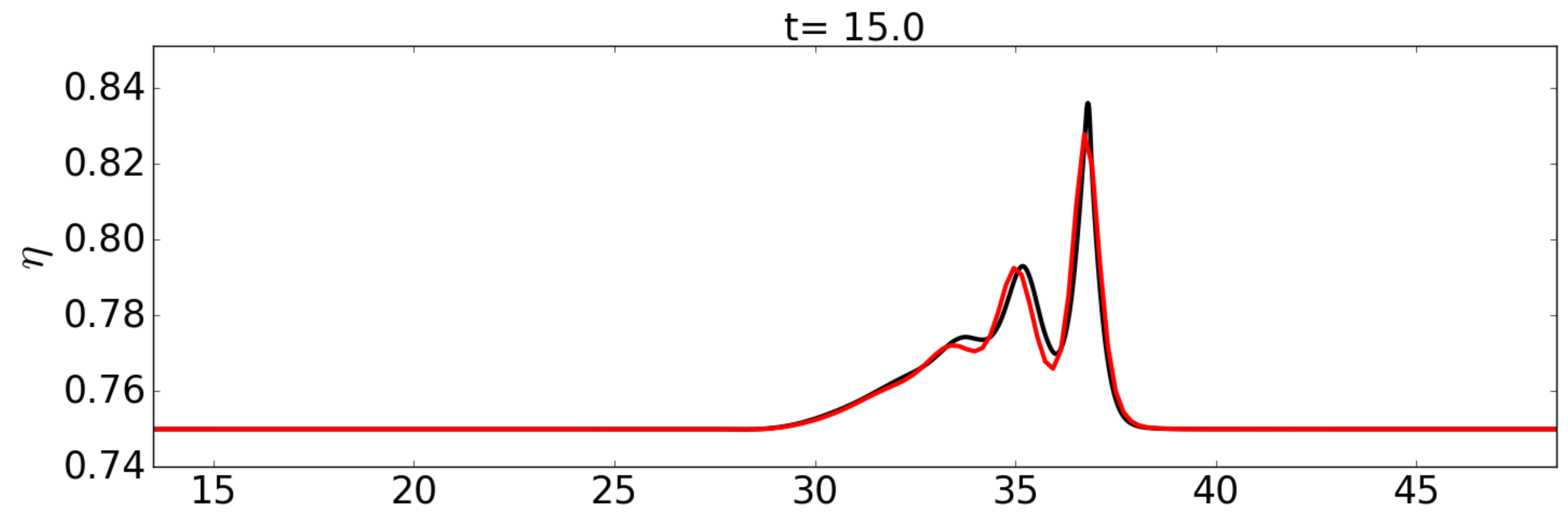} &
      \includegraphics[scale=0.14]{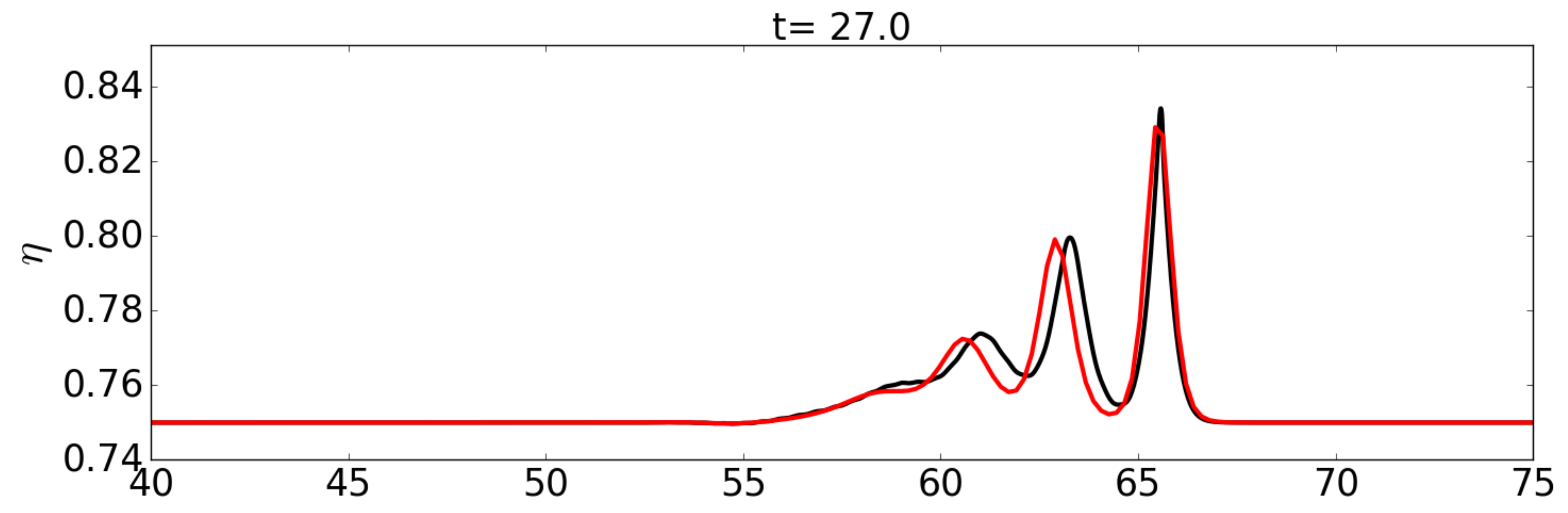} &
      \includegraphics[scale=0.14]{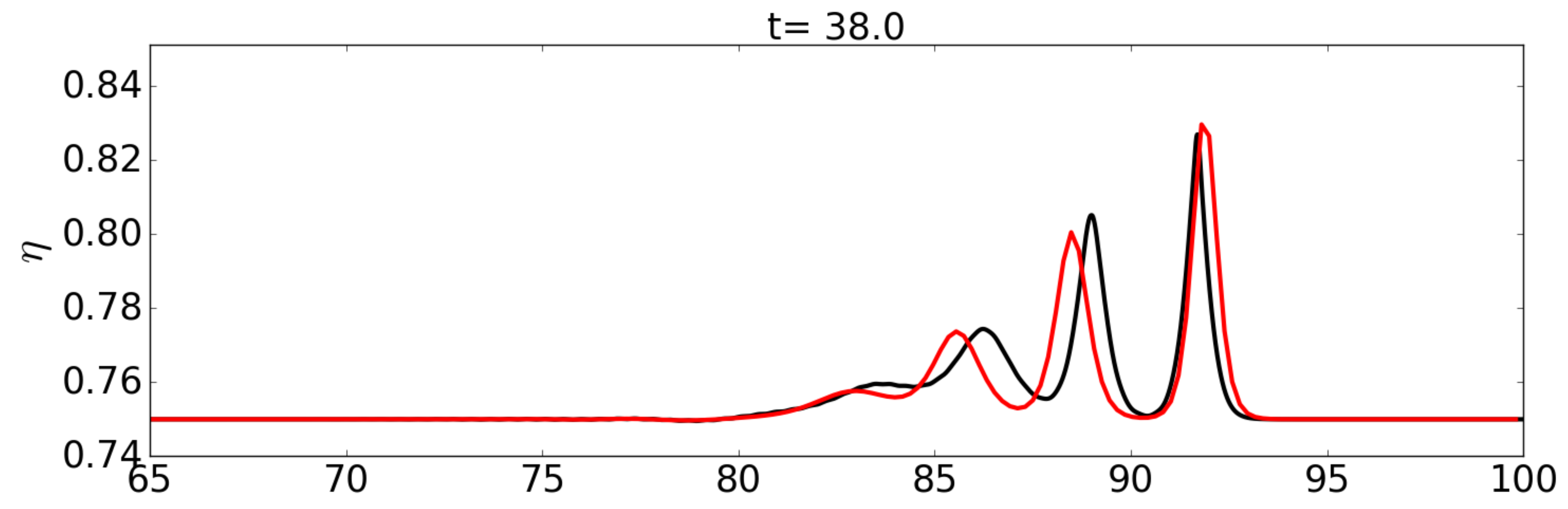}
      \\
      \includegraphics[scale=0.14]{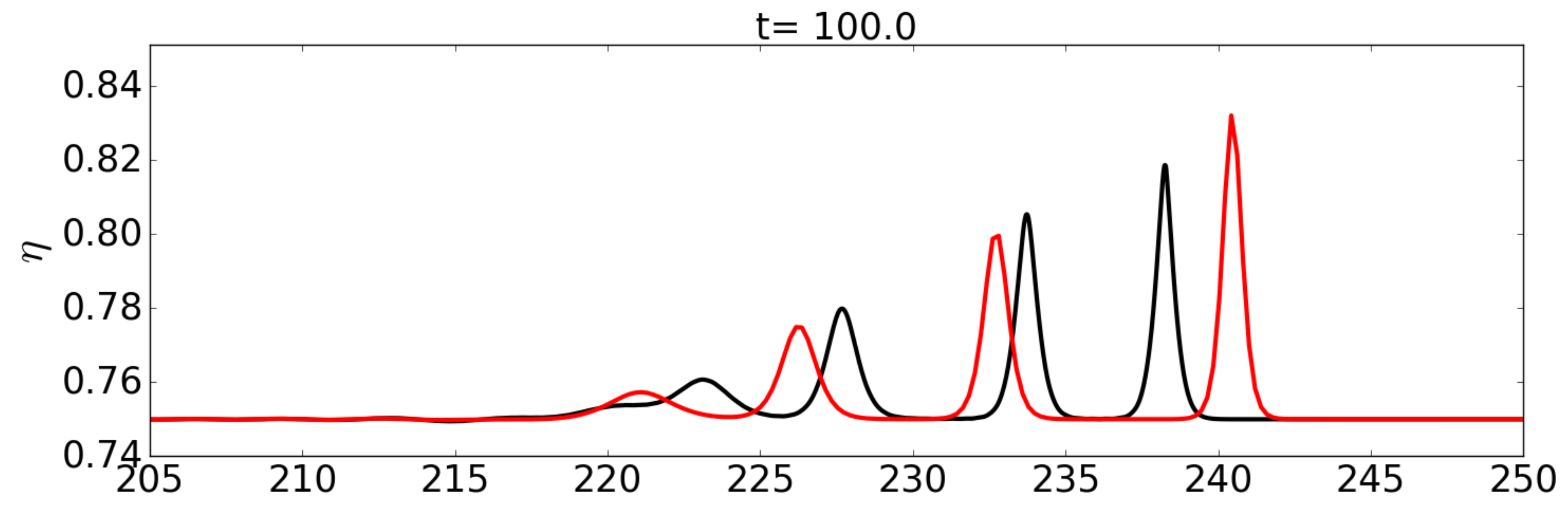} &
      \includegraphics[scale=0.14]{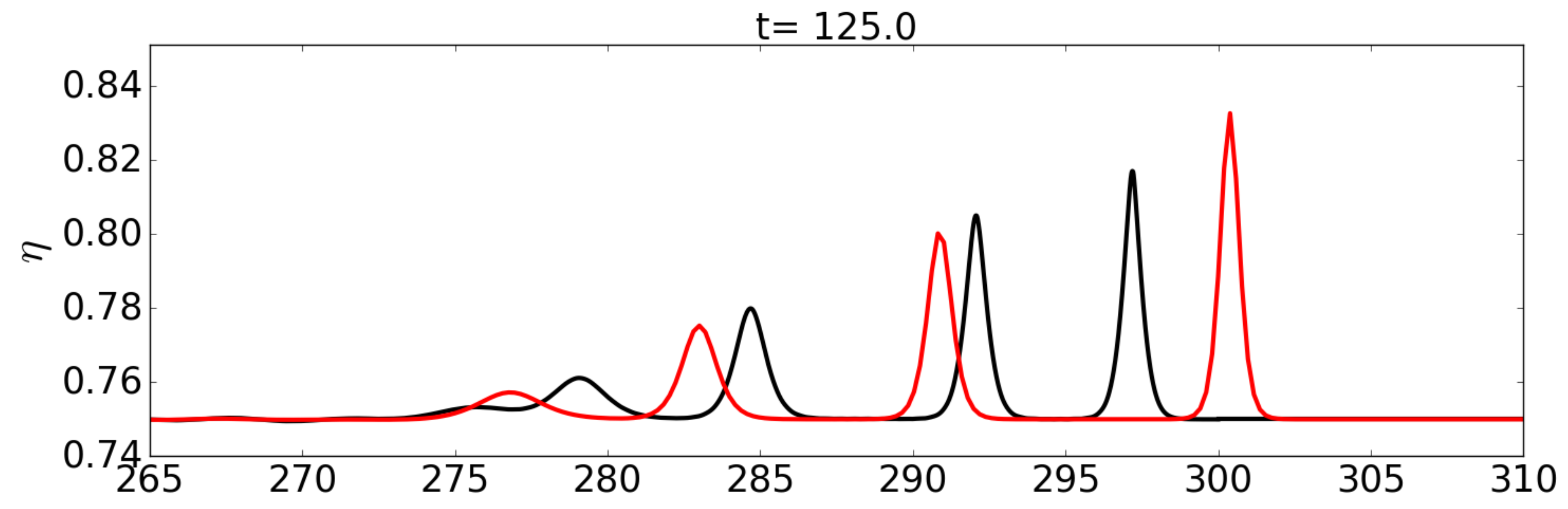} &
      \includegraphics[scale=0.14]{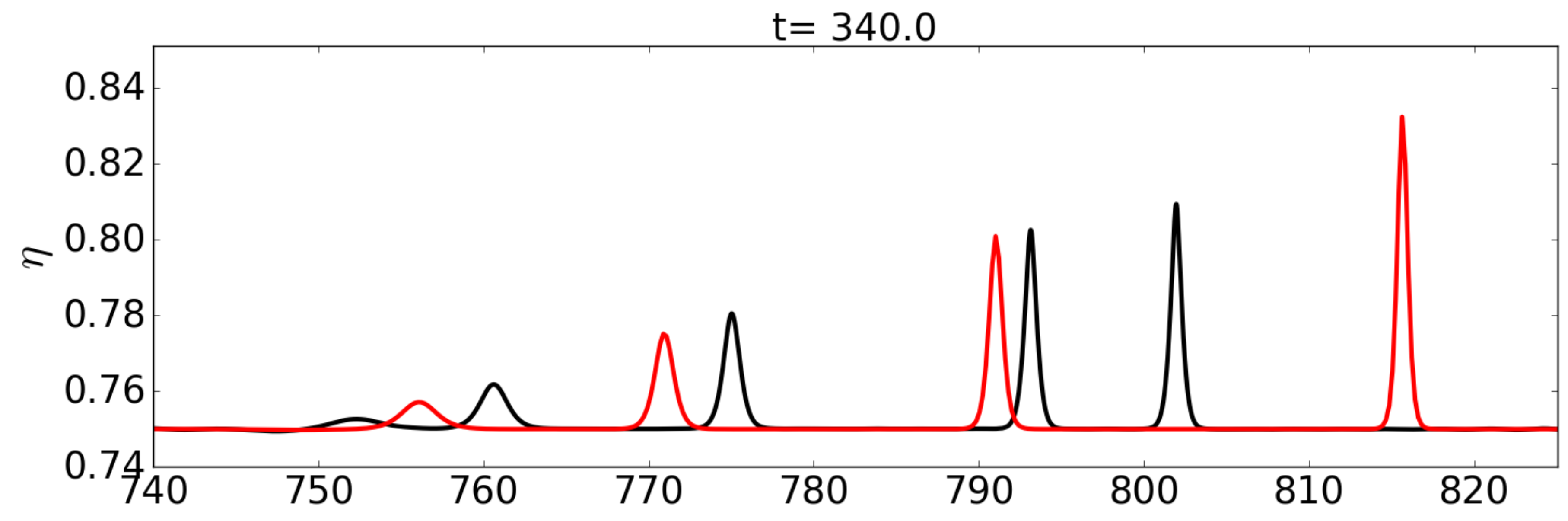}
    \end{tabular}
  }
  \caption{
    In red, solution of the KdV-type equation \eqref{sw_KdV}.
    In black, $y$-averaged solution of the shallow water equations \eqref{shallow_water_equations} with periodic bathymetry.
    The bathymetry is given by \eqref{bathymetry} with $\bB$ as indicated in the subfigures. 
    The initial condition is given by \eqref{eta0} with $\alpha=2$ and $\mwl$ and $\epsilon$ as 
    indicated in the subfigures.
    \label{fig:swes_vs_sw_kdv_diffractons}}
\end{figure}

First we consider the same scenario as in Figure \ref{fig:peregrine_regime1};
the bathymetry is given by \eqref{bathymetry} with $\bB=0.01$ and the initial condition 
is given by \eqref{eta0} with $\alpha=2$, $\mwl=0.015$ and $\epsilon=0.001$. 
The results are shown in Figure \ref{fig:swes_vs_sw_kdv_diffractons_regime1}. 
We see even closer agreement between the two models than in Figure \ref{fig:peregrine_regime1}.

Next we consider the same scenario as in Figure \ref{fig:peregrine_regime2}, 
which is also used in Figure \ref{fig:three_diffractons}. 
In this case, the bathymetry is given by \eqref{bathymetry} with $\bB=0.5$ and 
the initial condition is given by \eqref{eta0} with $\alpha=2$, $\mwl=0.75$ and $\epsilon=0.05$. 
The results are shown in Figure \ref{fig:swes_vs_sw_kdv_diffractons_regime2}.
Recall that for Peregrine's model (refer back to Figure
\ref{fig:peregrine_regime2}) in this case the dispersion inherently present in
water waves and bathymetric dispersion are both important.  But the shallow
water equations and the model \eqref{sw_KdV} both account for only bathymetric
dispersion, so we see much better agreement here.

Finally, in Figure \ref{fig:swes_vs_sw_kdv_diffractons_v2},
we consider the last scenario but with an initial wave that is twice as tall.
We see that the agreement between the models is worse (and the agreement with
\eqref{peregrine} would be even worse).  Both models \eqref{peregrine} and \eqref{sw_KdV}
include only a linear dispersive term, whereas
in \S\ref{sec:speed-amplitude} we 
observed that the speed-amplitude relationship of bathymetric solitary waves is somewhat
nonlinear.  Furthermore, for very large initial data, the shallow water solution will contain
shocks \cite{ketcheson2020effRH}, which cannot be represented by the models \eqref{peregrine} or \eqref{sw_KdV}.
We conclude that for sufficiently large initial data and long times, neither of these models
will remain close to the shallow water solution.

\section{Comparison of dispersive effects}\label{sec:about_dispersive_effects}
In this work we have used theoretical and numerical tools to analyze a new
class of solitary waves.  
Our objective in the remainder of the paper is to determine the feasibility
of observing them in experiments.
In this section we compare bathymetric dispersion with that inherently present in water wave models. 
To do that, we compare our model \eqref{sw_KdV} with KdV,
and determine conditions under which bathymetric dispersion should be
much stronger than the dispersion in KdV. We refer to the dispersion present in the KdV
equation as `KdV dispersion'. 

\begin{figure}[!h]
  \centering
  \subfloat[Regime considered in Figure \ref{fig:peregrine_regime2}. Here, $\mwl=0.75$.
    \label{fig:dispCoeff_eta_0p75}]{\includegraphics[scale=0.32]{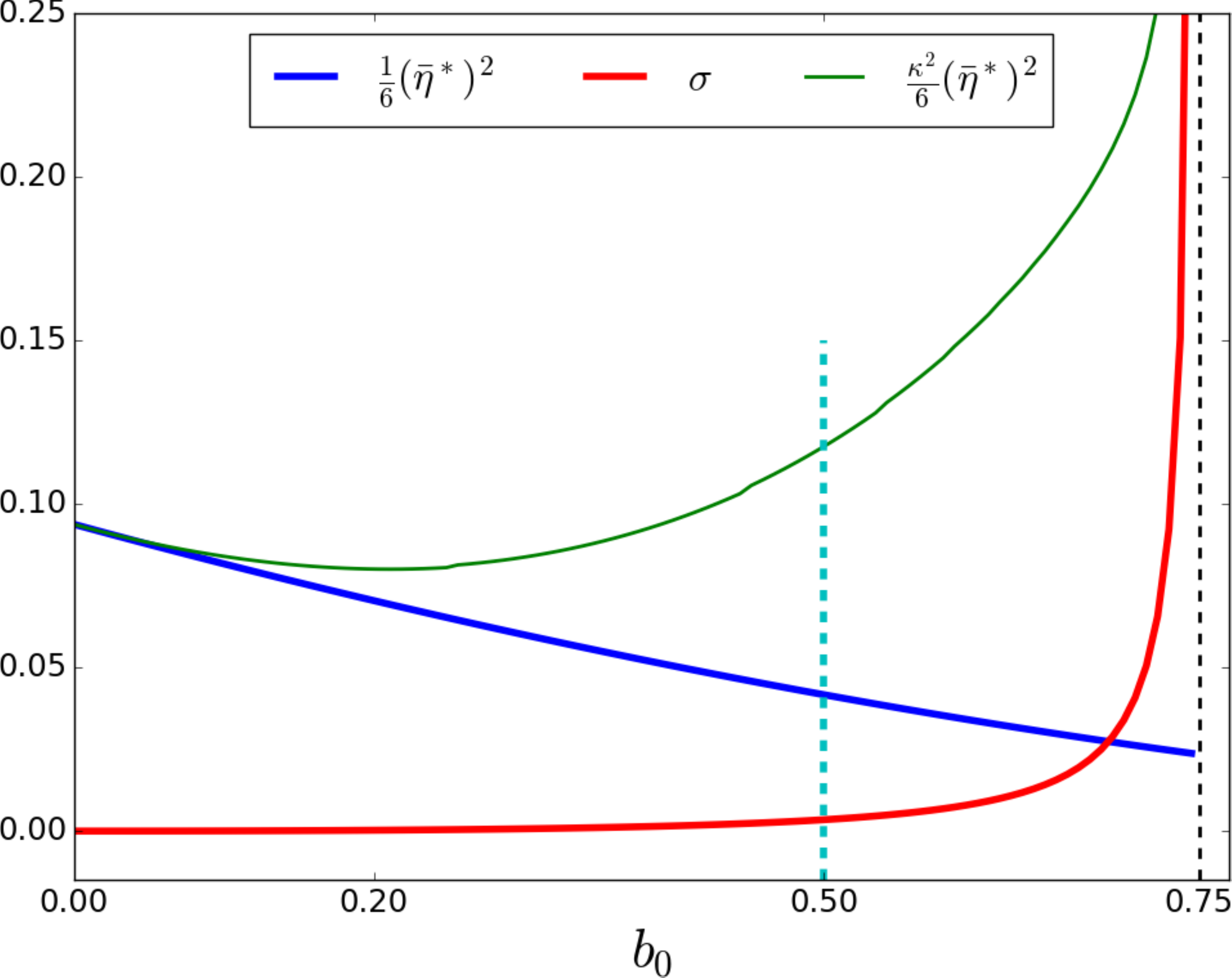}} \qquad
  \subfloat[Regime considered in Figure \ref{fig:peregrine_regime1}. Here, $\mwl=0.015$.
    \label{fig:dispCoeff_eta_0p015}]{\includegraphics[scale=0.32]{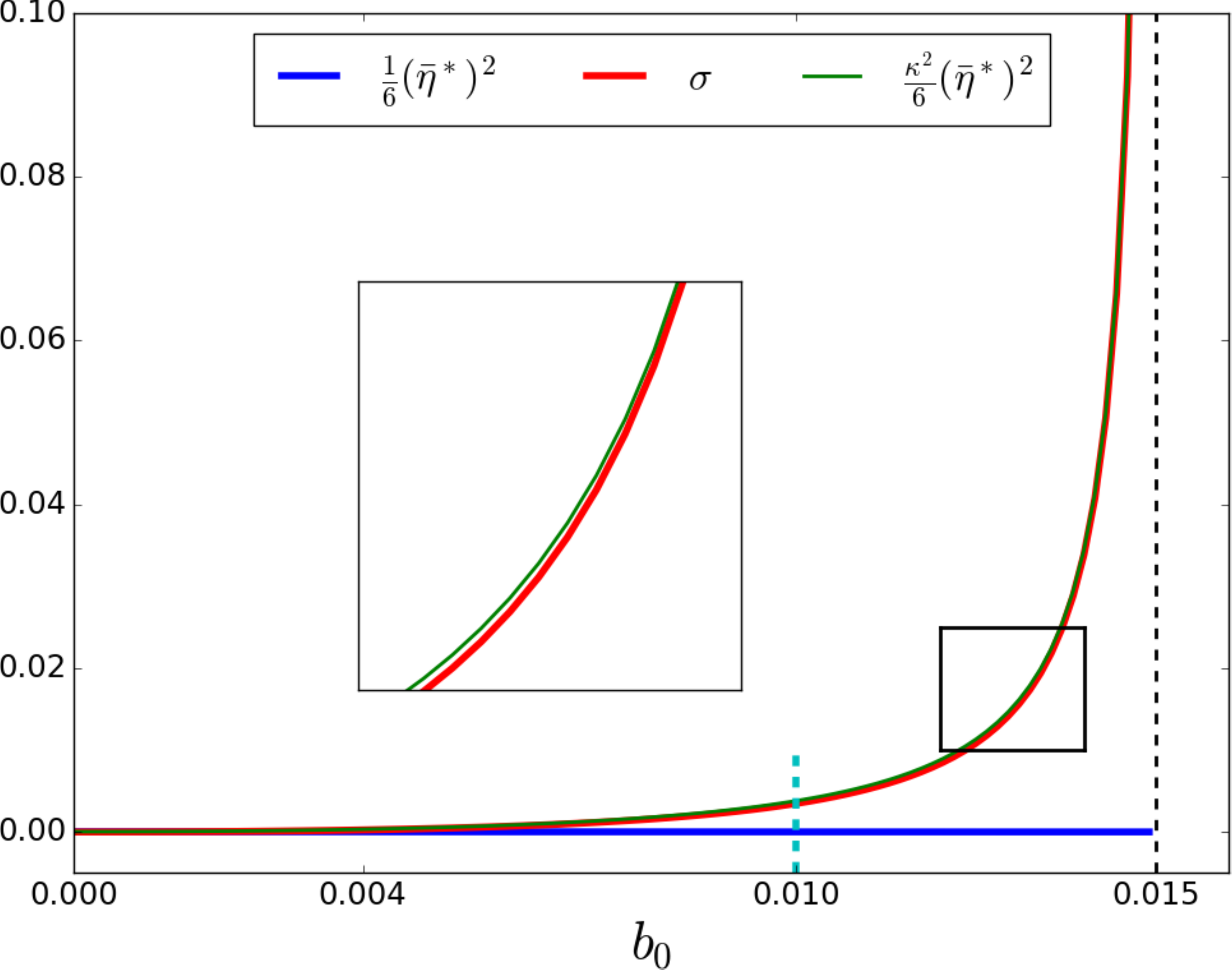}}
  \caption{Coefficients from the $\mathcal{O}(k^3)$ terms in the dispersion relations \eqref{dispRelations},
    and that of Peregrine's model \eqref{peregrine}.
    The blue and red curves correspond to KdV equation \eqref{KdV}
    and our model \eqref{sw_KdV}, respectively.
    The dispersive coefficient of Peregrine's model is plotted in green.
    \label{fig:dispCoeff}}
\end{figure}

Consider a flat channel and a weakly nonlinear regime in
which the (KdV) equation \cite{korteweg1895xli} is applicable.
The right-going KdV equation is given by
\begin{align}\label{KdV}
  \eta_t+\sqrt{g\mwl}\eta_x+\frac{3}{2}\sqrt{\frac{g}{\mwl}}(\eta-\mwl)\eta_x+\frac{1}{6}(\mwl)^2\sqrt{g\mwl}\eta_{xxx}=0,
\end{align}
where as usual $\eta(x,t)$ is the surface elevation and
$\mwl$ is the undisturbed surface elevation.
The KdV model has been validated experimentally, for instance in \cite{hammack1974korteweg}, wherein
water waves (over flat bathymetry) were observed to form solitary waves of the kind
predicted by \eqref{KdV} after propagating over a relatively long distance.
To explore the qualitative difference between bathymetric and KdV dispersion,
we study the dispersion relation of
\eqref{sw_KdV} and \eqref{KdV}, which are given by
\begin{subequations}\label{dispRelations}
  \begin{align}
    \omega_\text{Hom} &
    = \sqrt{g\bMwl}k\left[1 - \sigma(0) k^2 \right], \label{disp2} \\
    \omega_\text{KdV} &
    = \sqrt{g\mwl}k\left[1 -\frac{1}{6}(\mwl k)^2 \right], \label{dispRelation_KdV} 
  \end{align}
\end{subequations}
respectively. Note that \eqref{disp2} can also be obtained directly from \eqref{homogenized_linear_system}.
Here $\sigma(0)$ is given by \eqref{sw_KdV_dispCoeff} 
(with $\gamma=0$ for small-amplitude waves) and
$\bMwl$ is the average undisturbed depth for the non-rectangular channel.
It is natural to take the depth of the flat channel $\mwl$ equal to $\bMwl$,
in which case the $\mathcal{O}(k)$ terms for the two models agree.
The dominant dispersive effect arises from the term of $\mathcal{O}(k^3)$
in \eqref{dispRelations}.
In Figure \ref{fig:dispCoeff_eta_0p75} we compare the coefficient of this term in the two models,
taking $\mwl=0.75$ and a range of bathymetry profiles given by \eqref{bathymetry} with $\bB \in [0,0.75)$.
In the figure, the blue and the red plots are $\frac{(\bMwl)^2}{6}$ and $\sigma(0)$ respectively.
As one might expect, when the value of 
$\bB$ is small, the bathymetric
dispersion is small compared to the KdV dispersion.
On the other hand, if $\bB$ 
is close to the mean water level, bathymetric dispersion is stronger and
can be of the same order or much larger than KdV dispersion.
Thus, at least for small-amplitude, long-wavelength waves,
the two dispersive effects can be made comparable or either one can be made
dominant depending on the parameter $\bB$.

As discussed in Section \ref{sec:peregrine}, neither the KdV equation \eqref{KdV}
nor our model \eqref{sw_KdV} include both types of dispersive effects.
Peregrine's model \eqref{peregrine} includes both
sources of dispersion. 
In Figure \ref{fig:dispCoeff_eta_0p75}, we also plot (in green) the coefficient of dispersion
appearing in \eqref{peregrine}. 
Note that the dispersion predicted by Peregrine's model coincides with
that predicted by KdV and by our model in the limits when $b_0$ is close to
$0$ and $\eta^*$, respectively. This behavior is expected: when $b_0\approx 0$,
the bathymetry is almost flat so the main source of dispersion is that predicted
by KdV. On the other hand, when $b_0\approx \eta^*=0.75$, bathymetric dispersion is dominant;
see \eqref{sw_KdV_dispCoeff} and \eqref{hom_coeffs2}. 
In Figure \ref{fig:dispCoeff_eta_0p75}, we mark (with a dashed cyan line) $b_0=0.5$, which corresponds to the situation studied in
Figure \ref{fig:peregrine_regime2}. In this case, the dispersion in Peregrine's model and bathymetric dispersion
are significantly different, leading to the different solutions depicted in Figure \ref{fig:peregrine_regime2}.

On the other hand, for the situation we considered in Figure \ref{fig:peregrine_regime1} with
$\bB=0.01$ and $\mwl=0.015$, the bathymetric dispersion is much stronger. 
Consequently, the dispersive effects in our model \eqref{sw_KdV}
and Peregrine's model are similar, which explains the agreement in the simulations shown in
Figure \ref{fig:peregrine_regime1}. In Figure \ref{fig:dispCoeff_eta_0p015}, we plot the corresponding dispersive
coefficients of the three models (KdV \eqref{KdV}, our model \eqref{sw_KdV} and Peregrine's model \eqref{peregrine}). 

From equations \eqref{disp2} and \eqref{dispRelation_KdV}, it is possible to 
find a channel configuration that leads to bathymetric dispersion that has the
same magnitude as the dispersive effects appearing in
the classical KdV equation. We get that if 
\begin{align}\label{groove_cond}
  \bB = \frac{8(\bMwl)^2}{\Omega^2}
  \left[-4\bMwl + \sqrt{16(\bMwl)^2+1}\right]
\end{align}
then the shallow water equations \eqref{shallow_water_equations} with bathymetry 
given by \eqref{bathymetry} are a close approximation
(up to $\mathcal{O}(k^3)$) to the classical KdV equation \eqref{KdV}. 

\begin{remark}[About the effect of dissipation]
  As we concluded in this section, either source of dispersion (KdV or bathymetric dispersion) 
  can be made dominant. Since the dissipative effects in the propagation of water waves in flat 
  channels do not prevent the formation of solitons (see for instance \cite{hammack1974korteweg}), 
  it is reasonable to believe that bathymetric solitary waves can also be observed in a physical experiment.
  An important difference between these two scenarios is the presence of friction at the interface
  between the flat sections of the bathymetry. More detailed studies (or physical experiments) are
  needed to determine if this extra source of dissipation might prevent the formation of bathymetric solitary waves.
\end{remark}

\section{Conclusions}\label{sec:conclusions}
We have shown that bathymetric variation in an infinite periodic domain leads
to an effective dispersion of water waves, and have related this to the already-studied
phenomenon of dispersion of waves in non-rectangular channels.  This
dispersion is distinct from the dispersion accounted for in wave models like
KdV, and can on its own lead to solitary wave formation, which we call bathymetric solitary waves, 
even when the dominant behavior would normally be wave breaking.
Weakly nonlinear plane waves in this setting approximately satisfy a KdV-type equation.
This KdV-type model leads to soliton waves that approximate small amplitude bathymetric solitary waves. 
However, it is important to remark that bathymetric solitary waves are truly two-dimensional waves 
that travel in one direction. Bathymetric solitary waves, therefore, behave similar to the solitons 
emerging from the derived KdV-type model \eqref{sw_KdV} only when the amplitude is small enough. 
More strongly nonlinear waves exhibit more pronounced two-dimensional structure, have a nonlinear
speed-amplitude relation and evidently involve nonlinear dispersion.

We have shown in \S\ref{sec:peregrine} and \S\ref{sec:about_dispersive_effects} that the model by
\cite{peregrine1968long} agrees with the KdV equation \eqref{KdV}, which considers the inherently
dispersive nature of water waves, and with our model \eqref{sw_KdV}, which considers bathymetric dispersion,
in certain asymptotic regimes.
In general, however, the combination of these two types of dispersion is non trivial.
In particular, as shown in Figure \ref{fig:dispCoeff}, the coefficient of dispersion in Peregrine's model
\cite{peregrine1968long} is not simply the sum of the coefficients of dispersion in KdV and in our model. 
A proper analysis to identify the range of validity and agreement between Peregrine's model,
KdV and our model requires the solution of the elliptic PDE \eqref{elliptic_pde_peregrine}.
Doing so could also provide insights about how these two types of dispersive effects are combined together. 
This investigation is an area of future work.

\begin{figure}[!h]
  \begin{centering}
    \includegraphics[scale=0.22]{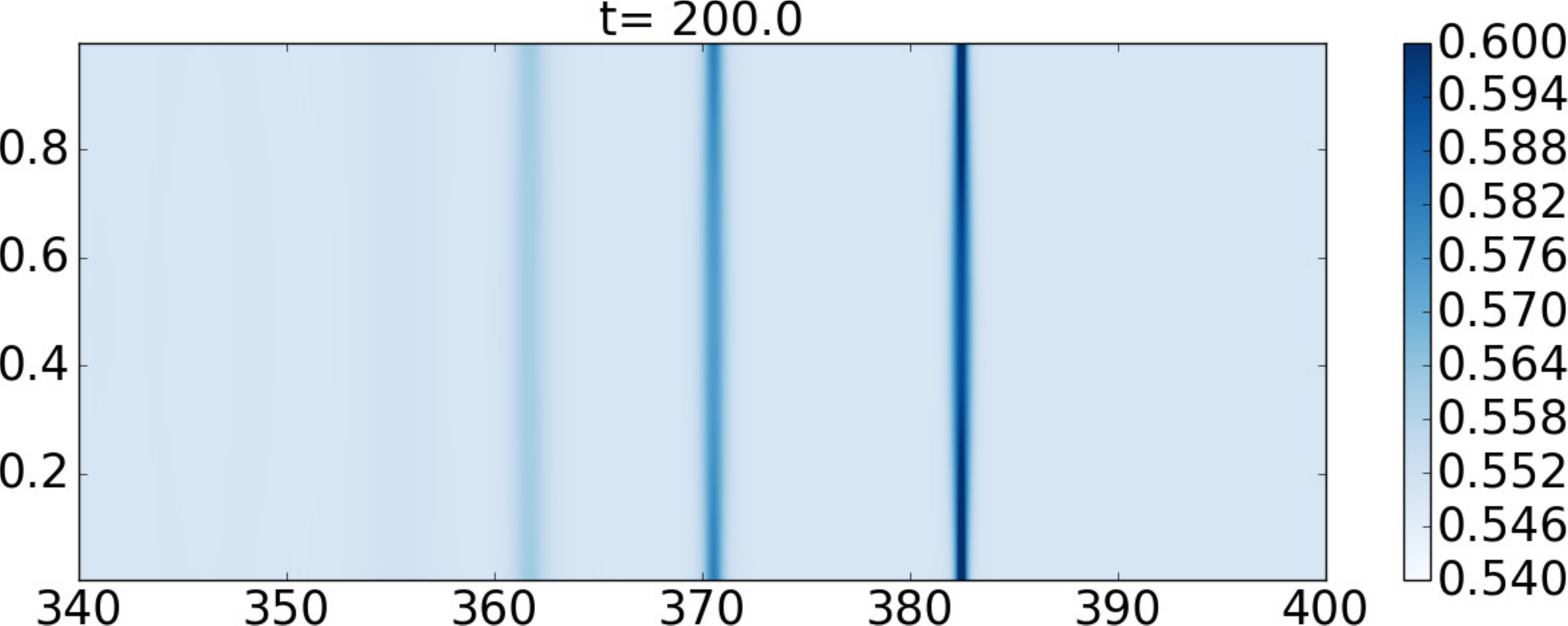}\qquad
    \includegraphics[scale=0.22]{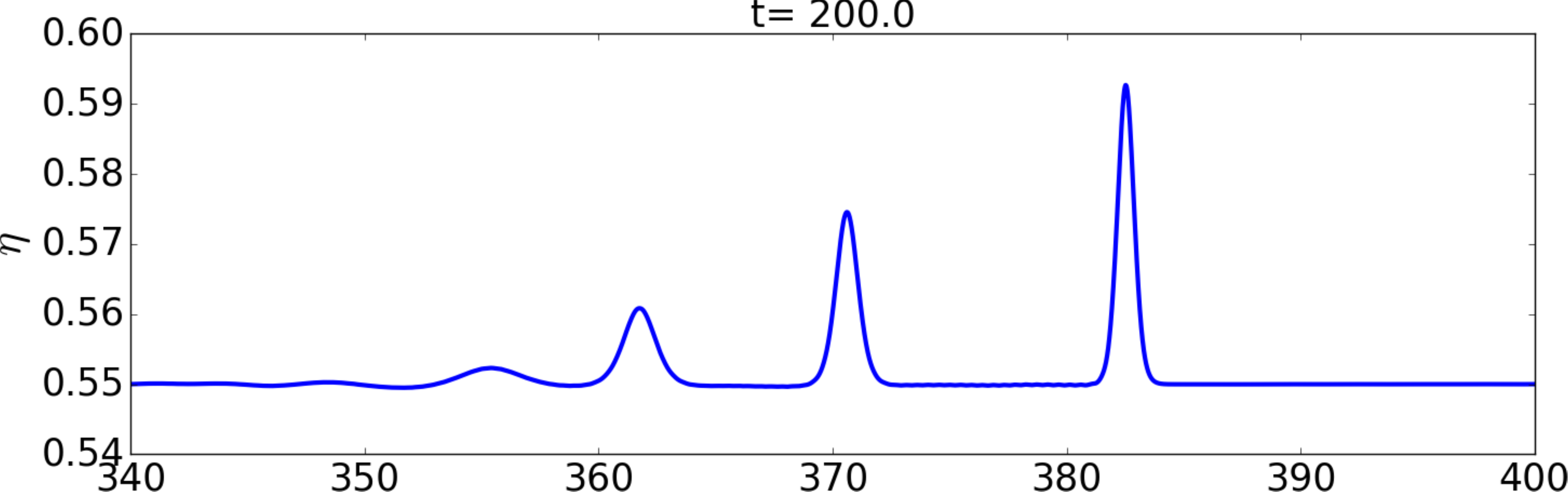}
    \par
  \end{centering}
  \caption{Solitary waves at $t=200$ in a channel like the one shown in Figure \ref{fig:channel}.
    The initial condition is given by \eqref{eta0}
    with $\epsilon=0.05$, $\mwl=0.75$ and $\alpha=2$.
    In the left panel we show the surface plots and in the right panel we show slices along $y=0.5$.
    \label{fig:channel_results}}
\end{figure}

Although we focus on piecewise-constant bathymetry, similar phenomena appear
in computational experiments with other kinds of bathymetry.
To demonstrate this we consider a channel with inclined walls,
like the one shown in Figure \ref{fig:channel}.
A similar problem was studied in \cite{chassagne2019dispersive} where the authors considered
the dispersive Green-Naghdi model \cite{green1976derivation} and showed that the changes in the
bathymetry enhance the dispersive effects.
Here we show that solitary waves arise in this setting also as solutions of the (dispersionless)
shallow water equations \eqref{shallow_water_equations}.
In Figure \ref{fig:channel_results}, we show the solution at $t=200$.
In the left panel we show the surface plot and in the right panel we plot a slice along $y=0.5$.

It is natural to ask whether experimental observation of these waves is feasible.
We have conducted computational experiments (not shown here) using the 3D Navier-Stokes
equations.  Preliminary results indicate that for scenarios like those studied in this work,
wave breaking is almost entirely avoided and an initial pulse breaks into multiple peaks, which
then evolve into solitary waves. Further and more detailed numerical investigation of these waves is 
part of ongoing investigation and will be published elsewhere.

\section{Acknowledgments}
We thank Prof. James Kirby for bringing to our attention the literature on waves in non-rectangular channels,
and the anonymous referees for many suggestions that improved this work.
This work was funded by King Abdullah University of Science and Technology (KAUST) in Thuwal, Saudi Arabia.
For computer time, this research used the resources of the Supercomputing Laboratory at KAUST.

\section*{Declaration of Interests}
The authors declare that they have no known competing financial interests or personal 
relationships that could appear to have influenced the work reported in this paper.

\bibliographystyle{abbrvnat}
\bibliography{shallow_water_diffractons}

\end{document}